\documentclass[12pt, a4paper]{article}

\usepackage[
  top=2.0cm, bottom=2.0cm,
  left=2.0cm, right=2.0cm,
  headheight=14pt
]{geometry}

\usepackage[T1]{fontenc}
\usepackage[utf8]{inputenc}
\usepackage{amsmath}             
\usepackage{newtxtext}          
\usepackage{newtxmath}          
\usepackage[scaled=0.92]{helvet}
\usepackage{microtype}          
\usepackage{textcomp}           

\usepackage{setspace}
\usepackage{bm}                 

\usepackage{booktabs}           
\usepackage{multirow}
\usepackage{array}              
\usepackage{siunitx}
\usepackage{threeparttable}     
\usepackage{tabularx}           
\usepackage{longtable}          
\usepackage{pdflscape}          

\renewcommand{\arraystretch}{1.22}
\usepackage{graphicx}
\usepackage{standalone}
\usepackage{subcaption}
\usepackage{float}
\usepackage{afterpage}          
\usepackage[section]{placeins}  

\usepackage[
  font      = small,            
  labelfont = {bf, sf, small},  
  labelsep  = period,           
  skip      = 6pt,              
  justification = justified,    
  singlelinecheck = false,      
]{caption}
\usepackage[dvipsnames]{xcolor}
\definecolor{linkblue}{HTML}{1A73E8}

\usepackage[
  backend      = biber,
  style        = authoryear,
  maxcitenames = 2,
  mincitenames = 1,
  sortcites    = true,
  natbib       = true,
]{biblatex}
\usepackage{xspace}
\usepackage{enumitem}
\setlist{
  nosep,
  topsep      = 5pt,
  itemsep     = 2pt,
  parsep      = 1pt,
  leftmargin  = 1.8em,
}                               

\usepackage{tikz}
\usetikzlibrary{shapes, arrows.meta, positioning, fit, backgrounds, calc}
\usepackage{pgfplots}
\pgfplotsset{compat=1.18}

\usepackage{titlesec}

\titleformat{\section}
  {\Large\bfseries\sffamily}             
  {\thesection}{0.8em}{}                  
\titleformat{\subsection}
  {\large\bfseries\sffamily}             
  {\thesubsection}{0.7em}{}               
\titleformat{\subsubsection}
  {\normalsize\bfseries\sffamily}        
  {\thesubsubsection}{0.6em}{}            
\titleformat{\paragraph}[runin]
  {\normalsize\bfseries\sffamily}       
  {}{0em}{}                               

\titlespacing*{\section}      {0pt}{22pt plus 4pt minus 2pt}{10pt plus 2pt minus 1pt}
\titlespacing*{\subsection}   {0pt}{16pt plus 3pt minus 2pt}{8pt plus 2pt minus 1pt}
\titlespacing*{\subsubsection}{0pt}{12pt plus 2pt minus 1pt}{6pt plus 1pt}
\titlespacing*{\paragraph}    {0pt}{10pt plus 2pt minus 1pt}{0.6em}

\makeatletter
\def\input@path{{sections/}{tables/}{figures/}}
\makeatother
\graphicspath{{figures/}}

\newcommand{\rmse}{\textsc{rmse}\xspace}
\newcommand{\mae}{\textsc{mae}\xspace}
\newcommand{\da}{\textsc{da}\xspace}
\newcommand{\mse}{\textsc{mse}\xspace}

\newcommand{\hfour}{\ensuremath{h=4}\xspace}
\newcommand{\htwentyfour}{\ensuremath{h=24}\xspace}
\newcommand{\nmodels}{nine\xspace}
\newcommand{\Nmodels}{Nine\xspace}
\newcommand{\nassets}{12\xspace}
\newcommand{\nseeds}{3\xspace}
\newcommand{\nhorizons}{2\xspace}
\newcommand{\ntrials}{5\xspace}

\newcommand{\autoformer}{Autoformer\xspace}
\newcommand{\dlinear}{DLinear\xspace}
\newcommand{\itransformer}{iTransformer\xspace}
\newcommand{\lstm}{LSTM\xspace}
\newcommand{\moderntcn}{ModernTCN\xspace}
\newcommand{\nhits}{N-HiTS\xspace}
\newcommand{\patchtst}{PatchTST\xspace}
\newcommand{\timesnet}{TimesNet\xspace}
\newcommand{\timexer}{TimeXer\xspace}

\newcommand{\eg}{e.g.,\xspace}

\newcommand{\pvalue}{\ensuremath{p}\nobreakdash-value\xspace}
\newcommand{\chisq}{\ensuremath{\chi^2}\xspace}
\newcommand{\Hzero}{\ensuremath{H_0}\xspace}
\newcommand{\cdfive}{\ensuremath{\mathrm{CD}_{0.05}}\xspace}
\newcommand{\nem}{Nemenyi\xspace}
\newcommand{\dm}{\textsc{dm}\xspace}

\newcommand{\icc}{\textsc{icc}\xspace}         
\newcommand{\jt}{\textsc{jt}\xspace}           
\newcommand{\stoufferz}{\ensuremath{Z_\mathrm{S}}\xspace}  

\newcommand{\crypto}{cryptocurrency\xspace}
\newcommand{\forex}{forex\xspace}
\newcommand{\indices}{equity indices\xspace}

\newcommand{\bX}{\ensuremath{\mathbf{X}}}
\newcommand{\by}{\ensuremath{\mathbf{y}}}
\newcommand{\byhat}{\ensuremath{\hat{\mathbf{y}}}}
\newcommand{\btheta}{\ensuremath{\boldsymbol{\theta}}}
\newcommand{\R}{\ensuremath{\mathbb{R}}}
\newcommand{\bigO}{\ensuremath{\mathcal{O}}}

\usepackage{fancyhdr}
\fancypagestyle{plain}{
  \fancyhf{}%
  \fancyfoot[C]{\small\sffamily\thepage}%
  
}

\usepackage{changepage}          
\renewenvironment{abstract}{%
  \vspace{0.8em}%
  \noindent\rule{\textwidth}{0.6pt}%
  \vspace{0.4em}%
  \begin{adjustwidth}{1.5em}{1.5em}%
  \small%
  \noindent{\sffamily\bfseries Abstract.}\quad\ignorespaces
}{%
  \end{adjustwidth}%
  \vspace{0.4em}%
  \noindent\rule{\textwidth}{0.6pt}%
  \vspace{1.2em}%
}

\renewcommand{\footnoterule}{%
  \kern-3pt\hrule width 0.35\textwidth height 0.4pt\kern 2.6pt%
}

\newcommand{\maybeincludegraphics}[2][]{%
  \IfFileExists{#2}{\includegraphics[#1]{#2}}{%
    \fbox{\parbox{0.85\textwidth}{\centering\small\itshape[Figure placeholder: \detokenize{#2}]}}%
  }%
}

\usepackage[
  colorlinks  = true,
  linkcolor   = linkblue,
  citecolor   = linkblue,
  urlcolor    = linkblue,
  bookmarks   = true,
  pdfstartview= FitH,
]{hyperref}
\DeclareFieldFormat{bibhyperref}{\bibhyperref{#1}}

\savebibmacro{cite}
\renewbibmacro*{cite}{%
  \printtext[bibhyperref]{%
    \restorebibmacro{cite}%
    \usebibmacro{cite}}}

\begin{document}

\begin{center}
  {\sffamily\bfseries\LARGE
    A Controlled Comparison of Deep Learning\\[2pt]
    Architectures for Multi-Horizon Financial Forecasting\par}

  \vspace{6pt}
  {\sffamily\large\itshape Evidence from 918 Experiments\par}

  \vspace{18pt}
  {\sffamily\large Nabeel Ahmad Saidd\par}
  \vspace{2pt}
  {\sffamily\small Dr.\ A.P.J.\ Abdul Kalam Technical University, Lucknow, India\par}
  \vspace{2pt}
  {\sffamily\small\texttt{\href{mailto:nabeelahmadsaidd@gmail.com}{nabeelahmadsaidd@gmail.com}}\par}
\end{center}

\vspace{10pt}
\thispagestyle{plain}

\begin{abstract}
Reliable comparison of deep learning architectures for financial forecasting requires controlled evaluation that separates architectural effects from hyperparameter tuning, random initialization, forecast horizon, and asset characteristics. We present a five-stage benchmark designed to provide a systematic and reproducible comparison through equal-budget Bayesian hyperparameter optimization, category-level configuration freezing, three-seed final training, held-out metric aggregation, and rank-based statistical validation. Nine architectures spanning Transformer, MLP/linear, convolutional, and recurrent families are evaluated using direct multi-step prediction from five OHLCV features across 12 assets (four cryptocurrency, four foreign exchange, and four equity-index instruments), with chronological 70/15/15 splits of 30,000 hourly windows per asset at 4- and 24-hour forecasting horizons. The benchmark comprises 918 runs, including 270 optimization trials and 648 final training runs. Across 24 asset--horizon evaluation points, ModernTCN achieves a mean rank of 1.333 and the lowest root mean squared error (RMSE) in 18 points (75.0\%), followed by PatchTST with a mean rank of 2.000. The global ranking differences are strongly supported by an Iman--Davenport test ($F(8,184)=101.36$, $p<10^{-15}$), with 33 of 36 Holm-adjusted pairwise comparisons statistically significant. Architectural rankings also exhibit strong cross-horizon consistency across all 12 assets ($\rho\geq0.683$; Stouffer $p=3.47\times10^{-10}$), while forecasting errors increase at the longer horizon. Architecture accounts for 99.90\% of raw error variance and 68.33\% after within-slot normalization excluding LSTM, whereas seed variance contributes at most 0.04\% (0.02\% without LSTM). Mean directional accuracy reaches 50.08\%. Overall, the findings establish a statistically validated comparison of nine forecasting architectures and identify ModernTCN and PatchTST as leading performers under the evaluated experimental conditions, providing a reproducible benchmark for systematic architecture selection in multi-asset financial time-series forecasting.
\end{abstract}

\medskip
\noindent{\sffamily\bfseries Keywords:} financial time-series forecasting; deep learning; model benchmarking; multi-horizon forecasting

\smallskip
\noindent{\sffamily\bfseries JEL Classification:} C45, C52, C53, G17
\clearpage
\section{Introduction}
\label{sec:introduction}

\subsection{Motivation}
\label{sec:motivation}

Multi-horizon price forecasting is central to modern finance: portfolio allocation relies on expected return estimates at multiple horizons, risk management depends on volatility projections, and algorithmic trading requires predictive signals whose quality degrades with the forecast window.  As \parencite{Cont2001} and \parencite{Fama1970} documented, financial time series exhibit non-stationarity, heavy-tailed returns, volatility clustering, leverage effects, and abrupt regime transitions---properties that make accurate multi-step prediction exceptionally difficult.

Deep learning architectures for temporal sequence modelling have proliferated rapidly over the past five years.  Transformer-based models now span a wide range of inductive biases: auto-correlation in the frequency domain, as in \parencite{Wu2021}; patch-based tokenisation with channel independence, as in \parencite{Nie2023}; inverted variate-wise attention, as in \parencite{Liu2024itransformer}; and exogenous-variable-aware cross-attention with learnable global tokens, as in \parencite{Wang2024timexer}.  Recent work has also designed architectures explicitly around regime adaptation and non-stationarity, as in RADIAN \parencite{Saidd2026RADIAN}.  At the same time, decomposition-based linear mappings have been shown to match or surpass Transformer performance on standard benchmarks \citep{Zeng2023}, while modern temporal convolutional architectures exploit large-kernel depthwise convolutions \citep{Luo2024moderntcn} and FFT-based 2D reshaping \citep{Wu2023timesnet} to capture multi-scale temporal structure.  Hierarchical MLP designs with multi-rate pooling, as in \parencite{Challu2023}, offer yet another approach to direct multi-step forecasting.
This architectural diversity raises a practical question: \emph{which architecture should a practitioner deploy for a given financial forecasting task, at which horizon, and for which asset class?}  A reliable answer requires controlled experimentation that isolates architectural merit from confounding factors---a requirement that existing benchmarks have not met, as Section~\ref{sec:benchmarking_gap} demonstrates.

\subsection{Research Gap}
\label{sec:research_gap}

Despite the growing body of forecasting studies, five persistent methodological shortcomings undermine published model comparisons:

\begin{enumerate}[label=\textbf{G\arabic*.}, leftmargin=1.5cm]
  \item \textbf{Uncontrolled hyperparameter budgets.}  Many benchmarks allocate different tuning effort to different models---or skip hyperparameter optimisation entirely---confounding tuning luck with architectural merit.  Prior work has introduced architectures with custom-tuned configurations while evaluating competitors under default or unspecified settings \citep{Zeng2023, Wu2023timesnet}, preventing fair attribution of performance differences.

  \item \textbf{Single-seed evaluation.}  The vast majority of comparative studies report results from a single random initialisation.  Seed-induced variance has been shown to exceed algorithmic differences in standard machine learning benchmarks \citep{Bouthillier2021}, with analogous effects documented in reinforcement learning \citep{Henderson2018}.  Without multi-seed replication, it is impossible to distinguish genuine architectural advantage from stochastic variation in weight initialisation.

  \item \textbf{Single-horizon analysis.}  Most comparisons evaluate a single forecasting horizon, precluding investigation of how architectural inductive biases interact with prediction difficulty as the forecast window extends.  Prior work has benchmarked recurrent networks at fixed horizons without characterising degradation behaviour, leaving cross-horizon generalisation unaddressed \citep{Hewamalage2021}.

  \item \textbf{Absent pairwise statistical correction.}  Even studies that report omnibus significance tests (\eg Friedman) rarely apply post-hoc pairwise corrections with family-wise error control, leaving the significance of individual ranking differences unquantified.  The M4 competition, reported by \parencite{Makridakis2018}, provided aggregate rankings but did not report pairwise tests among deep learning participants.

  \item \textbf{Narrow asset-class coverage.}  Benchmarks focusing on a single asset class---whether cryptocurrency \citep{Sezer2020}, equities, or standard forecasting datasets (ETTh, Weather, Electricity)---cannot assess whether ranking conclusions generalise across the structurally distinct dynamics of different financial markets.  Cross-class evaluation is necessary for reliable deployment guidance.
\end{enumerate}

Collectively, these gaps prevent reliable inference about the relationship between architectural inductive biases and financial time-series structure, and deprive practitioners of evidence-based model selection guidance---despite the finding (Section~\ref{sec:seed_robustness}) that architecture choice, not random initialisation, explains over 99\% of total forecast variance.

\subsection{Hypotheses and Objectives}
\label{sec:hypotheses}

Four testable hypotheses structure the empirical investigation.  Each states a clear null and a designated statistical test:

\begin{description}[leftmargin=1.2cm, labelwidth=1cm, labelsep=0.2cm, font=\bfseries]
  \item[H1.] \textbf{Ranking Non-Uniformity.}  The global performance ranking of the nine architectures is significantly non-uniform across evaluation points.\\
  \emph{\Hzero:} All models have equal expected rank.\\
  \emph{Test:} Rank-based leaderboard analysis across 24 evaluation points (12~assets $\times$ 2~horizons); win-rate counts; mean rank gaps between tiers.\\
  \emph{Evidence:} Sections~\ref{sec:global_rankings} and~\ref{sec:statistical_tests}.

  \item[H2.] \textbf{Cross-Horizon Ranking Stability.}  Top-ranked architectures at $\hfour$ maintain their relative superiority at $\htwentyfour$, despite absolute error amplification.\\
  \emph{\Hzero:} Rankings at $\hfour$ and $\htwentyfour$ are independent ($\rho_S = 0$).\\
  \emph{Test:} Spearman rank correlation between $\hfour$ and $\htwentyfour$ model rankings per asset; percentage degradation analysis.\\
  \emph{Evidence:} Section~\ref{sec:horizon_degradation}.

  \item[H3.] \textbf{Variance Dominance.}  Architecture choice explains a significantly larger proportion of total forecast variance than random seed initialisation.\\
  \emph{\Hzero:} Model and seed factors contribute equally to variance.\\
  \emph{Test:} Two-factor sum-of-squares variance decomposition; comparison of variance proportions across panels (raw, z-normalised all models, z-normalised modern only).\\
  \emph{Evidence:} Section~\ref{sec:seed_robustness}.

  \item[H4.] \textbf{Non-Monotonic Complexity--Performance.}  The relationship between trainable parameter count and forecasting error is non-monotonic---architectural inductive bias matters more than raw model capacity.\\
  \emph{\Hzero:} Monotonic negative correlation ($\rho_S = -1$) between parameter count and \rmse rank.\\
  \emph{Test:} Spearman rank correlation between parameter count and mean \rmse rank; visual inspection of complexity--performance scatter.\\
  \emph{Evidence:} Section~\ref{sec:complexity}.
\end{description}

The primary objective is to provide a controlled, statistically validated comparison of nine deep learning architectures across three asset classes at two forecasting horizons under identical experimental conditions, yielding evidence-based deployment guidance.

\subsection{Contributions}
\label{sec:contributions}

Five specific contributions advance the state of knowledge:

\begin{enumerate}[label=\textbf{C\arabic*.}]
  \item \textbf{Protocol-controlled fair comparison (addresses G1).}  \Nmodels architectures spanning four families (Transformer, MLP, CNN, RNN) are evaluated under a single five-stage protocol: fixed-seed Bayesian HPO (\ntrials Optuna TPE trials, seed~42), configuration freezing per asset class, identical chronological 70/15/15 splits, a common OHLCV feature set, and rank-based performance evaluation across \nassets instruments, three asset classes, and two horizons, totalling 918~runs (648~final training runs~$+$ 270~HPO trials).

  \item \textbf{Multi-seed robustness quantification (addresses G2).}  Final training is replicated across three independent seeds (123, 456, 789).  A two-factor variance decomposition shows that architecture choice explains 99.90\% of total forecast variance while seed variation accounts for 0.01\%, establishing model selection as the dominant lever for accuracy improvement and confirming that three-seed replication is sufficient.

  \item \textbf{Cross-horizon generalisation analysis (addresses G3).}  Identical models are evaluated at $\hfour$ and $\htwentyfour$ with matched protocol, characterising architecture-specific degradation (Table~\ref{tab:horizon_degradation_btc}) and identifying which inductive biases scale with prediction difficulty.

  \item \textbf{Asset-class-specific deployment guidance (addresses G5).}  Category-level analysis across \crypto, \forex, and \indices (Table~\ref{tab:category_metrics}) shows that top-tier rankings (\moderntcn, \patchtst) hold across all three classes, while mid-tier orderings (\dlinear, \nhits, \timexer) are category-dependent.  A per-asset best-model matrix (Table~\ref{tab:per_asset_best_models}) further reveals niche advantages: \nhits achieves the lowest error on lower-capitalisation cryptocurrency assets, providing actionable asset-level deployment guidance.

  \item \textbf{Open, deterministic benchmarking framework (addresses G1--G5).}  The complete pipeline---source code, configuration files, raw market data, processed datasets, all trained models, and evaluation outputs---is released under an open licence.  Accompanying documentation enables any researcher to reproduce every experiment via a unified command-line interface.
\end{enumerate}

\subsection{Paper Organisation}
\label{sec:outline}

The remainder of this paper is organised as follows.  Section~\ref{sec:literature_review} surveys related work and positions this study relative to existing benchmarks.  Section~\ref{sec:experimental_design} presents the unified experimental protocol.  Section~\ref{sec:results} reports the empirical findings structured by the four hypotheses.  Section~\ref{sec:discussion} interprets the results, provides economic context, and discusses limitations.  Section~\ref{sec:conclusion} summarises contributions and offers deployment recommendations.  Section~\ref{sec:reproducibility} provides a reproducibility statement.  Supplementary results are collected in the Appendix.

\section{Related Work}
\label{sec:literature_review}

This section positions the present study within the broader landscape of financial time-series forecasting.  It traces the evolution from classical econometric models through four families of deep learning architectures, reviews multi-step forecasting strategies, and identifies the methodological gaps this benchmark addresses.

\subsection{Classical and Statistical Approaches}
\label{sec:classical_forecasting}

The ARIMA family developed by \parencite{Box1970} remains a cornerstone of time-series analysis, capturing linear temporal dependencies through differencing and lagged-error terms.  Exponential smoothing methods developed by \parencite{Hyndman2008} offer computationally efficient trend--seasonality decomposition, while the ARCH and GARCH frameworks developed by \parencite{Engle1982} and \parencite{Bollerslev1986}, respectively, provide the standard toolkit for modelling conditional heteroscedasticity in financial returns.

These methods share three fundamental limitations.  First, they assume linearity in the conditional mean or variance, yet financial returns exhibit nonlinear phenomena---leverage effects, long memory, and regime transitions \citep{Cont2001}---that violate these assumptions.  Second, classical models are inherently univariate (or require explicit cross-variable specification), limiting their ability to exploit joint OHLCV information.  Third, multi-step forecasting under these frameworks typically proceeds recursively, compounding prediction errors at longer horizons \citep{Taieb2012}.  These limitations motivate the use of deep learning architectures that learn nonlinear, multivariate mappings directly from data.

\subsection{Deep Learning Architectures for Time-Series}
\label{sec:dl_time_series}

Deep learning approaches to time-series forecasting have evolved along four architectural families, each encoding distinct assumptions about temporal structure.  The specific models included in this benchmark are reviewed below, with emphasis on their temporal inductive biases.

\paragraph{Recurrent architectures.}
Long Short-Term Memory networks, introduced by \parencite{Hochreiter1997} and extended by \parencite{Gers2000}, use gated recurrence to address vanishing gradients, maintaining a cell state that selectively retains or discards information.  While effective for short-range dependencies, sequential processing limits parallelisation and hinders learning of very long-range patterns.  Surveys by \parencite{Hewamalage2021} and \parencite{Lim2021} confirm that LSTMs served as the default neural forecasting baseline during 2017--2021.  In this benchmark, \lstm represents the recurrent family, providing a classical reference point.  Its inductive bias is autoregressive: the hidden state compresses all past information into a fixed-dimensional vector, relying on recurrence to propagate long-range context.

\paragraph{Transformer-based architectures.}
\parencite{Vaswani2017} introduced the self-attention mechanism, which enables direct modelling of pairwise dependencies across arbitrary temporal lags and overcomes the sequential bottleneck of recurrence.  Four Transformer variants are evaluated:

\begin{itemize}[nosep]
  \item \autoformer, introduced by \parencite{Wu2021}, replaces canonical attention with an auto-correlation mechanism operating in the frequency domain at $\bigO(L \log L)$ complexity, coupled with progressive trend--seasonal decomposition.  Its inductive bias assumes that dominant temporal patterns manifest as periodic auto-correlations detectable via spectral analysis.

  \item \patchtst, introduced by \parencite{Nie2023}, segments input series into patches, treats each as a token, and applies a Transformer encoder with channel-independent processing and RevIN normalisation \citep{Kim2022} for distribution-shift mitigation.  Its inductive bias prioritises local temporal coherence within patches while using attention for global inter-patch dependencies.

  \item \itransformer, introduced by \parencite{Liu2024itransformer}, inverts the attention paradigm: each variate's full temporal trajectory serves as a token, and attention operates across the variate dimension, directly capturing cross-variable interactions.  Its inductive bias assumes that inter-variate relationships are the primary source of predictive information.

  \item \timexer, introduced by \parencite{Wang2024timexer}, separates target and exogenous variables, embedding the patched target alongside learnable global tokens and applying cross-attention to query inverted exogenous representations.  Its inductive bias explicitly separates autoregressive dynamics from exogenous covariate influence.
\end{itemize}

\paragraph{MLP and linear architectures.}
\parencite{Zeng2023} challenged the necessity of attention for time-series forecasting, demonstrating that \dlinear---a decomposition-based model with dual independent linear layers mapping seasonal and trend components, without hidden layers, activations, or attention---can match or exceed Transformer performance on standard benchmarks.  Its inductive bias assumes that temporal patterns are adequately captured by linear projections of decomposed components.

\nhits, introduced by \parencite{Challu2023}, employs a hierarchical stack of MLP blocks with multi-rate pooling: each block operates at a different temporal resolution, produces coefficients via a multi-layer perceptron, and interpolates them to the forecast horizon through basis-function expansion.  Its inductive bias prioritises multi-scale temporal structure through hierarchical signal decomposition.  Together, these results raise the question of whether attention contributes meaningfully to forecasting accuracy---a question this benchmark addresses.

\paragraph{Convolutional architectures.}
\parencite{BaiTCN2018} showed that temporal convolutional networks (TCNs) apply causal, dilated convolutions to capture long-range dependencies through hierarchical receptive fields.  The two convolutional models in this benchmark---\timesnet and \moderntcn---each depart from the standard TCN template in distinct ways:

\timesnet, introduced by \parencite{Wu2023timesnet}, transforms the forecasting problem from 1D to 2D by identifying dominant FFT-based periods, reshaping the sequence into 2D tensors indexed by period length and intra-period position, and applying Inception-style 2D convolutions.  Its inductive bias assumes that temporal dynamics decompose into inter-period and intra-period variations best captured through spatial convolutions.

\moderntcn, introduced by \parencite{Luo2024moderntcn}, employs large-kernel depthwise convolutions with structural reparameterisation (dual branches merged at inference), multi-stage downsampling, and optional RevIN normalisation.  Its inductive bias holds that local temporal patterns at multiple scales, captured through large receptive fields with efficient depthwise operations, suffice for accurate forecasting without frequency-domain or attention mechanisms.

A key question is whether these architectural differences yield \emph{consistent} and \emph{statistically significant} performance differences on financial data, or whether the experimental protocol dominates observed rankings.  The present study disentangles these effects through the controlled protocol described in Section~\ref{sec:experimental_design}.

\subsection{Multi-Step Forecasting Strategies}
\label{sec:multistep_strategies}

As \parencite{Taieb2012} and \parencite{Chevillon2007} explain, multi-step-ahead prediction admits three principal strategies.  The \emph{recursive} (iterated) strategy applies a one-step model iteratively, feeding predictions back as inputs; this approach is straightforward but accumulates errors geometrically with the horizon length.  The \emph{direct} strategy trains independent output heads for each future step, avoiding error accumulation at the cost of ignoring inter-step temporal coherence.  The \emph{multi-input multi-output} (MIMO) strategy produces all horizon steps in a single forward pass, preserving inter-step dependencies without iterative error propagation.

This benchmark adopts \emph{direct multi-step forecasting}: each model outputs all $h$ forecast steps simultaneously ($h \in \{4, 24\}$) in a single forward pass, matching the native output design of all nine architectures.  No model feeds predictions back as inputs.  Two separate experiments per horizon use distinct lookback windows ($w = 24$ for $\hfour$; $w = 96$ for $\htwentyfour$), enabling isolation of horizon-dependent degradation from architecture-dependent effects.  Three considerations motivate this choice: (i)~it avoids the error-accumulation confound of recursive strategies; (ii)~it matches every architecture's native mode, preventing protocol mismatch; and (iii)~it enables clean cross-horizon comparison (H2) by ensuring that $\hfour$ and $\htwentyfour$ results differ only in task difficulty.

\subsection{Benchmarking Practices and Identified Gaps}
\label{sec:benchmarking_gap}

Several prior studies have compared deep learning architectures for time-series forecasting, but persistent methodological limitations constrain the conclusions that can be drawn.  The most relevant benchmarks are reviewed below, mapped to the five gaps from Section~\ref{sec:research_gap}.

Large-scale forecasting competitions have advanced standardised evaluation methodologies.  The M4 competition, reported by \parencite{Makridakis2018}, introduced a common evaluation protocol across 100{,}000 series but focused on macroeconomic and demographic data, did not include modern deep learning architectures (released after 2020), and did not apply pairwise statistical corrections (\textbf{G1}, \textbf{G4}, \textbf{G5}).  The M5 competition, reported by \parencite{Makridakis2022}, extended to retail sales forecasting with hierarchical structure but again excluded recent Transformer, TCN, and MLP architectures (\textbf{G5}).

Within the deep learning literature, \parencite{Hewamalage2021} conducted a comprehensive benchmark of recurrent networks that excluded Transformer- and CNN-based alternatives and evaluated a single horizon (\textbf{G3}, \textbf{G5}).  \parencite{Zeng2023} demonstrated the competitiveness of linear models against Transformers, but with default hyperparameters for competitors and a single seed (\textbf{G1}, \textbf{G2}).  \parencite{Wu2023timesnet} benchmarked \timesnet against multiple baselines but with uncontrolled HPO budgets and single-seed evaluation (\textbf{G1}, \textbf{G2}).  \parencite{Nie2023} introduced \patchtst with strong benchmark results, but on non-financial datasets and without multi-seed replication (\textbf{G2}, \textbf{G5}).  \parencite{Liu2024itransformer} evaluated \itransformer across standard time-series benchmarks but without pairwise statistical tests or multi-horizon degradation analysis (\textbf{G3}, \textbf{G4}).  Within the financial forecasting literature specifically, \parencite{Sezer2020} noted the absence of controlled experimental comparisons without providing one.

Table~\ref{tab:prior_studies} summarises these prior studies and their gap coverage.  The central finding is that \emph{no prior study simultaneously addresses G1--G5}: every existing benchmark leaves at least two gaps uncontrolled.  The present study fills this compound gap by providing the first controlled, multi-seed, multi-horizon, multi-asset-class comparison with full statistical validation for financial time-series forecasting.

\begin{table}[htbp]
  \centering
  \caption{Summary of prior comparative studies in time-series forecasting.
    Columns indicate the number of models evaluated, number of datasets or asset
    classes, horizons tested, whether multi-seed evaluation was performed, and
    whether post-hoc pairwise statistical tests were applied.}
  \label{tab:prior_studies}
  \small
  \setlength{\tabcolsep}{3pt}%
  \begin{tabular}{%
    p{2.9cm}
    @{\hspace{5pt}} c
    @{\hspace{5pt}} c
    @{\hspace{5pt}} c
    @{\hspace{5pt}} c
    @{\hspace{5pt}} c
    @{\hspace{5pt}} c
  }
    \toprule
    \textbf{Study} & \textbf{Models} & \textbf{Datasets} & \textbf{Horizons}
      & \textbf{Multi-Seed} & \textbf{Pairwise Tests} & \textbf{Open Code} \\
    \midrule
    \citep{Makridakis2018} (M4)   & Many     & 100K series & Multiple & No  & No  & Partial \\
    \citep{Hewamalage2021}        & RNN only & 6           & Multiple & No  & No  & Partial \\
    \citep{Zeng2023}              & 6        & 9           & 4        & No  & No  & Yes     \\
    \citep{Wu2023timesnet}        & 8        & 8           & 4        & No  & No  & Yes     \\
    \citep{Nie2023}               & 7        & 8           & 4        & No  & No  & Yes     \\
    \citep{Liu2024itransformer}   & 8        & 7           & 4        & No  & No  & Yes     \\
    \midrule
    \textbf{Present study}        & \textbf{9} & \textbf{12 (3 classes)} & \textbf{2}
      & \textbf{Yes (3)} & \textbf{Yes} & \textbf{Yes} \\
    \bottomrule
  \end{tabular}
\end{table}

\section{Experimental Design}
\label{sec:experimental_design}

This section presents the complete experimental protocol as a unified, replicable specification.  Every design choice is stated with its rationale.  A reader equipped with the accompanying repository can reproduce every reported number by following this section sequentially.

\subsection{Formal Problem Definition}
\label{sec:problem_definition}

Let $\bX_t \in \R^{w \times d}$ denote a multivariate input window of $w$ consecutive hourly observations with $d = 5$ OHLCV features (Open, High, Low, Close, Volume), ending at time index~$t$.  The forecasting task is to learn a parametric mapping
\begin{equation}
  f_{\btheta}: \R^{w \times d} \rightarrow \R^{h},
  \quad \byhat_{t+1:t+h} = f_{\btheta}(\bX_t),
  \quad h \in \{4, 24\},
  \label{eq:task}
\end{equation}
where $\byhat_{t+1:t+h}$ is the predicted vector of future Close prices and $\btheta$ collects all learnable parameters.  Two horizon configurations are evaluated as \emph{completely separate experiments}, with no shared weights or intermediate results:

\begin{itemize}[nosep]
  \item \textbf{Short-term} ($\hfour$): lookback window $w = 24$ hours, predicting 4 hours ahead.
  \item \textbf{Long-term} ($\htwentyfour$): lookback window $w = 96$ hours, predicting 24 hours ahead.
\end{itemize}

All models employ \emph{direct multi-step forecasting}: the entire horizon vector $\byhat \in \R^h$ is produced in a single forward pass.  No model feeds predictions back as inputs, avoiding the error accumulation of recursive strategies described by \parencite{Taieb2012} and \parencite{Chevillon2007}.  The training objective is mean squared error:
\begin{equation}
  \mathcal{L}(\btheta) = \frac{1}{n} \sum_{i=1}^{n} \|\by_i - \byhat_i\|^2,
  \label{eq:loss}
\end{equation}
and the model selection criterion is minimum validation \mse.  This common loss function and selection criterion apply identically to all architectures, eliminating confounds from differential training objectives.

\subsection{Data}
\label{sec:data}

\subsubsection{Asset Universe}
\label{sec:asset_universe}

The benchmark spans \nassets financial instruments across three asset classes, each with structurally distinct market microstructure:

\begin{itemize}
  \item \textbf{Cryptocurrency} (4 assets): BTC/USDT, ETH/USDT, BNB/USDT, ADA/USDT.  These instruments trade continuously (24/7), exhibit high volatility, and are subject to rapid regime changes driven by speculative activity and regulatory events.

  \item \textbf{Forex} (4 assets): EUR/USD, USD/JPY, GBP/USD, AUD/USD.  Major currency pairs are characterised by high liquidity, short-term mean-reverting tendencies, and sensitivity to macroeconomic announcements and central-bank policy.

  \item \textbf{Equity indices} (4 assets): Dow Jones, S\&P 500, NASDAQ 100, DAX.  These indices track broad equity markets, exhibiting trending behaviour, lower intra-day volatility relative to cryptocurrency, and session-based trading hours.
\end{itemize}

All instruments are sampled at H1 (1-hour) frequency, providing uniform temporal resolution across asset classes.  For hyperparameter optimisation, one representative asset per class is designated: BTC/USDT (cryptocurrency), EUR/USD (forex), and Dow Jones (equity indices).  Optimised configurations are frozen and applied to all assets within the corresponding class, preventing asset-level overfitting while preserving category-level calibration (Section~\ref{sec:stage2_freeze}).

\subsubsection{Feature Specification}
\label{sec:feature_spec}

All models receive identical input tensors comprising five raw market features: Open, High, Low, Close, and Volume (OHLCV).  This deliberate restriction to unprocessed market data isolates the contribution of architectural design from feature-engineering confounds.  The technical-indicator ablation study of \textcite{Saidd2026Indicators} motivates treating this as an empirical design choice rather than assuming that additional indicators are universally beneficial; that study asks how forecast performance changes when engineered indicators are added, whereas the present study holds the representation fixed to compare architectures.  No technical indicators, lagged returns, calendar variables, or external covariates are introduced.  The sole forecast target is the Close price at each horizon step, ensuring that performance differences reflect architecture rather than feature availability.

\subsubsection{Preprocessing}
\label{sec:preprocessing}

The preprocessing pipeline transforms raw market data into normalised, windowed tensors through four stages:

\begin{enumerate}
  \item \textbf{Loading.}  Raw hourly OHLCV records are ingested from CSV files containing datetime, open, high, low, close, and volume columns.

  \item \textbf{Truncation.}  To ensure comparable dataset sizes across instruments with different histories, the most recent 30{,}000 time steps are retained for every asset prior to windowing.

  \item \textbf{Normalisation.}  Standard $z$-score normalisation (zero mean, unit variance) is fitted \emph{exclusively on the training partition} and applied unchanged to validation and test partitions.  This prevents leakage of future distributional information into the scaling statistics.  After inference, predictions are inverse-scaled to the original price domain before metric computation.

  \item \textbf{Windowing.}  Rolling windows of length $w + h$ are constructed.  Two configurations are employed: $(w, h) = (24, 4)$ for short-term and $(w, h) = (96, 24)$ for long-term forecasting.  Each window yields an input matrix $\bX \in \R^{w \times 5}$ and a target vector $\by \in \R^{h}$ (Close prices only).
\end{enumerate}

\subsubsection{Chronological Splits}
\label{sec:splits}

All partitions are strictly chronological to prevent future-data leakage:
\begin{itemize}[nosep]
  \item \textbf{Training}: first 70\% of samples (approximately 21{,}000 windows per asset per horizon).
  \item \textbf{Validation}: next 15\% of samples (approximately 4{,}500 windows).
  \item \textbf{Test}: final 15\% of samples (approximately 4{,}500 windows).
\end{itemize}

No shuffling is performed at any stage, preserving the temporal ordering essential for financial time-series.  Identical splits are applied to all models, ensuring that every architecture receives exactly the same training, validation, and test observations for each (asset, horizon) pair.  Split boundaries and sample counts are recorded alongside the processed datasets (Table~\ref{tab:dataset_summary}).  Return distributional statistics---mean return, volatility, skewness, excess kurtosis, first-order autocorrelation, and ADF unit-root test \pvalue---for all twelve assets are reported in Table~\ref{tab:distributional_stats}; all return series are stationary ($p < 0.001$) with heavy tails (excess kurtosis 15--96) and near-zero mean returns consistent with weak-form market efficiency.

\begin{table}[htbp]
  \centering
  \caption{Dataset summary.  All assets use H1 (hourly) frequency.  The most
    recent 30{,}000 windowed samples are retained per (asset, horizon) pair,
    split chronologically into 70\%/15\%/15\% train/val/test partitions.
    Window lengths: $w=24$ for $h=4$ and $w=96$ for $h=24$.  Features: OHLCV
    (5 channels); target: close price.}
  \label{tab:dataset_summary}
  \setlength{\tabcolsep}{6pt}%
  \begin{tabular}{%
    l
    l
    l
    r
    r
    r
    r
  }
    \toprule
    \textbf{Category} & \textbf{Asset} & \textbf{Date Range}
      & \textbf{Train} & \textbf{Val} & \textbf{Test} & \textbf{Total} \\
    \midrule
    \multirow{4}{*}{Crypto}
      & BTC/USDT  & 2021-03 -- 2026-02 & 21{,}000 & 4{,}500 & 4{,}500 & 30{,}000 \\
      & ETH/USDT  & 2021-03 -- 2026-02 & 21{,}000 & 4{,}500 & 4{,}500 & 30{,}000 \\
      & BNB/USDT  & 2021-03 -- 2026-02 & 21{,}000 & 4{,}500 & 4{,}500 & 30{,}000 \\
      & ADA/USDT  & 2021-05 -- 2026-02 & 21{,}000 & 4{,}500 & 4{,}500 & 30{,}000 \\
    \midrule
    \multirow{4}{*}{Forex}
      & EUR/USD   & 2017-12 -- 2026-02 & 21{,}000 & 4{,}500 & 4{,}500 & 30{,}000 \\
      & USD/JPY   & 2017-12 -- 2026-02 & 21{,}000 & 4{,}500 & 4{,}500 & 30{,}000 \\
      & GBP/USD   & 2017-12 -- 2026-02 & 21{,}000 & 4{,}500 & 4{,}500 & 30{,}000 \\
      & AUD/USD   & 2017-12 -- 2026-02 & 21{,}000 & 4{,}500 & 4{,}500 & 30{,}000 \\
    \midrule
    \multirow{4}{*}{Indices}
      & Dow Jones   & 2019-01 -- 2026-02 & 21{,}000 & 4{,}500 & 4{,}500 & 30{,}000 \\
      & S\&P~500    & 2019-01 -- 2026-02 & 21{,}000 & 4{,}500 & 4{,}500 & 30{,}000 \\
      & NASDAQ~100  & 2019-01 -- 2026-02 & 21{,}000 & 4{,}500 & 4{,}500 & 30{,}000 \\
      & DAX         & 2019-02 -- 2026-02 & 21{,}000 & 4{,}500 & 4{,}500 & 30{,}000 \\
    \midrule
    \multicolumn{2}{l}{\textbf{Total per horizon}} & ---
      & 252{,}000 & 54{,}000 & 54{,}000 & 360{,}000 \\
    \bottomrule
  \end{tabular}
\end{table}

\begin{table}[htbp]
  \centering
  \caption{Distributional statistics of hourly log returns for all twelve
    assets.  All series reject the Augmented Dickey--Fuller unit-root null
    at $p < 0.001$, confirming return stationarity.}
  \label{tab:distributional_stats}
  \setlength{\tabcolsep}{5pt}%
  \begin{threeparttable}
  \begin{tabular}{%
    l
    l
    r
    r
    r
    r
    r
    l
  }
    \toprule
    \textbf{Cat.} & \textbf{Asset}
      & $\bar{\mu}_r$ (${\times}10^{-4}$)
      & $\sigma_r$ (\%)
      & Skew
      & Kurt
      & ACF(1)
      & ADF $p$ \\
    \midrule
    \multirow{4}{*}{Crypto}
      & BTC/USDT  & $+0.4$ & 0.783 & $-0.47$ & $+35.8$ & $-0.043$ & $<$0.001 \\
      & ETH/USDT  & $+0.3$ & 0.973 & $-0.56$ & $+27.3$ & $-0.026$ & $<$0.001 \\
      & BNB/USDT  & $+0.8$ & 1.107 & $+0.08$ & $+35.0$ & $-0.077$ & $<$0.001 \\
      & ADA/USDT  & $\phantom{+}0.0$ & 1.187 & $-0.05$ & $+19.2$ & $-0.085$ & $<$0.001 \\
    \midrule
    \multirow{4}{*}{Forex}
      & EUR/USD   & $\phantom{+}0.0$ & 0.109 & $-0.01$ & $+15.2$ & $-0.010$ & $<$0.001 \\
      & USD/JPY   & $+0.1$ & 0.119 & $-0.98$ & $+46.6$ & $-0.002$ & $<$0.001 \\
      & GBP/USD   & $\phantom{+}0.0$ & 0.115 & $-1.66$ & $+95.8$ & $-0.018$ & $<$0.001 \\
      & AUD/USD   & $\phantom{+}0.0$ & 0.141 & $-0.07$ & $+17.9$ & $-0.020$ & $<$0.001 \\
    \midrule
    \multirow{4}{*}{Indices}
      & Dow Jones   & $+0.2$ & 0.225 & $-0.57$ & $+60.8$ & $-0.012$ & $<$0.001 \\
      & S\&P~500    & $+0.2$ & 0.235 & $-1.13$ & $+90.2$ & $-0.026$ & $<$0.001 \\
      & NASDAQ~100  & $+0.3$ & 0.286 & $-0.70$ & $+47.7$ & $-0.014$ & $<$0.001 \\
      & DAX         & $+0.2$ & 0.262 & $-1.74$ & $+83.0$ & $-0.011$ & $<$0.001 \\
    \bottomrule
  \end{tabular}
  \begin{tablenotes}[flushleft]\small
    \item All return distributions exhibit excess kurtosis (heavy tails) consistent with the
    stylised facts of financial markets; equities and forex show negative skewness (downside
    asymmetry), while crypto skewness is near zero.  Negative ACF(1) values indicate
    mild mean-reversion in hourly returns.  Volatility spans two orders of magnitude
    across asset classes (crypto $\approx 1\%$, forex $\approx 0.1\%$, indices $\approx 0.2\%$),
    reflecting differences in leverage, liquidity, and trading hours.
  \end{tablenotes}
  \end{threeparttable}
\end{table}

\begin{figure}[htbp]
  \centering
  \includegraphics[width=\textwidth]{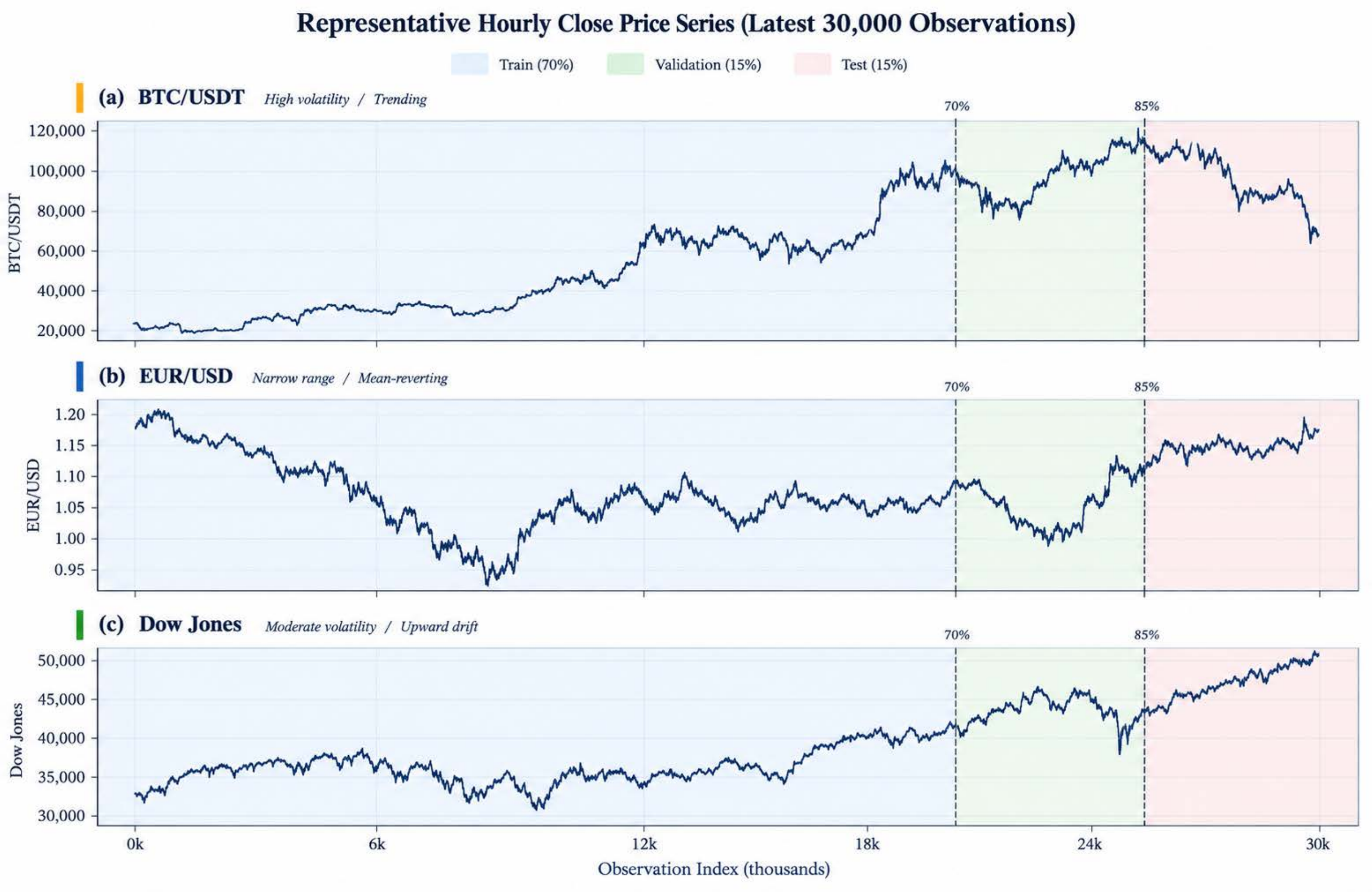}
  \caption{Representative hourly Close-price time series for one asset per class: BTC/USDT (cryptocurrency), EUR/USD (forex), and Dow Jones (equity indices).  Vertical dashed lines indicate chronological train/validation/test boundaries (70/15/15 split).  The three classes exhibit qualitatively different dynamics: high-volatility trending behaviour (cryptocurrency), low-volatility mean-reversion around a narrow range (forex), and moderate-volatility upward drift (equity indices).  All series comprise the most recent 30{,}000 hourly observations.}
  \label{fig:sample_data}
\end{figure}

\subsection{Model Architectures}
\label{sec:model_architectures}

\Nmodels architectures spanning four families are evaluated.  All models conform to a unified interface: input shape $(B, w, d)$ with $d = 5$ OHLCV features; output shape $(B, h)$, where $B$ is the batch size.  No model-specific feature engineering or data augmentation is permitted.  Table~\ref{tab:model_architectures} summarises the key architectural properties.

\paragraph{Transformer family (4 models).}
\autoformer, introduced by \parencite{Wu2021}, replaces self-attention with an auto-correlation mechanism operating in the frequency domain at $\bigO(L \log L)$ complexity, incorporating progressive series decomposition to separate trend and seasonal components.

\patchtst, introduced by \parencite{Nie2023}, segments input series into patches, treats each as a token, and applies a Transformer encoder with channel-independent processing and RevIN normalisation \citep{Kim2022}.

\itransformer, introduced by \parencite{Liu2024itransformer}, inverts the attention paradigm: each variate's full temporal trajectory serves as a token, and attention is computed across the variate dimension.

\timexer, introduced by \parencite{Wang2024timexer}, separates target and exogenous variables, embedding patched target representations alongside learnable global tokens and applying cross-attention to query inverted exogenous embeddings.

\paragraph{MLP/linear family (2 models).}
\dlinear, introduced by \parencite{Zeng2023}, applies moving-average decomposition to separate seasonal and trend components, then maps each through an independent linear layer---without hidden layers, activations, or attention.  It has the fewest parameters of any model in the benchmark (approximately 1{,}000 at $\hfour$).

\nhits, introduced by \parencite{Challu2023}, employs a hierarchical stack of MLP blocks with multi-rate pooling.  Each block operates at a different temporal resolution and interpolates coefficients to the forecast horizon through basis-function expansion.

\paragraph{Convolutional family (2 models).}
\timesnet, introduced by \parencite{Wu2023timesnet}, transforms forecasting from 1D to 2D by identifying dominant FFT-based periods, reshaping the sequence into 2D tensors, and applying Inception-style 2D convolutions to capture intra-period and inter-period patterns.

\moderntcn, introduced by \parencite{Luo2024moderntcn}, employs large-kernel depthwise convolutions with structural reparameterisation, multi-stage downsampling, and optional RevIN normalisation for multi-scale temporal pattern extraction.

\paragraph{Recurrent family (1 model).}
\lstm, introduced by \parencite{Hochreiter1997}, serves as the classical baseline: a multi-layer stacked LSTM encoder extracts the final hidden state, which a two-layer MLP head projects to the forecast vector.

\begin{table}[htbp]
  \centering
  \caption{Summary of the nine deep learning architectures evaluated.  All
    models receive OHLCV input of shape $(B,\,w,\,5)$ and produce direct
    multi-step forecasts of shape $(B,\,h)$.  Family abbreviations:
    RNN~= recurrent, MLP~= multi-layer perceptron,
    TCN~= temporal convolutional network, TF~= Transformer.}
  \label{tab:model_architectures}
  \small
  \setlength{\tabcolsep}{4pt}%
  \begin{tabular}{%
    l
    c
    p{5.5cm}
    l
  }
    \toprule
    \textbf{Model} & \textbf{Family} & \textbf{Key Mechanism} & \textbf{Reference} \\
    \midrule
    \autoformer   & TF  & Auto-correlation \& series decomposition        & \citep{Wu2021} \\
    \dlinear      & MLP & Linear layers with trend--seasonal decomposition & \citep{Zeng2023} \\
    \itransformer & TF  & Inverted attention on variate tokens             & \citep{Liu2024itransformer} \\
    \lstm         & RNN & Gated recurrent cells + dense MLP projection head & \citep{Hochreiter1997} \\
    \moderntcn    & TCN & Large-kernel depthwise convolutions with patching & \citep{Luo2024moderntcn} \\
    \nhits        & MLP & Hierarchical interpolation with multi-rate pooling & \citep{Challu2023} \\
    \patchtst     & TF  & Channel-independent patch-based Transformer      & \citep{Nie2023} \\
    \timesnet     & TCN & 2D temporal variation via FFT + Inception blocks  & \citep{Wu2023timesnet} \\
    \timexer      & TF  & Exogenous-variable-aware cross-attention          & \citep{Wang2024timexer} \\
    \bottomrule
  \end{tabular}
\end{table}

\subsection{Five-Stage Experimental Pipeline}
\label{sec:pipeline}

\begin{figure}[htbp]
  \centering
  \includegraphics[width=\textwidth]{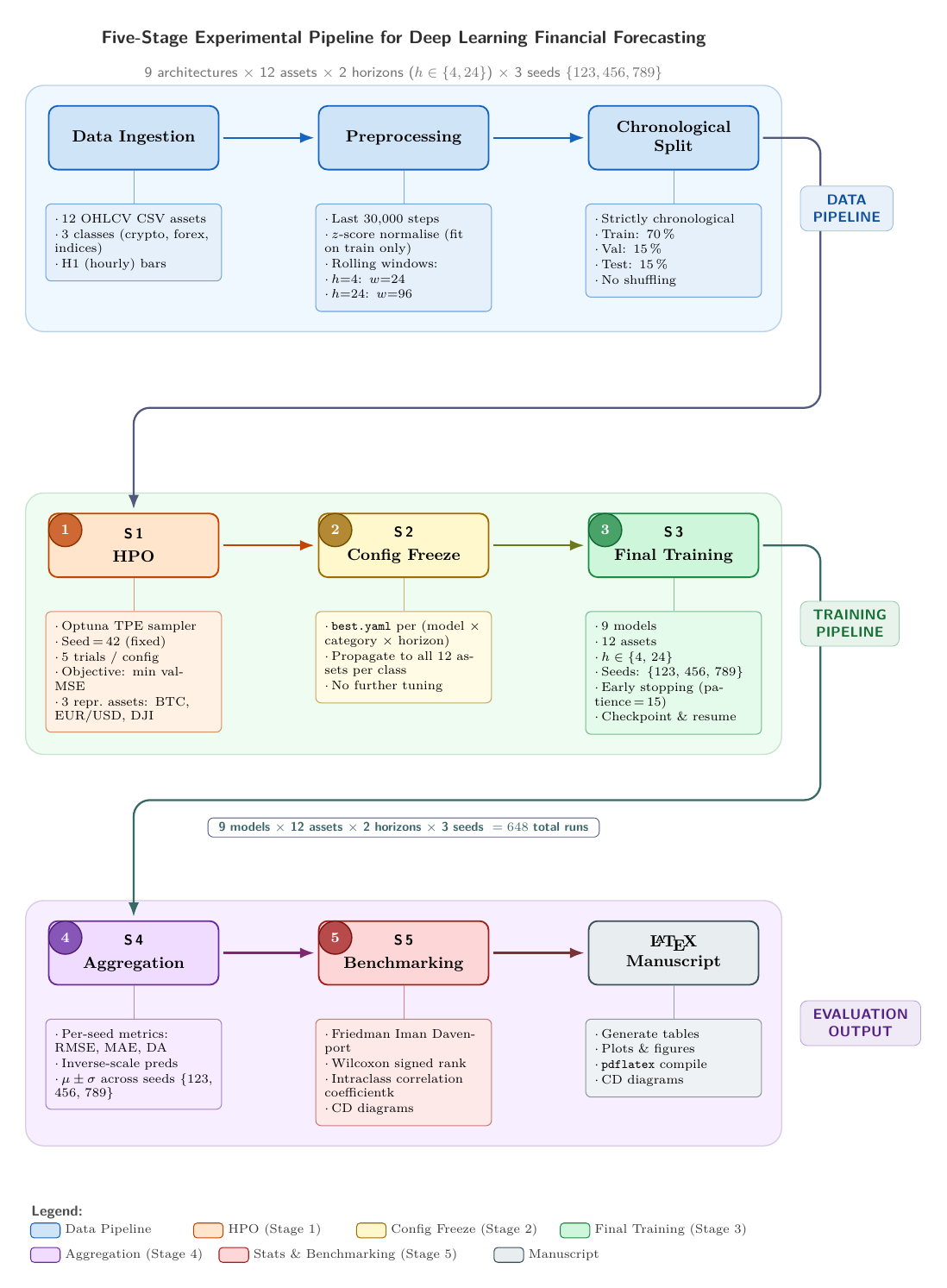}
 \caption{Five-stage experimental pipeline: fixed-seed Bayesian HPO; configuration freezing; multi-seed training with early stopping; metric aggregation (mean~$\pm$~std); and rank-based benchmarking (918 runs).}
 \label{fig:pipeline}
\end{figure}

The experimental pipeline comprises five sequential stages, each designed to eliminate a specific source of confounding.  Figure~\ref{fig:pipeline} provides a schematic overview.

\subsubsection{Stage 1: Fixed-Seed Hyperparameter Optimisation}
\label{sec:stage1_hpo}

Hyperparameter optimisation is performed using the Optuna framework introduced by \parencite{Akiba2019}, with the Tree-structured Parzen Estimator (TPE) sampler developed by \parencite{Bergstra2011}.  To ensure fairness, the following settings are held constant across all architectures:

\begin{itemize}[nosep]
  \item \textbf{Deterministic seed:} 42 (identical random state for all HPO runs).
  \item \textbf{Trial budget:} \ntrials trials per (model, category, horizon) configuration.
  \item \textbf{Sampler:} TPE with median pruner (2 startup trials, 5 warm-up steps).
  \item \textbf{Objective:} Minimise validation \mse.
  \item \textbf{Training budget per trial:} 50 epochs, batch size~256.
\end{itemize}

HPO is conducted exclusively on one representative asset per category (BTC/USDT, EUR/USD, Dow Jones), preventing overfitting to individual assets while capturing category-level dynamics.  This design is motivated by two considerations: (i)~tuning on all 12 assets would multiply computational cost by $12\times$ without a mechanism for cross-asset generalisation, and (ii)~freezing configurations at the category level ensures the same inductive prior governs all assets within a class.  Table~\ref{tab:hpo_search_spaces} provides the full search space specification.

The controlled \ntrials-trial budget is a deliberate choice that prioritises comparative fairness over exhaustive peak-performance search.  Three considerations justify this constraint.  First, in financial time-series forecasting---where the signal-to-noise ratio is low and non-stationarity is pervasive---extensive HPO risks selecting configurations that overfit to transient market regimes and fail to generalise to the test set.  To mitigate this risk, search ranges were drawn from commonly used configurations in the literature, and identical budgets were applied to all models to ensure a symmetrical evaluation framework.

Second, model parameter counts were kept in a comparable range across architectures (with the exception of \dlinear, which is intentionally lightweight by design).  Maintaining comparable capacity reduces search-space imbalance and prevents capacity-driven advantages from confounding the architectural comparison.

Third, the empirical evidence corroborates this design: the variance decomposition and rank stability across horizons (Section~\ref{sec:results}) show that architectural identity is the dominant factor in performance variance.  This evidence does not prove that five trials are sufficient to locate every model's optimum; in particular, the mid-tier ordering may be sensitive to additional trials.  Consequently, this protocol prioritises cross-architectural fairness and statistical validity, ensuring that observed performance differences reflect structural merits rather than differential tuning intensity.  An exhaustive search for all models is outside the scope of this benchmark and is retained as future work (Section~\ref{sec:future_work}).

\begin{table}[htbp]
  \centering
  \caption{Hyperparameter search spaces per model.  All models share learning
    rate $\in [5\times10^{-4},\,5\times10^{-3}]$ (log-uniform) and batch
    size $\in\{64,128\}$.  Only architecture-specific parameters are shown.
    HPO uses Optuna TPE with \ntrials trials per (model $\times$ horizon
    $\times$ asset class) on representative assets (BTC/USDT, EUR/USD,
    Dow Jones) only.}
  \label{tab:hpo_search_spaces}
  \small
  \setlength{\tabcolsep}{6pt}%
  \renewcommand{\arraystretch}{0.92}%
  \begin{tabular}{%
    l
    l
    p{5.2cm}
  }
    \toprule
    \textbf{Model} & \textbf{Hyperparameter} & \textbf{Range / Choices} \\
    \midrule
    \multirow{5}{*}{\autoformer}
      & Model dimension          & $\{64, 128\}$ \\
      & Attention heads          & $\{4, 8\}$ \\
      & Enc./dec.\ layers        & $\{1, 2\}$ each \\
      & Feedforward dimension    & $\{64, 128\}$ \\
      & Dropout rate             & $[0.0, 0.2]$, step~$0.05$ \\
    \midrule
    \multirow{2}{*}{\dlinear}
      & Moving-average kernel    & $[3, 51]$, step~$2$ \\
      & Per-channel mapping      & $\{\text{true}, \text{false}\}$ \\
    \midrule
    \multirow{4}{*}{\itransformer}
      & Model dimension          & $\{96, 112\}$ \\
      & Feedforward dimension    & $\{128, 256\}$ \\
      & Encoder layers           & $[2, 4]$ \\
      & Attention heads          & $\{4, 8, 16\}$ \\
    \midrule
    \multirow{4}{*}{\lstm}
      & Hidden state size        & $\{64, 128\}$ \\
      & Recurrent layers         & $[1, 3]$ \\
      & Projection head width    & $\{64, 128\}$ \\
      & Bidirectional            & $\{\text{true}, \text{false}\}$ \\
    \midrule
    \multirow{4}{*}{\moderntcn}
      & Patch size               & $\{8, 16\}$ \\
      & Channel dimensions       & $[32, 64, 96]$ \\
      & RevIN normalisation      & $\{\text{true}, \text{false}\}$ \\
      & Dropout / head dropout  & $[0.0, 0.2]$, step~$0.05$ \\
    \midrule
    \multirow{4}{*}{\nhits}
      & Number of blocks         & $[2, 3]$ \\
      & Hidden layer width       & $\{64, 96\}$ \\
      & Layers per block         & $[3, 6]$ \\
      & Pooling kernel sizes     & $\{[2,4,8],\,[4,8,12]\}$ \\
    \midrule
    \multirow{4}{*}{\patchtst}
      & Model dimension          & $\{64\}$ \\
      & Encoder layers           & $[1, 3]$ \\
      & Patch length             & $\{16, 24\}$ \\
      & Stride                   & $\{4, 8\}$ \\
    \midrule
    \multirow{3}{*}{\timesnet}
      & Model dimension          & $\{24, 32\}$ \\
      & Feedforward dimension    & $\{32, 64\}$ \\
      & Encoder layers           & $[2, 3]$ \\
    \midrule
    \multirow{4}{*}{\timexer}
      & Model dimension          & $\{64, 96\}$ \\
      & Feedforward dimension    & $\{128, 256\}$ \\
      & Encoder layers           & $[2, 3]$ \\
      & Attention heads          & $\{4, 8\}$ \\
    \bottomrule
  \end{tabular}
\end{table}

\subsubsection{Stage 2: Configuration Freezing}
\label{sec:stage2_freeze}

The best-performing configuration for each (model, category, horizon) triple---determined by validation \mse in Stage~1---is recorded and frozen.  These frozen configurations are applied unchanged to all assets within the corresponding category; no further tuning is permitted.  This eliminates asset-level overfitting and ensures that each architecture is evaluated on a single, category-level configuration.  The representative assets used for HPO are therefore also evaluation assets, which creates a possible evaluation-bias concern even though it is not chronological leakage.  We address this concern post hoc using the stored per-asset rank matrices in Table~\ref{tab:hpo_nonhpo_ranks}.  Table~\ref{tab:best_hyperparameters} presents the selected configurations.

\begin{table}[htbp]
  \centering
  \caption{Frozen best hyperparameters selected via Optuna TPE (\ntrials
    trials, seed~42, objective: minimum validation \mse).  Only key
    architecture-specific parameters are shown; all configurations also
    include learning rate and batch size.  Shared entries across categories
    indicate that the representative-asset optimum was identical.}
  \label{tab:best_hyperparameters}
  \footnotesize
  \setlength{\tabcolsep}{5pt}%
  \resizebox{\textwidth}{!}{%
  \begin{tabular}{%
    l
    l
    c
    p{8cm}
  }
    \toprule
    \textbf{Model} & \textbf{Category} & \textbf{$h$} & \textbf{Key Parameters} \\
    \midrule
    \multirow{2}{*}{\autoformer}
      & Crypto/Forex & 4/24 & $d_{\mathrm{model}}{=}128$, heads${=}4$, enc${=}1$, dec${=}2$, $d_{\mathrm{ff}}{=}128$, drop${=}0.20$ \\
      & Indices      & 4/24 & $d_{\mathrm{model}}{=}128$, heads${=}4$, enc${=}1$, dec${=}2$, $d_{\mathrm{ff}}{=}128$, drop${=}0.10$ \\
    \midrule
    \multirow{2}{*}{\dlinear}
      & All          & 4    & kernel${=}21$, individual${=}\text{true}$ \\
      & Crypto/Forex & 24   & kernel${=}43$, individual${=}\text{true}$ \\
    \midrule
    \multirow{2}{*}{\itransformer}
      & Crypto       & 24   & $d_{\mathrm{model}}{=}112$, $d_{\mathrm{ff}}{=}128$, layers${=}2$, heads${=}16$, drop${=}0.15$ \\
      & Other        & 4    & $d_{\mathrm{model}}{=}96$,  $d_{\mathrm{ff}}{=}128$, layers${=}4$, heads${=}16$, drop${=}0.00$ \\
    \midrule
    \multirow{2}{*}{\lstm}
      & All          & 4    & hidden${=}128$, layers${=}1$, mlp${=}128$, bidir${=}\text{true}$, drop${=}0.15$ \\
      & All          & 24   & hidden${=}64$,  layers${=}1$, mlp${=}128$, bidir${=}\text{true}$, drop${=}0.10$ \\
    \midrule
    \multirow{2}{*}{\moderntcn}
      & Crypto       & 24   & patch${=}16$, dims${=}[32,64,96]$, RevIN, drop${=}0.20$, head-drop${=}0.20$ \\
      & Other        & 4/24 & patch${=}8$,  dims${=}[32,64,96]$, RevIN+affine, drop${=}0.10$, head-drop${=}0.10$ \\
    \midrule
    \multirow{2}{*}{\nhits}
      & Crypto       & 4    & blocks${=}2$, hidden${=}64$, layers${=}3$, pool${=}[4,8,12]$, drop${=}0.05$ \\
      & Crypto       & 24   & blocks${=}3$, hidden${=}96$, layers${=}4$, pool${=}[4,8,12]$, drop${=}0.00$ \\
    \midrule
    \multirow{2}{*}{\patchtst}
      & Crypto       & 4    & $d_{\mathrm{model}}{=}64$, layers${=}2$, patch${=}24$, stride${=}8$, drop${=}0.05$ \\
      & Other        & 24   & $d_{\mathrm{model}}{=}64$, layers${=}1$, patch${=}24$, stride${=}4$, drop${=}0.15$ \\
    \midrule
    \timesnet
      & All          & 4/24 & $d_{\mathrm{model}}{=}32$, $d_{\mathrm{ff}}{=}32$, layers${=}2$, top-$k{=}3$, drop${=}0.00$ \\
    \midrule
    \multirow{2}{*}{\timexer}
      & Crypto       & 4    & $d_{\mathrm{model}}{=}64$, $d_{\mathrm{ff}}{=}256$, layers${=}2$, heads${=}4$, drop${=}0.05$ \\
      & Other        & 4/24 & $d_{\mathrm{model}}{=}64$, $d_{\mathrm{ff}}{=}128$, layers${=}3$, heads${=}8$, drop${=}0.05$ \\
    \bottomrule
  \end{tabular}%
  }
\end{table}

\subsubsection{Stage 3: Multi-Seed Final Training}
\label{sec:stage3_training}

Final training is conducted for every (model, asset, horizon, seed) combination under the frozen configuration from Stage~2.  The following training protocol is applied identically to all runs:

\begin{itemize}[nosep]
  \item \textbf{Seeds:} 123, 456, 789 (three independent initialisations per configuration).
  \item \textbf{Maximum epochs:} 100.
  \item \textbf{Batch size:} As determined by HPO (typically 64 or 128).
  \item \textbf{Optimiser:} Adam with weight decay $10^{-4}$.
  \item \textbf{Learning rate scheduler:} ReduceLROnPlateau (patience~5, factor~0.5, minimum LR~$10^{-6}$).
  \item \textbf{Early stopping:} Patience~15, monitoring validation loss (minimum improvement threshold $\delta_{\min} = 10^{-4}$).
  \item \textbf{Gradient clipping:} $\ell_2$-norm clipped at 1.0.
  \item \textbf{Loss function:} \mse (Equation~\ref{eq:loss}).
\end{itemize}

Each seed controls all sources of randomness: Python's standard library, NumPy, PyTorch CPU and CUDA generators, cuDNN backend settings, and the Python hash seed (set via environment variable before process startup).  DataLoader workers derive their seeds deterministically from the primary seed.  Checkpointing occurs every epoch, retaining both the best model (lowest validation loss) and the latest model.  Interrupted training resumes from the last completed epoch, restoring optimiser, scheduler, and random number generator states.

\subsubsection{Stage 4: Metric Aggregation}
\label{sec:stage4_aggregation}

Evaluation metrics (\rmse, \mae, \da) are computed exclusively on the held-out test set.  Predictions are inverse-scaled to the original price domain using training-set scaler parameters before metric computation, ensuring that errors are expressed in economically meaningful units.  For each (model, asset, horizon) triple, the three seed-specific metrics are aggregated as mean~$\pm$~standard deviation, providing both a point estimate and an uncertainty bound.

\subsubsection{Stage 5: Benchmarking and Statistical Validation}
\label{sec:stage5_benchmark}

The final stage generates all comparative analyses:

\begin{itemize}[nosep]
  \item \textbf{Global leaderboard:} Models ranked by mean \rmse rank across all \nassets assets and \nhorizons horizons (24~evaluation points per model).
  \item \textbf{Category-level analysis:} Per-category aggregated metrics and rankings.
  \item \textbf{Cross-horizon degradation:} \rmse change from $\hfour$ to $\htwentyfour$ per model per asset.
  \item \textbf{Statistical validation:} Rank-based leaderboard analysis and two-factor variance decomposition of model vs.\ seed contributions (Section~\ref{sec:statistical_framework}).
\end{itemize}

\paragraph{Dual-plot convention.}  \lstm serves as a classical baseline, but its errors (one to two orders of magnitude above modern models) compress the visual scale in comparison plots, obscuring performance differences among the eight modern architectures.  Body figures therefore use the no-\lstm variant for finer discrimination; the all-models variant including \lstm appears in Appendix~\ref{sec:app_dual_plots}.  All tabular results always include all nine models.

\subsection{Evaluation Metrics}
\label{sec:metrics}

Three complementary metrics are computed on the held-out test set for every (model, asset, horizon, seed) configuration.  All predictions are inverse-transformed to the original price scale before computation, ensuring that error magnitudes are economically interpretable.

\begin{enumerate}[nosep]
  \item \textbf{\rmse (Root Mean Squared Error)}---the \emph{primary ranking metric}, penalising large deviations quadratically.  \rmse is appropriate for financial risk assessment, where large forecast errors carry disproportionate cost:
  \begin{equation}
    \text{RMSE} = \sqrt{\frac{1}{n}\sum_{i=1}^{n}(y_i - \hat{y}_i)^2}.
    \label{eq:rmse}
  \end{equation}

  \item \textbf{\mae (Mean Absolute Error)}---a secondary metric that is robust to outliers and provides a median-biased point estimate of forecast error:
  \begin{equation}
    \text{MAE} = \frac{1}{n}\sum_{i=1}^{n}|y_i - \hat{y}_i|.
    \label{eq:mae}
  \end{equation}

  \item \textbf{\da (Directional Accuracy)}---the fraction of horizon steps where the predicted direction of price change matches the realised direction, providing an economically interpretable measure of forecast quality relevant to trading-signal applications:
  \begin{equation}
    \text{DA} = \frac{1}{n}\sum_{i=1}^{n}\mathbb{1}\!\left[\text{sign}(\hat{y}_i - \hat{y}_{i-1}) = \text{sign}(y_i - y_{i-1})\right].
    \label{eq:da}
  \end{equation}
\end{enumerate}

All results are reported as mean~$\pm$~standard deviation across \nseeds seeds (123, 456, 789), enabling quantification of initialisation-induced uncertainty.  The concordance between \rmse and \mae rankings is examined in Section~\ref{sec:category_analysis} to verify that findings are robust to the choice of error metric.

\subsection{Statistical Validation Framework}
\label{sec:statistical_framework}

A two-tier framework is employed to characterise observed performance differences:

\begin{enumerate}[nosep]
  \item \textbf{Two-factor variance decomposition} $\rightarrow$ \textbf{H3}.  A sum-of-squares decomposition partitions total forecast variance into three components: model-attributable, seed-attributable, and residual interaction.  Reported as a percentage of total sum of squares across three panels (raw, z-normalised all models, z-normalised modern only).

  \item \textbf{Spearman rank correlation} $\rightarrow$ \textbf{H2, H4}.  Spearman's $\rho$ between $\hfour$ and $\htwentyfour$ model rankings per asset tests cross-horizon stability (H2).  Spearman's $\rho$ between parameter count and mean \rmse rank tests whether complexity--performance is monotonic (H4).  Both correlations are tested for significance against $\rho = 0$.
\end{enumerate}

\subsection{Experimental Scale}
\label{sec:experimental_scale}

The total experimental scale is:

\begin{itemize}[nosep]
  \item \textbf{HPO (Stage~1):} $9 \text{ models} \times 3 \text{ representative assets} \times 2 \text{ horizons} \times 5 \text{ trials} = 270$ trial runs.
  \item \textbf{Final training (Stage~3):} $9 \text{ models} \times 12 \text{ assets} \times 2 \text{ horizons} \times 3 \text{ seeds} = 648$ training runs.
  \item \textbf{Total:} $270 + 648 = \mathbf{918}$ experimental runs.
\end{itemize}

Each of the 648 final training runs produces a separate metrics evaluation on the test set, yielding 648 individual (model, asset, horizon, seed) performance records that form the basis of all analyses in Section~\ref{sec:results}.

\section{Results}
\label{sec:results}

This section presents findings from 918 experimental runs: 648 final training runs (9~models $\times$ 12~assets $\times$ 2~horizons $\times$ 3~seeds) plus 270 HPO trials.  Results are organised by the four hypotheses from Section~\ref{sec:hypotheses}.  All claims reference specific tables or figures; mechanistic interpretation is deferred to Section~\ref{sec:discussion}.

\subsection{Global Performance Rankings}
\label{sec:global_rankings}

Table~\ref{tab:global_leaderboard} presents the global leaderboard, ranking \nmodels models across four architectural families by mean \rmse rank across all \nassets assets and both horizons (24~evaluation points per model).  Three distinct performance tiers emerge:

\begin{itemize}[nosep]
  \item \textbf{Top tier:} \moderntcn (CNN; mean rank 1.333, median 1.0) and \patchtst (Transformer; mean rank 2.000, median 2.0).
  \item \textbf{Middle tier:} \itransformer (3.667), \timexer (4.292), \dlinear (4.958), and \nhits (5.250), spanning the Transformer and MLP / Linear families.
  \item \textbf{Bottom tier:} \timesnet (7.708), \autoformer (7.833), and \lstm (7.958).
\end{itemize}

The separation between tiers is substantial: the gap between the top tier (ranks 1--2) and bottom tier (ranks 7--9) spans more than 5.5 rank positions, and performance does not correlate uniformly with model family. Both the top and bottom tiers contain CNN and Transformer-based architectures, suggesting that specific implementation details (e.g., patching, large-kernel convolutions) matter more than broad architectural classes.

\begin{table}[htbp]
  \centering
  \caption{Global model ranking aggregated across all 12 assets and both
    horizons ($h\in\{4,24\}$), categorised by architectural family. Each
    (asset, horizon) pair contributes one rank based on mean \rmse over
    three seeds.  Mean and median ranks are computed over 24 evaluation slots.
    Win~Count indicates the number of slots in which a model achieved rank~1.
    Bold marks the best value per column.}
  \label{tab:global_leaderboard}
  \setlength{\tabcolsep}{6pt}%
  \begin{tabular}{%
    l%
    l
    S[table-format=1.3]
    S[table-format=1.1]
    l
  }
    \toprule
    \textbf{Model}
      & \textbf{Family}
      & {\textbf{Mean Rank}}
      & {\textbf{Median Rank}}
      & \textbf{Wins (of 24)} \\
    \midrule
    \textbf{\moderntcn} & CNN           & \bfseries 1.333 & \bfseries 1.0 & \textbf{18~(75.0\%)} \\
    \patchtst           & Transformer   & 2.000           & 2.0           & 3~(12.5\%)            \\
    \itransformer       & Transformer   & 3.667           & 3.0           & 0                     \\
    \timexer            & Transformer   & 4.292           & 4.0           & 0                     \\
    \dlinear            & MLP / Linear  & 4.958           & 5.0           & 0                     \\
    \nhits              & MLP / Linear  & 5.250           & 6.0           & 3~(12.5\%)            \\
    \timesnet           & CNN           & 7.708           & 8.0           & 0                     \\
    \autoformer         & Transformer   & 7.833           & 8.0           & 0                     \\
    \lstm               & RNN           & 7.958           & 9.0           & 0                     \\
    \bottomrule
  \end{tabular}
\end{table}

To assess whether inclusion of the HPO representative assets affects the
leaderboard, Table~\ref{tab:hpo_nonhpo_ranks} compares the stored per-asset
ranks for BTC/USDT, EUR/USD, and Dow Jones with the ranks for the other nine
assets.  The mean-rank vectors are highly concordant ($\rho=0.996$).  ModernTCN
and PatchTST remain the top two models in both groups; the largest changes are
small shifts among N-HiTS, DLinear, and LSTM.  Thus, the overlap does not
change the principal top-tier conclusion, although the comparison does not
replace an evaluation on assets excluded from HPO.

\begin{table}[htbp]
  \centering
  \begin{threeparttable}
  \caption{Robustness of model ranks to the inclusion of representative assets
    in HPO.  Mean ranks are computed from the stored per-asset rank matrices:
    six evaluation slots (three HPO assets $\times$ two horizons) for the HPO
    group and 18 slots (nine non-HPO assets $\times$ two horizons) for the
    non-HPO group.  Lower ranks are better.}
  \label{tab:hpo_nonhpo_ranks}
  \setlength{\tabcolsep}{8pt}%
  \begin{tabular}{l S[table-format=1.3] S[table-format=1.3] S[table-format=+1.3]}
    \toprule
    \textbf{Model} & {HPO assets} & {Non-HPO assets} & {$\Delta$ (non-HPO $-$ HPO)} \\
    \midrule
    \moderntcn    & 1.167 & 1.389 & +0.222 \\
    \patchtst     & 1.833 & 2.056 & +0.222 \\
    \itransformer  & 3.667 & 3.667 &  0.000 \\
    \timexer      & 4.167 & 4.333 & +0.167 \\
    \dlinear       & 4.667 & 5.056 & +0.389 \\
    \nhits          & 5.667 & 5.111 & -0.556 \\
    \timesnet      & 7.667 & 7.722 & +0.056 \\
    \autoformer    & 7.833 & 7.833 &  0.000 \\
    \lstm          & 8.333 & 7.833 & -0.500 \\
    \bottomrule
  \end{tabular}
  \begin{tablenotes}[flushleft]\footnotesize
    \item HPO assets are BTC/USDT, EUR/USD, and Dow Jones; all remaining
    assets form the non-HPO group.  The Spearman correlation between the two
    nine-model mean-rank vectors is $\rho=0.996$ (computed from the stored
    rank matrices).  The top tier is unchanged; the largest differences are
    confined to N-HiTS and LSTM, while the overall ordering is preserved.
  \end{tablenotes}
  \end{threeparttable}
\end{table}

In terms of win counts, \moderntcn achieves the lowest \rmse on 18 of 24~evaluation points (75.0\% win rate), while \nhits and \patchtst each win 3~points (12.5\%).  No other architecture achieves a single first-place finish.

Table~\ref{tab:per_asset_best_models} disaggregates these wins by asset and horizon. \moderntcn's dominance is concentrated in forex and equity indices (16 of 16~wins), while its cryptocurrency record is more contested: \nhits wins on ETH/USDT (both horizons) and ADA/USDT ($\hfour$), and \patchtst wins on BTC/USDT ($\hfour$; $\rmse = 731.05$ vs.\ \moderntcn's $731.63$, $\Delta = 0.08\%$) and ADA/USDT ($\htwentyfour$).  Notably, no model other than \moderntcn achieves the lowest \rmse on any forex or equity index asset at $\htwentyfour$, underscoring its superior long-horizon generalisation outside the cryptocurrency domain.

\begin{table}[htbp]
  \centering
  \begin{threeparttable}
  \caption{Best-performing model per asset and horizon, determined by lowest
    mean \rmse across three seeds.  \moderntcn achieves the lowest \rmse
    on 18 of 24~evaluation points (75\%); \nhits wins on 3~points
    (all in cryptocurrency); \patchtst wins on 3~points (one crypto,
    one crypto, one forex).  Bold highlights non-\moderntcn winners,
    revealing niche asset-specific advantages.}
  \label{tab:per_asset_best_models}
  \setlength{\tabcolsep}{7pt}%
  \begin{tabular}{l l l l}
    \toprule
    \textbf{Category} & \textbf{Asset}
      & \textbf{Best at $\hfour$}
      & \textbf{Best at $\htwentyfour$} \\
    \midrule
    \multirow{4}{*}{Crypto}
      & BTC/USDT    & \textbf{\patchtst}  & \moderntcn          \\
      & ETH/USDT    & \textbf{\nhits}     & \textbf{\nhits}     \\
      & BNB/USDT    & \moderntcn          & \moderntcn          \\
      & ADA/USDT    & \textbf{\nhits}     & \textbf{\patchtst}  \\
    \addlinespace
    \multirow{4}{*}{Forex}
      & EUR/USD     & \moderntcn          & \moderntcn          \\
      & USD/JPY     & \moderntcn          & \moderntcn          \\
      & GBP/USD     & \textbf{\patchtst}  & \moderntcn          \\
      & AUD/USD     & \moderntcn          & \moderntcn          \\
    \addlinespace
    \multirow{4}{*}{Indices}
      & DAX         & \moderntcn          & \moderntcn          \\
      & Dow Jones   & \moderntcn          & \moderntcn          \\
      & S\&P~500    & \moderntcn          & \moderntcn          \\
      & NASDAQ~100  & \moderntcn          & \moderntcn          \\
    \midrule
    \multicolumn{2}{l}{\textbf{Win count summary}}
      & \multicolumn{2}{l}{\moderntcn: 18\quad \nhits: 3\quad \patchtst: 3} \\
    \bottomrule
  \end{tabular}
  \begin{tablenotes}[flushleft]\footnotesize
    \item \nhits wins exclusively on lower-capitalisation cryptocurrency
      assets (ETH/USDT, ADA/USDT), suggesting that its hierarchical
      multi-rate pooling is particularly suited to the high-frequency,
      multi-scale volatility structure of altcoin markets.
    \item \patchtst wins at $\hfour$ on BTC/USDT ($\rmse = 731.05$
      vs.\ \moderntcn's $731.63$; $\Delta = 0.08\%$) and on
      GBP/USD, but cedes to \moderntcn at $\htwentyfour$ in both cases.
    \item No model other than \moderntcn wins on any forex or equity index
      asset at $\htwentyfour$, underscoring its superior long-horizon
      generalisation.
  \end{tablenotes}
  \end{threeparttable}
\end{table}

\begin{figure}[htbp]
  \centering
  \includegraphics[width=0.85\textwidth]{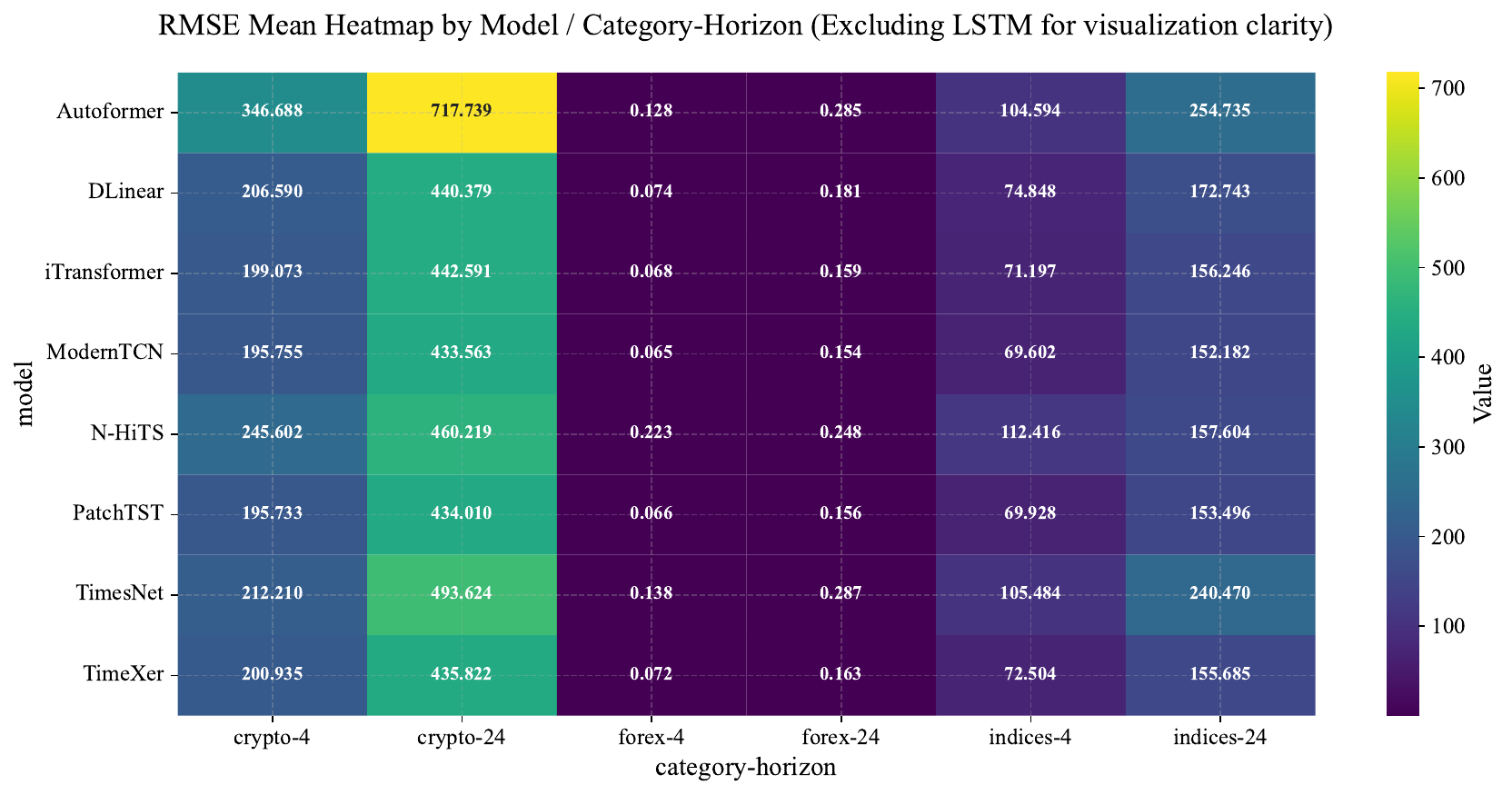}
  \caption{Global \rmse heatmap across eight modern architectures and 24 evaluation points (12~assets $\times$ 2~horizons).  Lighter cells indicate lower error.  \moderntcn and \patchtst consistently achieve the lowest \rmse values across all asset--horizon combinations.  \lstm is excluded for visual clarity; the full nine-model variant is provided in Appendix~\ref{sec:app_dual_plots}, Figure~\ref{fig:app_global_heatmap_all}.  Values represent mean \rmse across three seeds.}
  \label{fig:global_heatmap}
\end{figure}

\begin{figure}[htbp]
  \centering
  \includegraphics[width=0.7\textwidth]{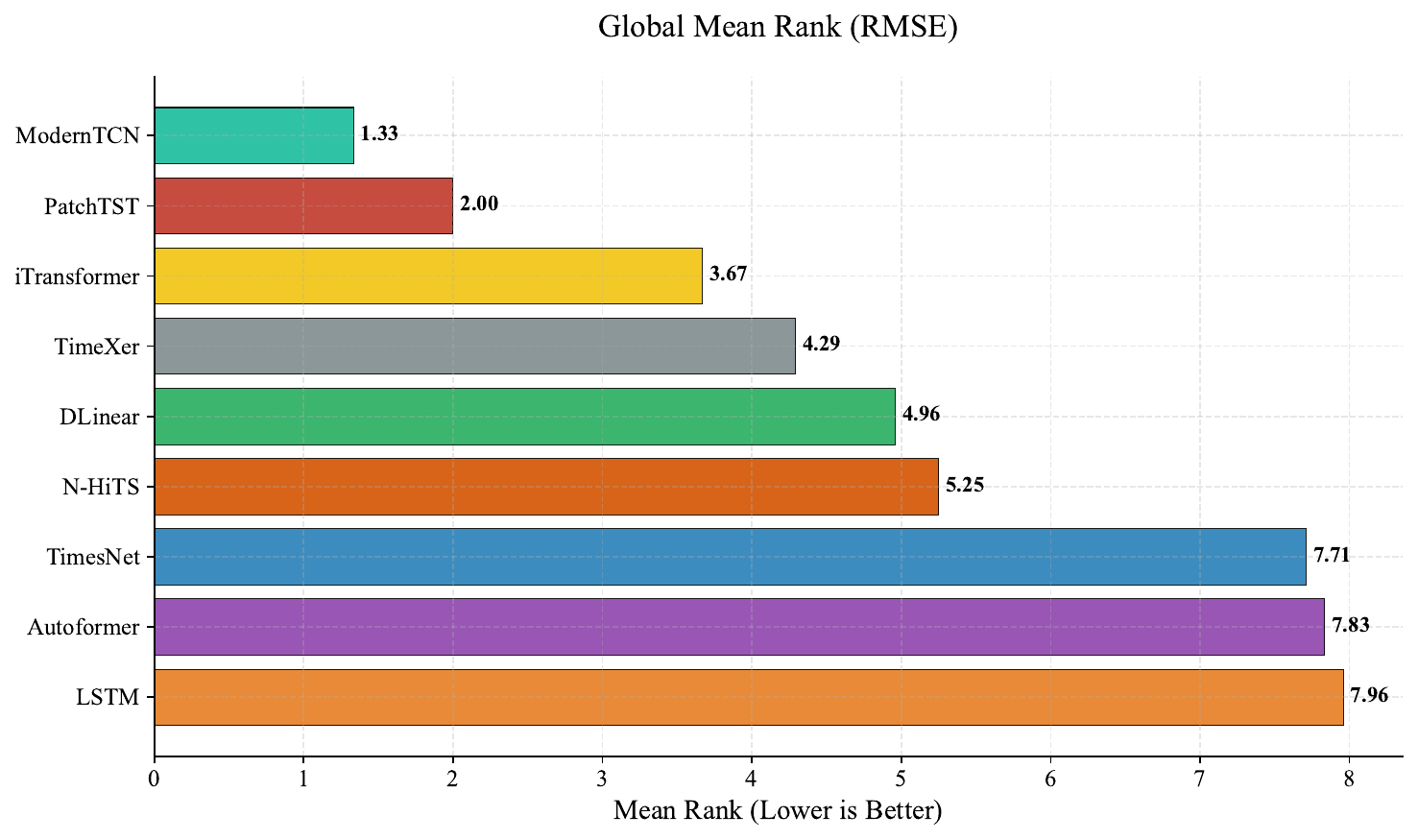}
  \caption{Global mean rank comparison across 24 evaluation points (12~assets $\times$ 2~horizons).  Lower values indicate better performance.  Three distinct tiers are visible: \moderntcn and \patchtst (ranks 1--2), a middle group of four models (ranks 3--6), and a bottom group comprising \timesnet, \autoformer, and \lstm (ranks 7--9).  Error bars represent rank standard deviation across evaluation points.}
  \label{fig:global_rank_comparison}
\end{figure}

Figure~\ref{fig:global_heatmap} displays the \rmse heatmap across all evaluation points.  The body panel excludes \lstm to reveal finer distinctions among the eight modern architectures (the all-models variant appears in Appendix~\ref{sec:app_dual_plots}).  \moderntcn and \patchtst consistently occupy the lowest-error cells.  Figure~\ref{fig:global_rank_comparison} presents the rank distribution, confirming the three-tier structure.

\subsection{Category-Level Analysis}
\label{sec:category_analysis}

Table~\ref{tab:category_metrics} reports category-level aggregated \rmse and \mae for each model, providing the asset-class dimension of H1.  Error magnitudes differ by several orders of magnitude across categories due to underlying price scales, but relative model ordering is preserved.

\sisetup{
  detect-weight          = true,
  text-series-to-math    = true,
  table-number-alignment = center
}

\begin{table}[htbp]
  \centering
  \caption{Category-level mean \rmse and \mae averaged across all assets within
    each category and across both horizons ($h\in\{4,24\}$).  Values are
    aggregated over 4~assets $\times$ 2~horizons $\times$ 3~seeds.  Bold
    marks the best (lowest) value per category and metric.  \lstm is
    included for completeness but excluded from ranking discussions
    due to convergence failures.}
  \label{tab:category_metrics}
  \footnotesize
  \setlength{\tabcolsep}{7pt}%
  \resizebox{\textwidth}{!}{%
  \begin{tabular}{%
    l%
    S[table-format=4.2, round-mode=places, round-precision=2] @{\hspace{2pt}}
    S[table-format=4.2, round-mode=places, round-precision=2]
    S[table-format=1.4, round-mode=places, round-precision=4] @{\hspace{2pt}}
    S[table-format=1.4, round-mode=places, round-precision=4]
    S[table-format=4.2, round-mode=places, round-precision=2] @{\hspace{2pt}}
    S[table-format=4.2, round-mode=places, round-precision=2]
  }
    \toprule
    & \multicolumn{2}{c}{\textbf{Crypto}}
    & \multicolumn{2}{c}{\textbf{Forex}}
    & \multicolumn{2}{c}{\textbf{Indices}} \\
    \cmidrule(lr){2-3} \cmidrule(lr){4-5} \cmidrule(lr){6-7}
    \textbf{Model}
      & {\rmse} & {\mae}
      & {\rmse} & {\mae}
      & {\rmse} & {\mae} \\
    \midrule
    \autoformer         & 532.21  & 393.83  & 0.2068 & 0.1551 & 179.66  & 136.24  \\
    \dlinear            & 323.48  & 220.87  & 0.1279 & 0.0940 & 123.80  &  90.67  \\
    \itransformer       & 320.83  & 217.99  & 0.1136 & 0.0785 & 113.72  &  79.13  \\
    \lstm               & 2398.94 & 2041.58 & 3.6679 & 3.5858 & 1548.56 & 1478.21 \\
    \textbf{\moderntcn} 
                        & \bfseries 314.66 & \bfseries 211.14 
                        & \bfseries 0.1098 & \bfseries 0.0750 
                        & \bfseries 110.89 & \bfseries 76.09 \\
    \nhits              & 352.91  & 255.77  & 0.2356 & 0.1977 & 135.01  & 101.04  \\
    \patchtst           & 314.87  & 211.70  & 0.1108 & 0.0761 & 111.71  &  77.04  \\
    \timesnet           & 352.92  & 249.41  & 0.2127 & 0.1596 & 172.98  & 131.61  \\
    \timexer            & 318.38  & 215.24  & 0.1174 & 0.0815 & 114.09  &  79.19  \\
    \bottomrule
  \end{tabular}%
  }%
\end{table}

\paragraph{Cryptocurrency.}
\moderntcn achieves the lowest mean \rmse (314.66), closely followed by \patchtst (314.87; $\Delta = 0.07\%$) and \timexer (318.38).  \itransformer (320.83) and \dlinear (323.48) are competitive within a 3\% range of the leader.  \lstm exhibits errors approximately $7.6\times$ higher than the best model (2{,}398.94), reflecting fundamental convergence difficulties under the standard training protocol.  \nhits (352.91), \autoformer (532.21), and \timesnet (352.92) form the lower-performing group.

\paragraph{Forex.}
\moderntcn leads (mean \rmse 0.1098), followed by \patchtst (0.1108; $\Delta = 0.9\%$) and \itransformer (0.1136).  \timexer (0.1174) and \dlinear (0.1279) remain competitive on an absolute scale.  \lstm (3.668) is approximately $33\times$ worse than the leader, while \nhits (0.2356), \autoformer (0.2068), and \timesnet (0.2127) form the lower tier.

\paragraph{Equity indices.}
\moderntcn achieves the lowest \rmse (110.89), with \patchtst (111.71; $\Delta = 0.7\%$), \itransformer (113.72), and \timexer (114.09) in close succession.  \dlinear (123.80) and \nhits (135.01) are moderately higher.  \timesnet (172.98), \autoformer (179.67), and \lstm (1{,}548.56) trail substantially.

\begin{figure}[htbp]
  \centering
  \includegraphics[width=0.75\textwidth]{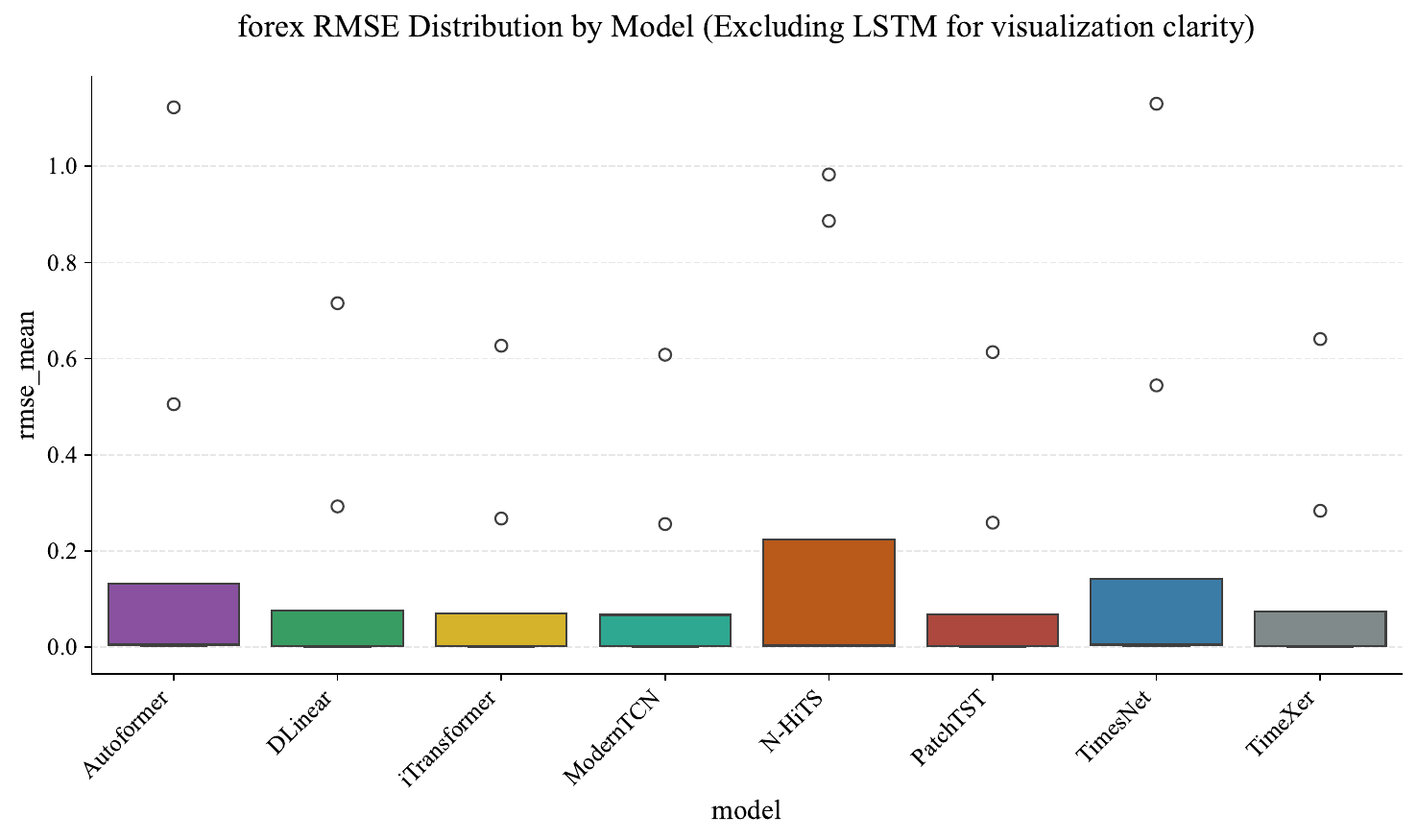}
  \caption{Category-level rank distributions across assets within each category, excluding \lstm for visual clarity.  \moderntcn exhibits the tightest rank distribution (consistently rank~1 across all categories), indicating stable cross-asset performance.  The full nine-model variant is provided in Appendix~\ref{sec:app_dual_plots}.  Boxes show interquartile range; whiskers extend to the most extreme rank observed.}
  \label{fig:category_boxplot}
\end{figure}

\begin{figure}[htbp]
  \centering
  \includegraphics[width=0.75\textwidth]{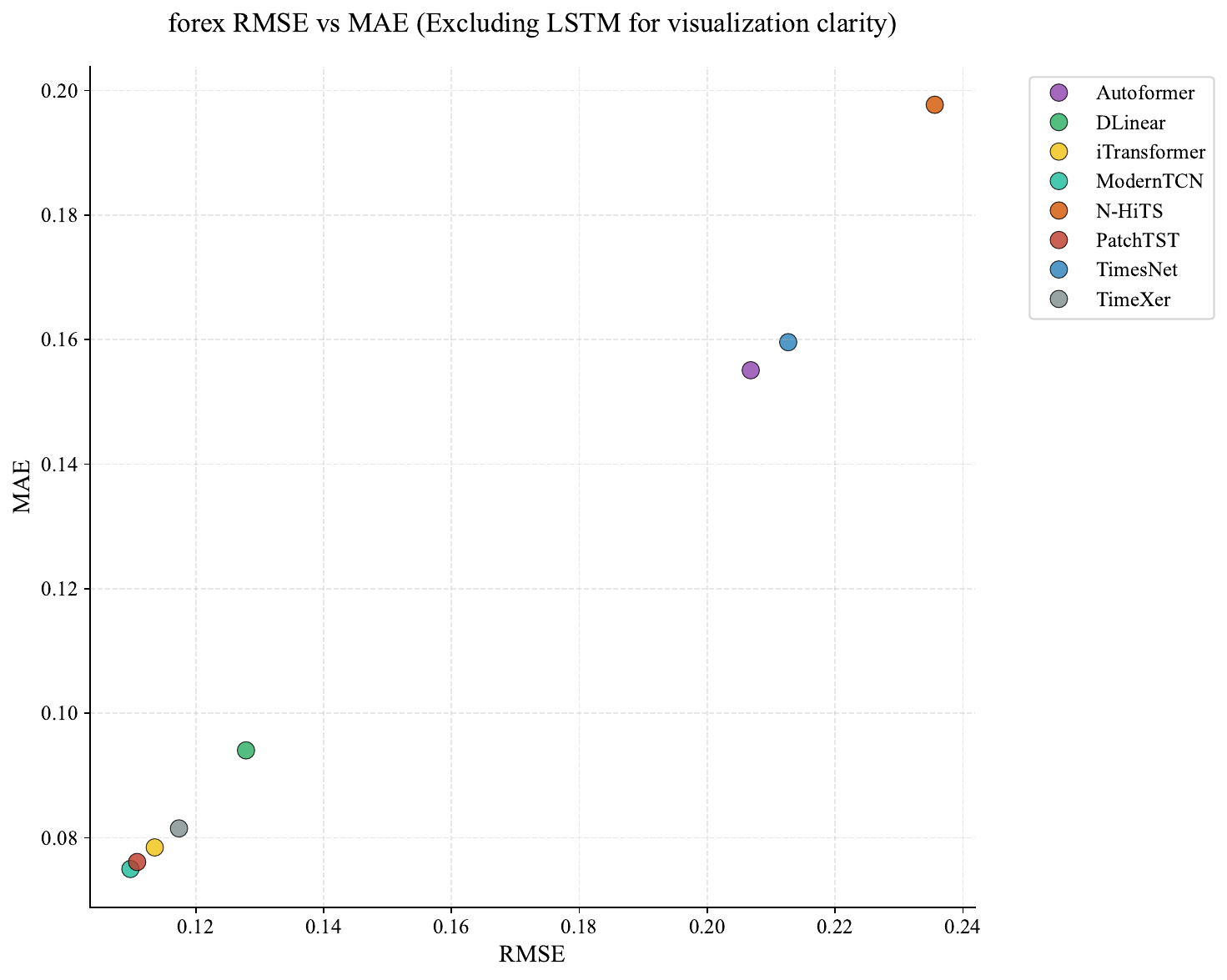}
  \caption{Category-level \rmse vs.\ \mae scatter plot for eight modern architectures.  Each point represents one model's mean error within a category.  The near-perfect linear correlation ($R^2 > 0.99$) confirms that model rankings are consistent across the two error metrics, indicating that findings based on \rmse generalise to \mae.  The full nine-model variant is provided in Appendix~\ref{sec:app_dual_plots}.}
  \label{fig:category_scatter}
\end{figure}

\paragraph{Cross-category ranking consistency.}
Across all three categories, \moderntcn and \patchtst consistently occupy the top two positions (Figure~\ref{fig:category_boxplot}).  This consistency is confirmed by the \rmse--\mae concordance in Figure~\ref{fig:category_scatter}: near-perfect linear correlation between the two error metrics shows that rankings are robust to metric choice.  Category dendrograms and per-category performance matrices appear in Appendix~\ref{sec:app_category_rankings} (Figures~\ref{fig:app_category_hierarchical_rankings} and~\ref{fig:app_category_performance_matrices}); \moderntcn and \patchtst cluster at the top of every dendrogram.

\subsection{Cross-Horizon Degradation}
\label{sec:horizon_degradation}

Table~\ref{tab:horizon_degradation_btc} presents \rmse at $\hfour$ and $\htwentyfour$ for each representative asset (BTC/USDT, EUR/USD, Dow Jones), along with the percentage degradation ($\Delta\% = 100 \times (\text{RMSE}_{h=24} - \text{RMSE}_{h=4}) / \text{RMSE}_{h=4}$).  Table~\ref{tab:horizon_ranking_shift} provides the corresponding rank shift $\Delta = r_{24} - r_4$ for each model, isolating relative ordering changes from absolute error magnitudes.


\begin{table}[htbp]
  \centering
  \caption{Horizon degradation for \textbf{BTC/USDT} (top), \textbf{EUR/USD}
    (middle), and \textbf{Dow Jones} (bottom).  \rmse values are three-seed
    means.  $\Delta\%$ denotes the relative \rmse increase from $h{=}4$ to
    $h{=}24$: $\Delta\% = 100\times(\mathrm{RMSE}_{24} -
    \mathrm{RMSE}_{4})/\mathrm{RMSE}_{4}$.  Bold marks the model with the
    lowest \rmse at $h{=}24$ across all nine architectures.}
  \label{tab:horizon_degradation_btc}
  \setlength{\tabcolsep}{9pt}%
  \begin{tabular}{l
    S[table-format=5.2, round-precision=2]
    S[table-format=5.2, round-precision=2]
    S[table-format=3.2, round-precision=2]}
    \toprule
    \textbf{Model}
      & {\rmse, $h{=}4$}
      & {\rmse, $h{=}24$}
      & {$\Delta\%$} \\
    \midrule
    \autoformer   & 1284.9  & 2670.0  & 107.8 \\
    \dlinear      &  772.0  & 1644.1  & 113.0 \\
    \itransformer &  743.5  & 1651.3  & 122.1 \\
    \lstm         & 8029.9  & 10878.7 &  35.5 \\
    \moderntcn    &  731.6  & \bfseries 1617.4  & 121.1 \\
    \nhits        &  930.5  & 1724.5  &  85.3 \\
    \patchtst     &  731.1  & 1619.1  & 121.5 \\
    \timesnet     &  793.6  & 1840.4  & 131.9 \\
    \timexer      &  750.3  & 1624.9  & 116.6 \\
    \bottomrule
  \end{tabular}
\end{table}


\begin{table}[htbp]

  \centering

  \caption{Horizon degradation for \textbf{EUR/USD}.  \rmse values are
    three-seed means.  $\Delta\%$ defined as in
    Table~\ref{tab:horizon_degradation_btc}.}

  \label{tab:horizon_degradation_eurusd}

  \setlength{\tabcolsep}{9pt}%

  \begin{tabular}{l
    S[table-format=1.5, round-precision=5]
    S[table-format=1.5, round-precision=5]
    S[table-format=3.2, round-precision=2]}

    \toprule

    \textbf{Model}
      & {\rmse, $h{=}4$}
      & {\rmse, $h{=}24$}
      & {$\Delta\%$} \\

    \midrule

    \autoformer   & 0.00284 & 0.00626 & 120.3 \\
    \dlinear      & 0.00159 & 0.00347 & 118.5 \\
    \itransformer & 0.00156 & 0.00345 & 121.0 \\
    \lstm         & 0.00179 & 0.00364 & 102.6 \\
    \moderntcn    & 0.00152 & \bfseries 0.00338 & 121.4 \\
    \nhits        & 0.00170 & 0.00353 & 107.4 \\
    \patchtst     & 0.00153 & 0.00344 & 124.9 \\
    \timesnet     & 0.00298 & 0.00627 & 110.3 \\
    \timexer      & 0.00161 & 0.00349 & 116.7 \\

    \bottomrule

  \end{tabular}

\end{table}
\begin{table}[htbp]
  \centering
  \caption{Horizon degradation for \textbf{Dow Jones}.  \rmse values are
    three-seed means.  $\Delta\%$ defined as in
    Table~\ref{tab:horizon_degradation_btc}.}
  \label{tab:horizon_degradation_usa30}
  \setlength{\tabcolsep}{9pt}%
  \begin{tabular}{l
    S[table-format=4.2, round-precision=2]
    S[table-format=4.2, round-precision=2]
    S[table-format=3.2, round-precision=2]}
    \toprule
    \textbf{Model}
      & {\rmse, $h{=}4$}
      & {\rmse, $h{=}24$}
      & {$\Delta\%$} \\
    \midrule
    \autoformer   &  160.7 &  410.7 & 155.6 \\
    \dlinear      &  121.3 &  279.1 & 130.1 \\
    \itransformer &  115.2 &  258.4 & 124.3 \\
    \lstm         & 1261.7 & 1688.5 &  33.8 \\
    \moderntcn    &  112.5 & \bfseries 249.5 & 121.7 \\
    \nhits        &  160.2 &  255.9 &  59.7 \\
    \patchtst     &  113.0 &  252.2 & 123.2 \\
    \timesnet     &  160.3 &  431.4 & 169.1 \\
    \timexer      &  118.4 &  258.3 & 118.1 \\
    \bottomrule
  \end{tabular}
\end{table}
\begin{table}[htbp]
  \centering
  \caption{Horizon ranking shift for the three representative assets.
    $r_4$ and $r_{24}$: model rank at $\hfour$ and $\htwentyfour$ respectively
    (lower is better; 1\,=\,best).
    $\Delta = r_{24} - r_4$: positive values indicate rank degradation
    (the model performs \emph{relatively worse} at longer horizons);
    negative values indicate rank improvement.
    Values in \textbf{bold} denote $|\Delta| \geq 2$.
    Rankings are based on mean \rmse across three seeds.}
  \label{tab:horizon_ranking_shift}
  \setlength{\tabcolsep}{6pt}%
  \begin{threeparttable}
  \begin{tabular*}{\textwidth}{%
    @{\extracolsep{\fill}}
    l
    S[table-format=1.2, round-precision=2]
    S[table-format=1.2, round-precision=2]
    S[table-format=+1.2, round-precision=2]
    @{\hspace{2em}\extracolsep{\fill}}
    S[table-format=1.2, round-precision=2]
    S[table-format=1.2, round-precision=2]
    S[table-format=+1.2, round-precision=2]
    @{\hspace{2em}\extracolsep{\fill}}
    S[table-format=1.2, round-precision=2]
    S[table-format=1.2, round-precision=2]
    S[table-format=+1.2, round-precision=2]
    @{}
  }
    \toprule
    & \multicolumn{3}{c}{\textbf{BTC/USDT}}
    & \multicolumn{3}{c}{\textbf{EUR/USD}}
    & \multicolumn{3}{c}{\textbf{Dow Jones}} \\
    \cmidrule(lr){2-4}\cmidrule(lr){5-7}\cmidrule(lr){8-10}
    \textbf{Model}
      & {$r_4$} & {$r_{24}$} & {$\Delta$}
      & {$r_4$} & {$r_{24}$} & {$\Delta$}
      & {$r_4$} & {$r_{24}$} & {$\Delta$} \\
    \midrule
    \autoformer   & 8 & 8 &  0 & 8 & 8 &  0 & 8 & 7 & -1 \\
    \dlinear      & 5 & 4 & -1 & 4 & 4 &  0 & 5 & 6 & +1 \\
    \itransformer & 3 & 5 & \bfseries +2 & 3 & 3 &  0 & 3 & 5 & \bfseries +2 \\
    \lstm         & 9 & 9 &  0 & 7 & 7 &  0 & 9 & 9 &  0 \\
    \moderntcn    & 2 & 1 & -1 & 1 & 1 &  0 & 1 & 1 &  0 \\
    \nhits        & 7 & 6 & -1 & 6 & 6 &  0 & 6 & 3 & \bfseries -3 \\
    \patchtst     & 1 & 2 & +1 & 2 & 2 &  0 & 2 & 2 &  0 \\
    \timesnet     & 6 & 7 & +1 & 9 & 9 &  0 & 7 & 8 & +1 \\
    \timexer      & 4 & 3 & -1 & 5 & 5 &  0 & 4 & 4 &  0 \\
    \bottomrule
  \end{tabular*}
  \begin{tablenotes}[flushleft]\footnotesize
    \item EUR/USD exhibits perfect rank stability ($\Delta = 0$ for all nine models),
    confirming that the forex ranking is invariant to horizon.
    iTransformer degrades by 2 positions in both BTC/USDT and Dow Jones,
    suggesting its inductive bias is less suited to longer-horizon financial prediction.
    N-HiTS improves by 3 positions in Dow Jones at $\htwentyfour$, the largest positive
    shift observed, consistent with its multi-rate pooling design capturing longer
    temporal structure in indices.
  \end{tablenotes}
  \end{threeparttable}
\end{table}

All models exhibit higher \rmse at $\htwentyfour$ relative to $\hfour$, consistent with the established understanding that prediction uncertainty grows with the forecast window.  However, degradation rates vary meaningfully across architectures and assets:

\paragraph{BTC/USDT.}
Degradation ranges from 85.3\% (\nhits) to 131.9\% (\timesnet) among modern architectures.  The top-tier models \moderntcn ($\Delta\% = 121.1$) and \patchtst ($\Delta\% = 121.5$) exhibit nearly identical degradation.  \lstm shows the lowest relative degradation (35.5\%), not because of strong long-horizon performance, but because its $\hfour$ errors are already substantially elevated (8{,}029.9 vs.\ 731.6 for \moderntcn).

\paragraph{EUR/USD.}
Degradation magnitudes are comparable: \moderntcn ($\Delta\% = 121.4$) and \patchtst ($\Delta\% = 124.9$) degrade similarly, while \nhits ($\Delta\% = 107.4$) shows the lowest degradation among modern architectures.

\paragraph{Dow Jones.}
\timesnet exhibits the highest degradation (169.1\%), followed by \autoformer (155.6\%).  \nhits shows notably lower degradation (59.7\%), suggesting that its hierarchical multi-rate pooling may capture multi-scale patterns that transfer across horizons.

\paragraph{Cross-horizon rank stability.}
Despite $2$--$2.5\times$ absolute error amplification, top-tier models maintain their relative ranking across horizons for all three representative assets: \moderntcn and \patchtst hold positions 1--2 at both $\hfour$ and $\htwentyfour$.  Rank shifts are concentrated in the middle tier; for example, \nhits improves by 3~ranks on Dow Jones (from rank~6 to rank~3), while \itransformer drops by 2~ranks on BTC/USDT (from rank~3 to rank~5).  EUR/USD rankings are perfectly stable across horizons.  A comprehensive heatmap of per-model per-asset degradation percentages appears in Appendix~\ref{sec:app_horizon_sensitivity} (Figure~\ref{fig:app_horizon_sensitivity_heatmap}).  Section~\ref{sec:qualitative_fidelity} provides complementary qualitative evidence through actual-versus-predicted overlays.

\begin{figure}[htbp]
  \centering
  \includegraphics[width=\textwidth]{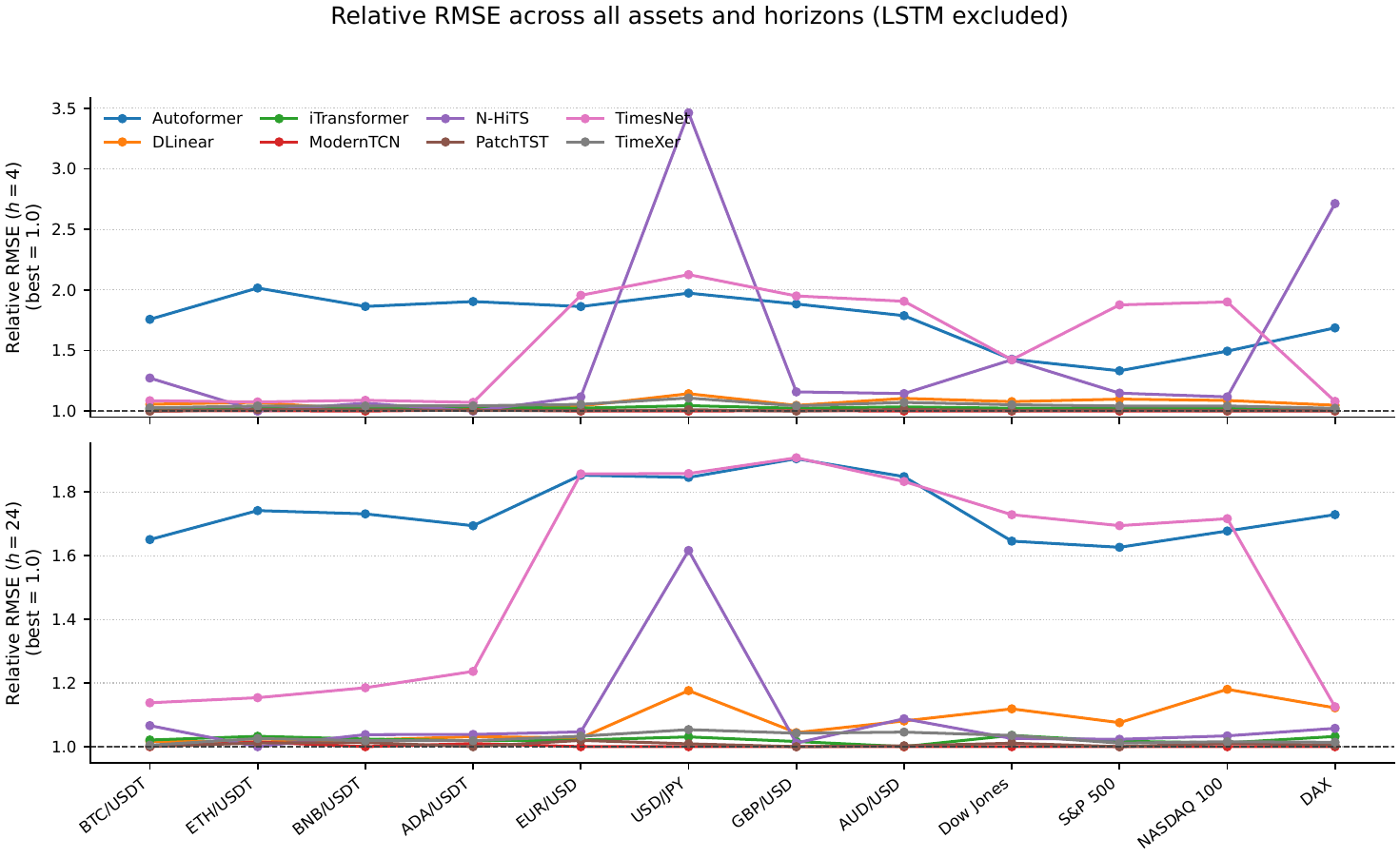}
  \caption{Relative \rmse across all twelve assets and both horizons for eight modern architectures (\lstm excluded).  Each point is the stored mean \rmse divided by the best model's mean \rmse within that asset--horizon slot, so 1.0 is the slot winner and values above 1.0 are relatively worse.  The scale-free presentation permits comparison across cryptocurrency, forex, and equity-index price magnitudes.}
  \label{fig:asset_model_comparison_grouped}
\end{figure}

Figure~\ref{fig:asset_model_comparison_grouped} extends this analysis to all twelve assets.  \moderntcn and \patchtst consistently occupy the lowest-error positions across all assets and both horizons, confirming that their inductive biases---large-kernel convolutions and patch-based self-attention---generalise robustly to extended temporal contexts.  Conversely, \timesnet and \autoformer show the most pronounced degradation, suggesting that their multi-periodicity mechanisms are susceptible to fidelity loss at longer horizons in high-noise financial domains.

\begin{figure}[htbp]
  \centering
  \includegraphics[width=0.75\textwidth]{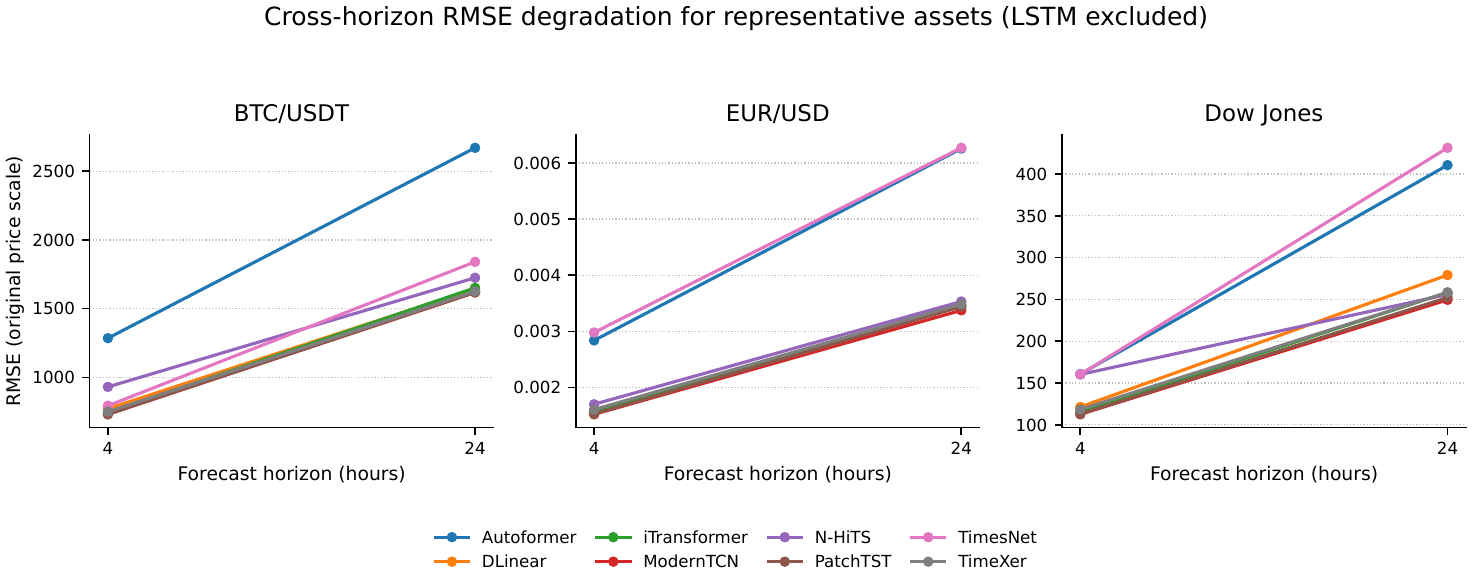}
  \caption{Cross-horizon \rmse degradation for eight modern architectures on the three representative assets used for HPO: BTC/USDT, EUR/USD, and Dow Jones (\lstm excluded).  Lines connect each model's stored mean \rmse at $\hfour$ and $\htwentyfour$; the three panels retain their original price scales.  All architectures exhibit absolute error growth, while the ordering of the top tier remains stable.  The full nine-model variant is provided in Appendix~\ref{sec:app_dual_plots}.}
  \label{fig:horizon_degradation}
\end{figure}

\subsection{Seed Robustness and Variance Decomposition}
\label{sec:seed_robustness}

The two-factor variance decomposition (Table~\ref{tab:variance_decomposition}) is reported in three panels: raw \rmse, z-score normalised \rmse across all nine models, and z-score normalised \rmse excluding the \lstm outlier.

\begin{table}[htbp]
  \centering
  \begin{threeparttable}
  \caption{Two-factor variance decomposition of forecast \rmse.
    Raw (untransformed) and z-normalised (within each asset--horizon slot)
    panels are shown with and without \lstm.  Seed variance is negligible
    ($<\!0.1\,\%$) in all cases.}
  \label{tab:variance_decomposition}
  \setlength{\tabcolsep}{9pt}%
  \begin{tabular}{%
    l%
    S[table-format=2.2]
    S[table-format=2.2]
    S[table-format=2.2]
  }
    \toprule
    \textbf{Factor}
      & {\textbf{Raw (\%)}}
      & {\textbf{$z$-norm, all (\%)}}
      & {\textbf{$z$-norm, no \lstm (\%)}} \\
    \midrule
    \textbf{Model (architecture)}  & \bfseries 99.90 & \bfseries 48.32 & \bfseries 68.33 \\
    Seed (initialisation)          &  0.01 &  0.04 &  0.02 \\
    Residual (model$\times$slot)   &  0.09 & 51.64 & 31.66 \\
    \bottomrule
  \end{tabular}
  \begin{tablenotes}[flushleft]\footnotesize
    \item \textbf{Raw} panel: pooled variance decomposition on untransformed \rmse;
      \lstm's errors ($7\times$--$33\times$ higher than best model) dominate
      the model sum of squares, inflating the Model fraction to 99.90\,\%.
    \item \textbf{$z$-norm} panels: \rmse standardised within each
      (asset, horizon) slot before ANOVA, removing price-magnitude scale
      effects across asset classes.
    \item Residual ($\approx\!32$--$52\,\%$) reflects model$\times$slot
      interaction: each architecture has a context-dependent advantage on
      different asset--horizon combinations.  Seed variance ($<\!0.1\,\%$)
      is negligible across all three panels, validating the three-seed protocol.
  \end{tablenotes}
  \end{threeparttable}
\end{table}

\paragraph{Raw panel.}
On the original price scale, architecture absorbs 99.90\% of total sum-of-squares variance, versus 0.01\% for seed and 0.09\% for the residual.  This extreme dominance is largely an artefact of \lstm's outlier \rmse values: a single model $7{-}33\times$ above the median inflates $\mathrm{SS}_{\text{model}}$ relative to all other terms.

\paragraph{Z-normalised panels: all models.}
After z-scoring each model's \rmse within each (asset, horizon) slot, the architecture factor falls to \textbf{48.32\%}, seed to \textbf{0.04\%}, and the residual rises to \textbf{51.64\%}.  The large residual reflects genuine heterogeneity in which model excels on a given asset--horizon combination---a meaningful signal rather than noise.

\paragraph{Z-normalised panel: modern models only.}
Excluding \lstm, architecture recovers to \textbf{68.33\%} (seed: 0.02\%; residual: 31.66\%), confirming that architecture remains the dominant factor even among competitive modern models, though asset--horizon context contributes substantially.

In all three panels, seed variance is negligible ($\leq 0.04\%$): rankings are stable with respect to random initialisation and three seeds suffice.

Per-asset variance decompositions corroborate the global result.  At $\hfour$, architecture accounts for 99.75\% of variance on BTC/USDT, 97.44\% on EUR/USD, and 99.08\% on Dow Jones; seed fractions are 0.02\%, 0.35\%, and 0.01\% respectively.  Even EUR/USD, with the highest relative seed contribution, shows seed variance two orders of magnitude below the architecture factor.  At $\htwentyfour$, the pattern tightens further: architecture explains 99.86\% on BTC/USDT (seed: 0.02\%), confirming that initialisation effects diminish---rather than amplify---at longer horizons.

\begin{figure}[htbp]
  \centering
  \begin{subfigure}[b]{0.48\textwidth}
    \includegraphics[width=\textwidth]{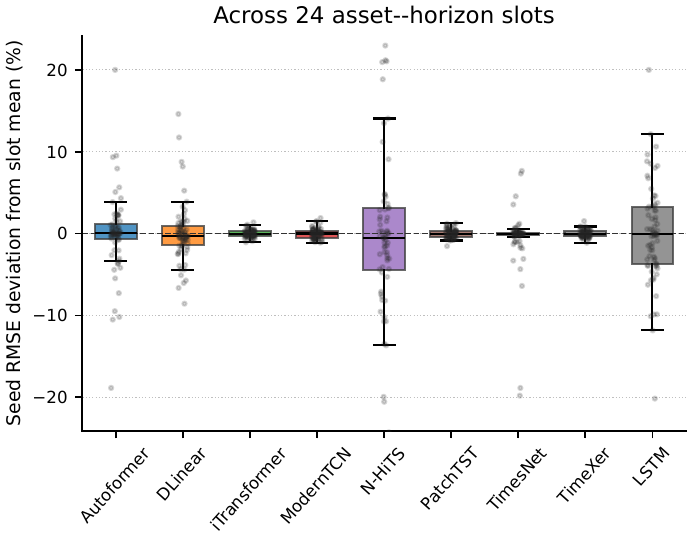}
    \caption{Seed RMSE deviations.}
    \label{fig:seed_violin}
  \end{subfigure}
  \hfill
  \begin{subfigure}[b]{0.48\textwidth}
    \includegraphics[width=\textwidth]{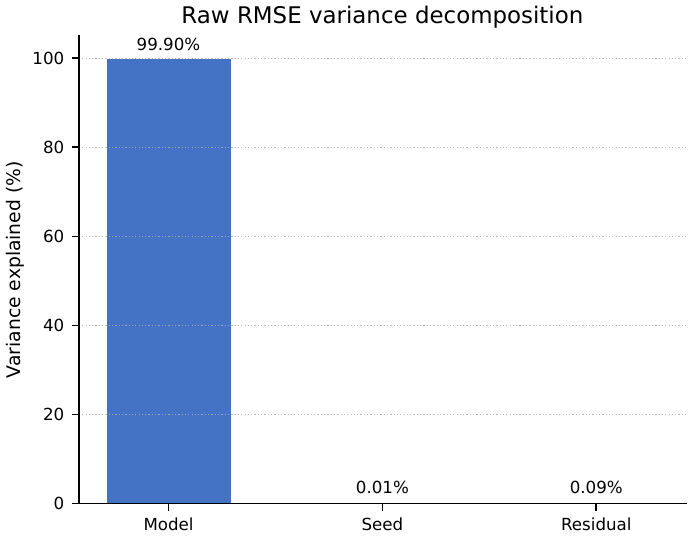}
    \caption{Raw variance shares.}
    \label{fig:variance_pie}
  \end{subfigure}
\caption{Seed robustness analysis regenerated from all stored per-seed metrics.  (a)~Boxplots and lightly overlaid observations show each seed's percentage deviation from the mean within its asset--horizon--model slot across 24 slots.  (b)~Annotated raw variance shares: architecture explains 99.90\%, seed 0.01\%, and residual 0.09\%.  After z-score normalisation within each slot, the architecture share falls to 48.3\% (68.3\% excluding \lstm); see Table~\ref{tab:variance_decomposition} for the full breakdown.}
  \label{fig:robustness}
\end{figure}

Figure~\ref{fig:robustness}(a) displays seed-to-seed \rmse deviations per model as boxplots with lightly overlaid observations.  Inter-seed variance is negligible relative to inter-model differences across all nine architectures: even \lstm, which has the highest absolute seed variance, shows seed-induced variation that is small relative to its distance from the nearest competitor.  Figure~\ref{fig:robustness}(b) provides an annotated bar chart of the raw variance decomposition; the z-normalised breakdown appears in Table~\ref{tab:variance_decomposition}.  Per-asset seed-variance box plots and scatter plots for all three representative assets appear in Appendix~\ref{sec:app_robustness} (Figures~\ref{fig:app_seed_boxplot_h4}--\ref{fig:app_scatter_seed_h24}).

\subsection{Directional Accuracy}
\label{sec:directional_accuracy}

Directional accuracy (\da) quantifies the fraction of forecasts that correctly predict the sign of the next price change.  Table~\ref{tab:directional_accuracy} reports mean \da\ per model, category, and horizon.

\sisetup{
  round-mode              = places,
  round-precision         = 2,
  detect-weight           = true,
  text-series-to-math     = true,
  table-number-alignment  = center
}

\begin{table}[htbp]
  \centering
  \begin{threeparttable}
  \caption{Mean directional accuracy (\da, \%) per model, asset class, and
    horizon, averaged across assets within each category and over three
    seeds. Values close to 50\% indicate no systematic directional bias.}
  \label{tab:directional_accuracy}
  \footnotesize
  \setlength{\tabcolsep}{6pt}

  \begin{tabular}{
    l
    S[table-format=2.2]
    S[table-format=2.2]
    S[table-format=2.2]
    S[table-format=2.2]
    S[table-format=2.2]
    S[table-format=2.2]
  }
    \toprule
    & \multicolumn{2}{c}{\textbf{Crypto}}
    & \multicolumn{2}{c}{\textbf{Forex}}
    & \multicolumn{2}{c}{\textbf{Indices}} \\
    \cmidrule(lr){2-3}\cmidrule(lr){4-5}\cmidrule(lr){6-7}
    \textbf{Model}
      & {$h{=}4$} & {$h{=}24$}
      & {$h{=}4$} & {$h{=}24$}
      & {$h{=}4$} & {$h{=}24$} \\
    \midrule
    \autoformer     & 50.41 & 49.96 & 50.06 & 50.11 & 50.01 & 49.98 \\
    \dlinear        & 49.70 & 49.94 & 50.13 & 50.12 & 49.97 & 49.96 \\
    \itransformer   & 49.58 & 50.22 & 50.04 & 49.99 & 49.90 & 50.18 \\
    \lstm           & 49.73 & 50.07 & 50.08 & 49.95 & 49.88 & 49.95 \\
    \moderntcn      & 50.04 & 49.96 & 49.96 & 50.03 & 50.34 & 50.03 \\
    \nhits          & 50.05 & 49.87 & 50.02 & 50.03 & 50.78 & 49.92 \\
    \patchtst       & 50.07 & 49.85 & 50.10 & 49.98 & 50.42 & 49.95 \\
    \timesnet       & 50.57 & 50.34 & 50.05 & 49.88 & 50.27 & 50.18 \\
    \timexer        & 49.86 & 49.94 & 49.98 & 50.08 & 50.92 & 50.20 \\
    \bottomrule
  \end{tabular}

  \begin{tablenotes}[flushleft]\footnotesize
    \item Mean \da\ across all 54 model--category--horizon combinations is
      50.08\%, indistinguishable from a fair-coin baseline. No combination
      deviates meaningfully from 50\%. Values below 50\,\% indicate slight
      down-trend bias arising from scaling artefacts, not genuine negative
      directional skill.
  \end{tablenotes}

  \end{threeparttable}
\end{table}

Across all 9~models $\times$ 3~categories $\times$ 2~horizons ($=54$ combinations), mean \da\ is \textbf{50.08\%}.  No combination deviates meaningfully from 50\% and no horizon trend is discernible.

MSE-trained deep learning architectures therefore show no established directional
advantage in this benchmark.  DA near 50\% is compatible with weak-form
efficiency at hourly resolution, but it can also arise structurally from
regression-to-the-mean when an MSE objective estimates a quasi-martingale
conditional mean.  The present design cannot distinguish these mechanisms, so
the result should not be read as a standalone test of the efficient-market
hypothesis.  Section~\ref{sec:economic_interpretation} discusses implications
for trading strategies and alternative objectives.

\subsection{Statistical Significance Tests}
\label{sec:statistical_tests}

Tables~\ref{tab:stat_tests_friedman}--\ref{tab:stat_tests_jt} report the full battery of statistical tests, providing formal confirmation of the descriptive findings in Sections~\ref{sec:global_rankings}--\ref{sec:directional_accuracy}.


\begin{table}[htbp]
  \centering
  \begin{threeparttable}
  \caption{Statistical significance tests — Panel~A: Friedman-Iman-Davenport
    omnibus tests.  \chisq: Friedman $\chi^2$ statistic (8~df); $F$:
    Iman-Davenport $F$-statistic; $n$: number of blocks (evaluation points).
    All tests reject \Hzero at $\alpha = 0.001$.}
  \label{tab:stat_tests_friedman}
  \setlength{\tabcolsep}{7pt}%
  \begin{tabular}{%
    l
    S[table-format=3.2]
    S[table-format=3.2]
    l
    l
    c
  }
    \toprule
    \textbf{Scope}
      & {\chisq(8)}
      & {$F$}
      & {df}
      & {$p$}
      & {$n$} \\
    \midrule
    Global (all 12 assets, $h\in\{4,24\}$) & 156.49 & 101.36 & (8,\;184) & $<\!10^{-15}$ & 24 \\
    Crypto (4~assets, $h\in\{4,24\}$)      &  49.23 &  23.34 & (8,\;56)  & $<\!10^{-15}$ &  8 \\
    Forex (4~assets, $h\in\{4,24\}$)       &  56.00 &  49.00 & (8,\;56)  & $<\!10^{-15}$ &  8 \\
    Indices (4~assets, $h\in\{4,24\}$)     &  60.40 & 117.44 & (8,\;56)  & $<\!10^{-15}$ &  8 \\
    \bottomrule
  \end{tabular}
  \end{threeparttable}
\end{table}

\begin{table}[htbp]
  \centering
  \begin{threeparttable}
  \caption{Statistical significance tests — Panel~B: Spearman rank
    correlations between model rankings at $\hfour$ and $\htwentyfour$, plus
    Stouffer combined test.  Rankings are based on mean \rmse across three
    seeds for each of the nine architectures ($n = 9$ per asset).
    All 12 per-asset correlations are significant at $\alpha = 0.05$.}
  \label{tab:stat_tests_spearman}
  \setlength{\tabcolsep}{8pt}%
  \begin{tabular}{%
    l
    l
    S[table-format=1.4]
    l
  }
    \toprule
    \textbf{Category} & \textbf{Asset}
      & {Spearman $\rho$}
      & {$p$-value} \\
    \midrule
    Crypto  & ADA/USDT          & 0.683 & 0.0424 \\
    Crypto  & BNB/USDT          & 0.917 & 0.0005 \\
    Crypto  & BTC/USDT          & 0.917 & 0.0005 \\
    Crypto  & ETH/USDT          & 0.783 & 0.0125 \\
    \addlinespace
    Forex   & AUD/USD           & 0.867 & 0.0025 \\
    Forex   & EUR/USD           & 1.000 & $<\!10^{-9}$ \\
    Forex   & GBP/USD           & 0.800 & 0.0096 \\
    Forex   & USD/JPY           & 0.950 & 0.0001 \\
    \addlinespace
    Indices & DAX               & 0.883 & 0.0016 \\
    Indices & Dow Jones         & 0.867 & 0.0025 \\
    Indices & S\&P~500          & 0.967 & $<\!10^{-4}$ \\
    Indices & NASDAQ~100        & 0.983 & $<\!10^{-5}$ \\
    \midrule
    \multicolumn{2}{l}{\textbf{Stouffer combined} ($n = 12$ assets)}
      & \multicolumn{1}{S[table-format=1.2]}{6.17}
      & $3.47\times10^{-10}$ \\
    \bottomrule
  \end{tabular}
  \begin{tablenotes}[flushleft]\footnotesize
    \item Stouffer combined $Z$-statistic aggregates the 12 per-asset
      one-sided Spearman $p$-values using the inverse-normal method.
      The global $p = 3.47\times10^{-10}$ confirms that cross-horizon
      rank stability is a systematic property of the benchmark.
  \end{tablenotes}
  \end{threeparttable}
\end{table}

\begin{table}[htbp]
  \centering
  \begin{threeparttable}
  \caption{Statistical significance tests — Panel~C: Intraclass Correlation
    Coefficient (\icc; two-way mixed, absolute agreement) for three
    representative assets at $\htwentyfour$.  High \icc values confirm
    negligible seed-to-seed variation relative to inter-model differences.}
  \label{tab:stat_tests_icc}
  \setlength{\tabcolsep}{8pt}%
  \begin{tabular}{%
    l
    S[table-format=1.4]
    S[table-format=4.1]
    l
  }
    \toprule
    \textbf{Asset ($h{=}24$)}
      & {\icc}
      & {$F$-statistic}
      & {$p$-value} \\
    \midrule
    BTC/USDT     & 0.9982 & 1650.2 & $<\!10^{-15}$ \\
    EUR/USD      & 0.9904 &  309.6 & $<\!10^{-15}$ \\
    S\&P~500     & 0.9987 & 2255.6 & $<\!10^{-15}$ \\
    \bottomrule
  \end{tabular}
  \begin{tablenotes}[flushleft]\footnotesize
    \item Each \icc computed over 9 models $\times$ 3 seeds.  $F$-statistic
      tests \Hzero: all seed means are equal.  Values above 0.99 indicate
      that $>99\%$ of inter-model variance is attributable to architecture
      rather than random initialisation.
  \end{tablenotes}
  \end{threeparttable}
\end{table}

\begin{table}[htbp]
  \centering
  \begin{threeparttable}
  \caption{Statistical significance tests — Panel~D:
    Jonckheere-Terpstra (\jt) test for a monotonic relationship between
    model complexity (parameter count) and \rmse rank.  A significant
    positive result would indicate that more parameters reliably yield
    better ranks.  Three complexity groups: $\leq\!30\mathrm{K}$,
    $30\mathrm{K}$--$200\mathrm{K}$, $>\!200\mathrm{K}$ parameters.}
  \label{tab:stat_tests_jt}
  \setlength{\tabcolsep}{8pt}%
  \begin{tabular}{%
    l
    c
    S[table-format=+1.4]
    S[table-format=1.4]
    c
  }
    \toprule
    \textbf{Category}
      & \textbf{Horizon}
      & {\jt $z$}
      & {$p$-value}
      & \textbf{Monotonic?} \\
    \midrule
    Crypto  & 4  & +0.348 & 0.364 & No \\
    Crypto  & 24 & -0.348 & 0.636 & No \\
    \addlinespace
    Forex   & 4  & +0.377 & 0.353 & No \\
    Forex   & 24 & -0.290 & 0.614 & No \\
    \addlinespace
    Indices & 4  & -1.248 & 0.894 & No \\
    Indices & 24 & -1.480 & 0.931 & No \\
    \bottomrule
  \end{tabular}
  \begin{tablenotes}[flushleft]\footnotesize
    \item No test achieves significance ($\alpha = 0.05$); all $p > 0.35$.
      Negative $z$ values indicate a tendency for \emph{fewer} parameters
      to yield better ranks, consistent with the Pareto frontier defined
      by \dlinear, \patchtst, and \moderntcn (Figure~\ref{fig:complexity}).
  \end{tablenotes}
  \end{threeparttable}
\end{table}

\paragraph{Global Friedman-Iman-Davenport test.}
The Friedman-Iman-Davenport test on all 24~evaluation points (12~assets $\times$ 2~horizons) yields $F(8,184) = 101.36$, $p < 10^{-15}$ (Friedman $\chisq(8) = 156.49$, $p = 8.67 \times 10^{-30}$), firmly rejecting the null hypothesis that all nine architectures perform equivalently.  Table~\ref{tab:stat_tests_friedman} indicates that exactly the same conclusion holds within every individual asset class: crypto ($F = 23.34$), forex ($F = 49.00$), and indices ($F = 117.44$), each with $p < 10^{-15}$.

\paragraph{Post-hoc Holm-Wilcoxon pairwise tests (global).}
Of the $\binom{9}{2} = 36$ pairwise Wilcoxon comparisons at the global level ($n = 24$ observations), 33~are statistically significant after Holm correction ($\alpha = 0.05$).  The three non-significant pairs are all \emph{intra-tier}: \timexer~vs.~\itransformer ($p_\text{Holm} = 0.480$), \autoformer~vs.~\timesnet ($p_\text{Holm} = 0.529$), and \dlinear~vs.~\nhits ($p_\text{Holm} = 0.529$) --- confirming that only neighbouring models within the same performance tier are statistically indistinguishable.

At the per-category level ($n = 8$: 4~assets $\times$ 2~horizons), no pairwise comparison reaches significance after Holm correction (all $p_\text{Holm} > 0.28$).  This reflects a \emph{power constraint}, not an absence of effect: with $n = 8$, the minimum achievable Wilcoxon $p$-value is $0.0078$; the Holm step-down procedure requires the most extreme raw $p$-value to beat $0.05/36 = 0.0014$, which is unattainable at this sample size.  The per-category result is thus consistent with---and subsumed by---the decisive global test.

\paragraph{Critical difference diagram.}
Figure~\ref{fig:cd_diagram} visualises the Holm-Wilcoxon significance structure as a critical difference (CD) diagram.  The horizontal axis represents mean rank across all $N = 24$ evaluation blocks ($k = 9$ models); lower values denote superior performance.  Each model is placed at its exact mean rank derived from \texttt{global\_ranking\_aggregated.csv}.  Thick horizontal bars connect pairs that are \emph{not} statistically distinguishable after Holm correction ($\alpha = 0.05$); all unlabelled pairs are significant.

Three intra-tier equivalence groups emerge directly from the pairwise test results.  Within the middle tier, \itransformer (rank~3.667) and \timexer (rank~4.292) are statistically indistinguishable ($p_\text{Holm} = 0.480$), as are \dlinear (rank~4.958) and \nhits (rank~5.250) ($p_\text{Holm} = 0.529$).  Within the bottom tier, \timesnet (rank~7.708) and \autoformer (rank~7.833) are statistically equivalent ($p_\text{Holm} = 0.529$).  All 33~remaining comparisons are statistically significant, including every cross-tier comparison.  In particular, the top-tier boundary is unambiguous: \moderntcn (rank~1.333) and \patchtst (rank~2.000) are each significantly superior to every model in the middle and bottom tiers ($p_\text{Holm} \leq 0.028$ in all cases).

The bracketed interval labelled $\cdfive = 2.451$ in the upper-left corner displays the Nemenyi critical difference computed as $\mathrm{CD} = q_{0.05}(9)\sqrt{k(k+1)/(6N)} = 3.102 \times \sqrt{90/144}$, where $q_{0.05}(9) = 3.102$ is the studentised range critical value at $\alpha = 0.05$ for $k = 9$ and $N = 24$.  This bracket is shown as a reference only; all significance claims in this paper are derived from the more powerful Holm-corrected Wilcoxon procedure rather than the Nemenyi threshold.  Notably, the full rank span from \moderntcn to \lstm ($\Delta r = 6.625$) exceeds $2.7 \times \cdfive$, confirming that the top-to-bottom separation is not a boundary case but an overwhelming statistical gap.

\begin{figure}[htbp]
  \centering
%
\begin{tikzpicture}[font=\small]

\draw[thick] (0,0) -- (12,0);

\foreach \r in {1,...,9}{
  \pgfmathsetmacro{\xt}{(\r-1)*1.5}
  \draw (\xt, 0.10) -- (\xt,-0.10);
  \node[below, font=\scriptsize] at (\xt,-0.10) {\r};
}

\foreach \xm in {0.5000, 1.5000, 4.0000, 4.9375, 5.9375,
                 6.3750, 10.0625, 10.2500, 10.4375}{
  \fill[black] (\xm,0) circle[radius=2.2pt];
}


\draw[gray!55,thin] (0.5000,0) -- (0.5000,1.8);
\node[above, inner sep=1pt, font=\footnotesize\bfseries]
  at (0.5000,1.8) {ModernTCN};

\draw[gray!55,thin] (1.5000,0) -- (1.5000,1.2);
\node[above, inner sep=1pt, font=\footnotesize]
  at (1.5000,1.2) {PatchTST};

\draw[gray!55,thin] (4.0000,0) -- (4.0000,1.8);
\node[above, inner sep=1pt, font=\footnotesize]
  at (4.0000,1.8) {iTransformer};

\draw[gray!55,thin] (4.9375,0) -- (4.9375,1.2);
\node[above, inner sep=1pt, font=\footnotesize]
  at (4.9375,1.2) {TimeXer};

\draw[gray!55,thin] (5.9375,0) -- (5.9375,0.65);
\node[above, inner sep=1pt, font=\footnotesize]
  at (5.9375,0.65) {DLinear};


\draw[gray!55,thin] (6.3750,0) -- (6.3750,-0.85);
\node[below, inner sep=1pt, font=\footnotesize]
  at (6.3750,-0.85) {N-HiTS};


\draw[gray!55,thin] (10.0625,0) -- (10.0625,-0.85) -- (13.3,-0.85);
\node[right, inner sep=2pt, font=\footnotesize] at (13.3,-0.85) {TimesNet};

\draw[gray!55,thin] (10.2500,0) -- (10.2500,-1.55) -- (13.3,-1.55);
\node[right, inner sep=2pt, font=\footnotesize] at (13.3,-1.55) {Autoformer};

\draw[gray!55,thin] (10.4375,0) -- (10.4375,-2.25) -- (13.3,-2.25);
\node[right, inner sep=2pt, font=\footnotesize] at (13.3,-2.25) {LSTM};


\draw[line width=2.8pt, black!70] (4.0000,0.32) -- (4.9375,0.32);

\draw[line width=2.8pt, black!70] (5.9375,0.32) -- (6.3750,0.32);

\draw[line width=2.8pt, black!70] (10.0625,0.32) -- (10.2500,0.32);

\draw[thick,<->,>=stealth] (0,2.52) -- (3.6761,2.52);
\draw[thin] (0,2.40) -- (0,2.65);
\draw[thin] (3.6761,2.40) -- (3.6761,2.65);
\node[above, font=\scriptsize] at (1.8381,2.52) {CD $= 2.451$};

\node[below, font=\scriptsize\itshape] at (6.0,-2.80) {Mean Rank};

\end{tikzpicture}
  \caption{Critical difference diagram for \nmodels forecasting architectures
           across $N = 24$ evaluation blocks (12~assets $\times$ 2~horizons).
           Models are placed at their exact mean \rmse rank; lower rank is better.
           Thick horizontal bars connect models that are \emph{not} statistically
           distinguishable at $\alpha = 0.05$ under Holm-corrected Wilcoxon tests
           (\texttt{holm\_wilcoxon.csv}):
           \itransformer--\timexer ($p_\text{Holm} = 0.480$),
           \dlinear--\nhits ($p_\text{Holm} = 0.529$), and
           \timesnet--\autoformer ($p_\text{Holm} = 0.529$).
           All other 33 pairwise comparisons are significant.
           The bracketed interval (upper left) shows the Nemenyi critical
           difference $\cdfive = 2.451$ ($k = 9$, $N = 24$, $q_{0.05} = 3.102$)
           for reference only.}
  \label{fig:cd_diagram}
\end{figure}

\paragraph{Cross-horizon Spearman rank correlations and Stouffer combination.}
Table~\ref{tab:stat_tests_spearman} reports the Spearman $\rho$ between model rankings at $\hfour$ and $\htwentyfour$ for all twelve assets.  All 12~correlations are positive and statistically significant ($p < 0.05$), with $\rho$ ranging from $0.683$ (ADA/USDT) to $1.000$ (EUR/USD).  The Stouffer combined statistic $\stoufferz = 6.17$ ($p = 3.47 \times 10^{-10}$) confirms that cross-horizon rank stability is a globally systematic property: no single architecture's ranking collapses between the two horizons.

\paragraph{Intraclass Correlation Coefficient (\icc).}
\icc(3,k) analysis for three representative assets at $\htwentyfour$ (Table~\ref{tab:stat_tests_icc}) yields ICC $> 0.990$ in all cases, with $F$-statistics ranging from 309.6 to 2255.6 ($p < 10^{-15}$).  These values indicate that more than 99\% of inter-model variance arises from architecture rather than random initialisation, corroborating the $\leq 0.04\%$ seed contribution in Table~\ref{tab:variance_decomposition}.  At $\hfour$, ICC remains high: 0.9966 (BTC/USDT, $F = 873.9$), 0.9668 (EUR/USD, $F = 88.4$), and 0.9864 (Dow Jones, $F = 218.2$), all with $p < 10^{-11}$.  EUR/USD's comparatively lower $\hfour$ ICC reflects the tighter model clustering on this low-volatility pair rather than genuine seed instability, as even 96.7\% remains far above conventional reliability thresholds.

\paragraph{Diebold-Mariano pairwise tests.}
For BTC/USDT at $\htwentyfour$, Holm-corrected Diebold-Mariano (\dm) tests show that the top cluster (\moderntcn, \patchtst, \timexer, \dlinear, \itransformer) is internally partially distinguishable.  Specifically, \moderntcn~vs.~\patchtst is not significant ($p_\text{Holm} = 0.453$), and \moderntcn~vs.~\timexer is borderline ($p_\text{Holm} = 0.094$), while all comparisons to \lstm, \autoformer, and \timesnet are highly significant ($p < 10^{-14}$).

EUR/USD at $\htwentyfour$ exhibits a sharper separation structure.  \moderntcn is statistically \emph{distinguishable} from \patchtst ($t_\text{DM} = -10.17$, $p_\text{Holm} = 2.77 \times 10^{-24}$) and from \timexer ($t_\text{DM} = -9.21$, $p_\text{Holm} = 3.30 \times 10^{-20}$)---unlike BTC/USDT, where these top-tier differences are not significant.  The non-significant pairs on EUR/USD are \dlinear~vs.~\patchtst ($p_\text{Holm} = 0.122$), \dlinear~vs.~\timexer ($p_\text{Holm} = 0.187$), \dlinear~vs.~\itransformer ($p_\text{Holm} = 0.187$), and \itransformer~vs.~\patchtst ($p_\text{Holm} = 0.717$).  Thus, on the most liquid and low-noise forex pair, \moderntcn's superiority is statistically unambiguous, whereas the middle cluster (\dlinear, \itransformer, \patchtst, \timexer) remains internally indistinguishable.  This asset-dependent DM separability suggests that the statistical power to discriminate the top two architectures depends on the signal-to-noise properties of the underlying market.

These results confirm that the top-tier boundary (\moderntcn and \patchtst) is not artefactual, but that fine-grained rankings within the three-to-five-model cluster should be interpreted with appropriate caution across individual assets.

\paragraph{Jonckheere-Terpstra test for complexity monotonicity.}
The Jonckheere-Terpstra (\jt) test for a monotonic rank-descending trend with increasing parameter count finds no significant relationship in any of the six category--horizon combinations tested (all $p > 0.35$; Table~\ref{tab:stat_tests_jt}).  Four of the six $z$-values are negative, indicating that fewer parameters tend to yield \emph{better} ranks on average.  This formally corroborates the non-monotonic complexity--performance finding (Section~\ref{sec:complexity}).

\paragraph{Directional accuracy z-tests.}
One-sample z-tests on directional accuracy confirm that no architecture's \da\ deviates significantly from the 50\% null (all Holm-corrected $p > 0.43$ for BTC/USDT at $\htwentyfour$), corroborating the aggregate finding in Section~\ref{sec:directional_accuracy}.  The test with the largest $|z|$ is \itransformer ($z = 0.263$, $p_\text{Holm} = 1.0$), confirming that no model possesses directional skill at this resolution.

\subsection{Complexity--Performance Relationship}
\label{sec:complexity}

\begin{figure}[htbp]
  \centering
  \begin{subfigure}[b]{0.48\textwidth}
    \includegraphics[width=\textwidth]{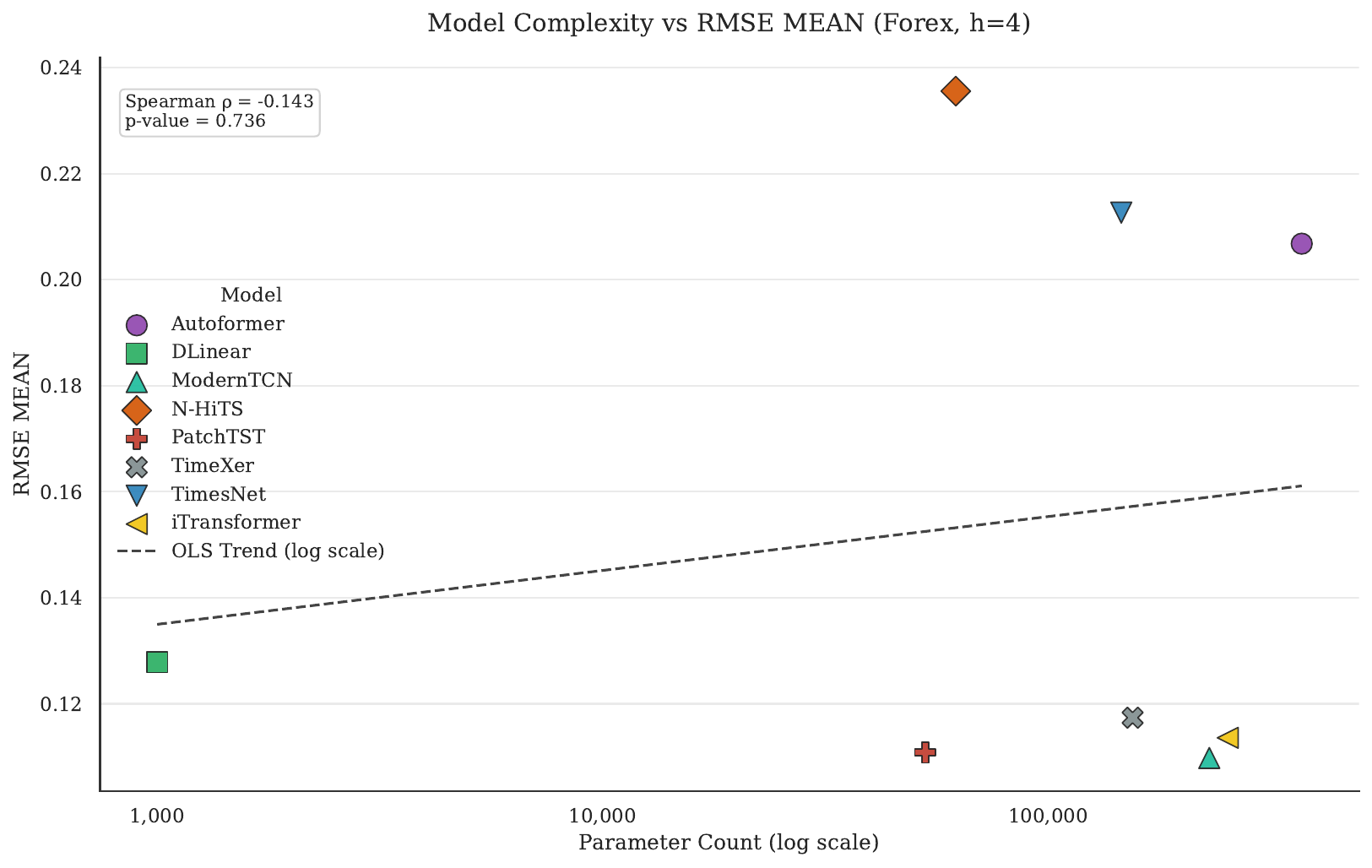}
    \caption{Horizon $h=4$.}
    \label{fig:complexity_h4}
  \end{subfigure}
  \hfill
  \begin{subfigure}[b]{0.48\textwidth}
    \includegraphics[width=\textwidth]{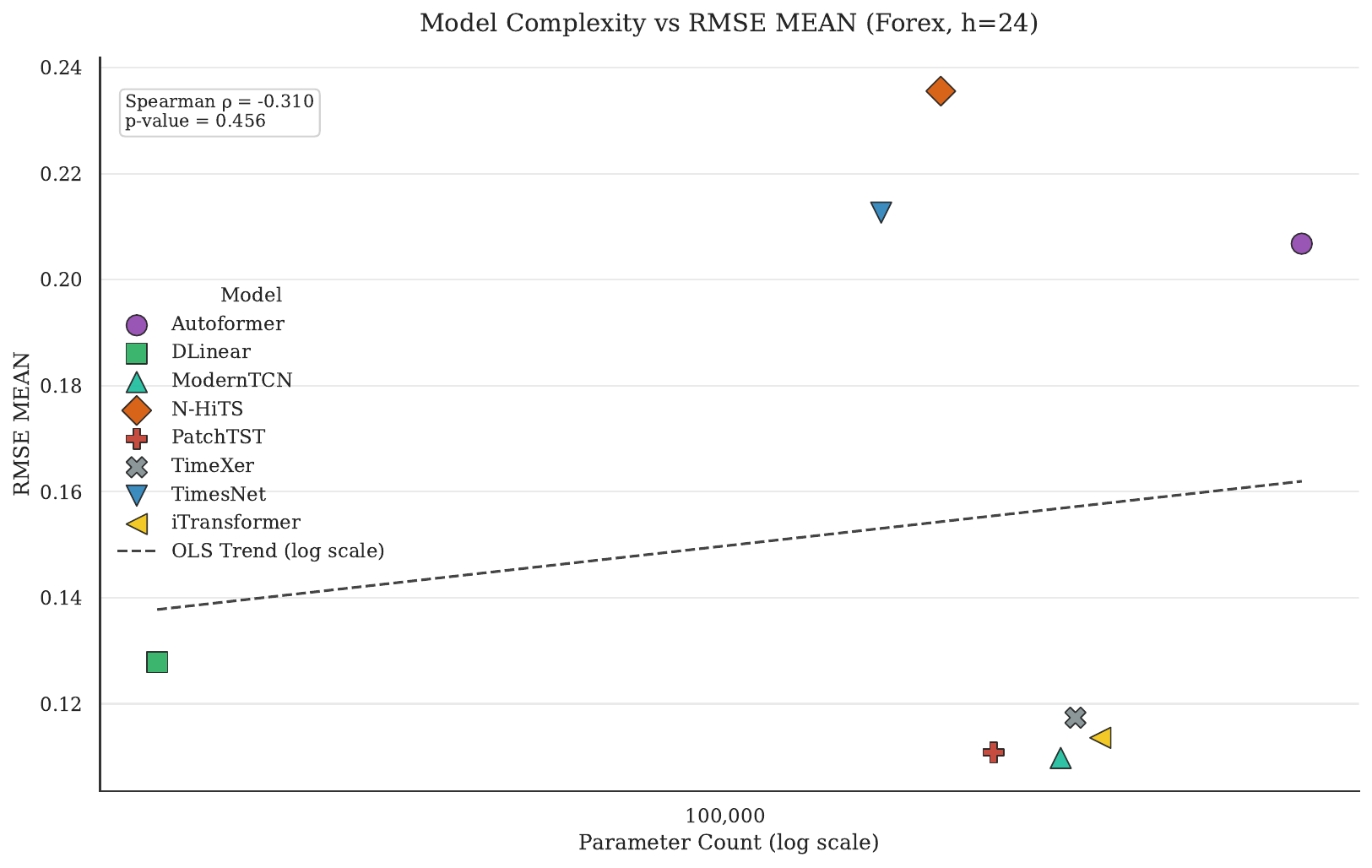}
    \caption{Horizon $h=24$.}
    \label{fig:complexity_h24}
  \end{subfigure}
  \caption{Complexity--performance trade-off (excluding \lstm). The horizontal axis represents the number of trainable parameters (log scale); the vertical axis represents the mean \rmse rank across all assets and seeds. The Pareto frontier is clearly defined by \dlinear, \patchtst, and \moderntcn.}
  \label{fig:complexity}
\end{figure}

The relationship between model complexity, measured by trainable parameter count, and forecasting performance is shown in Figure~\ref{fig:complexity}.  The analysis focuses on modern architectures, excluding the \lstm baseline.

The empirical results reveal several insights:

\paragraph{Pareto-efficient architectures.}
A distinct Pareto frontier is defined by \dlinear, \patchtst, and \moderntcn.  \dlinear (approx.\ 1{,}000 parameters) represents the extreme efficiency point, achieving mid-tier performance (rank~5) with minimal capacity.  \patchtst (approx.\ 103K) and \moderntcn (approx.\ 230K) occupy the optimal region, providing the lowest \rmse ranks globally by deploying parameters into inductive biases suited to financial time series.

\paragraph{Diminishing returns at high complexity.}
Beyond approximately $2.5 \times 10^5$ parameters, returns diminish sharply.  \autoformer (approx.\ 438K) and \itransformer (approx.\ 253K) do not improve commensurately with their larger capacity. \autoformer's rank is consistently lower than the simpler \dlinear, suggesting that excess capacity without appropriate temporal decomposition may lead to overfitting on volatile OHLCV features.

\paragraph{Horizon consistency.}
Comparing Figure~\ref{fig:complexity_h4} and Figure~\ref{fig:complexity_h24} shows that the efficiency profile remains stable across horizons.  Absolute errors increase at $h=24$, but relative model positions on the complexity--performance plane are preserved, indicating that architectural efficiency is a robust property of the model design.

\subsection{Qualitative Forecast Fidelity}
\label{sec:qualitative_fidelity}

Quantitative metrics efficiently rank architectures but compress multi-dimensional forecast behaviour into scalar summaries.  This subsection complements the tabular evidence with actual-versus-predicted overlays, organised along three dimensions: short-horizon tracking ($\hfour$), medium-horizon behaviour ($\htwentyfour$, step~12), and long-horizon degradation ($\htwentyfour$, step~24).  All plots use seed~123; analogous patterns hold across seeds given the near-zero seed variance established in Section~\ref{sec:seed_robustness}.

\paragraph{Short-horizon tracking fidelity ($h = 4$, steps 1 and 4).}

\begin{figure}[htbp]
  \centering
  \begin{subfigure}[b]{0.48\textwidth}
    \includegraphics[width=\textwidth]{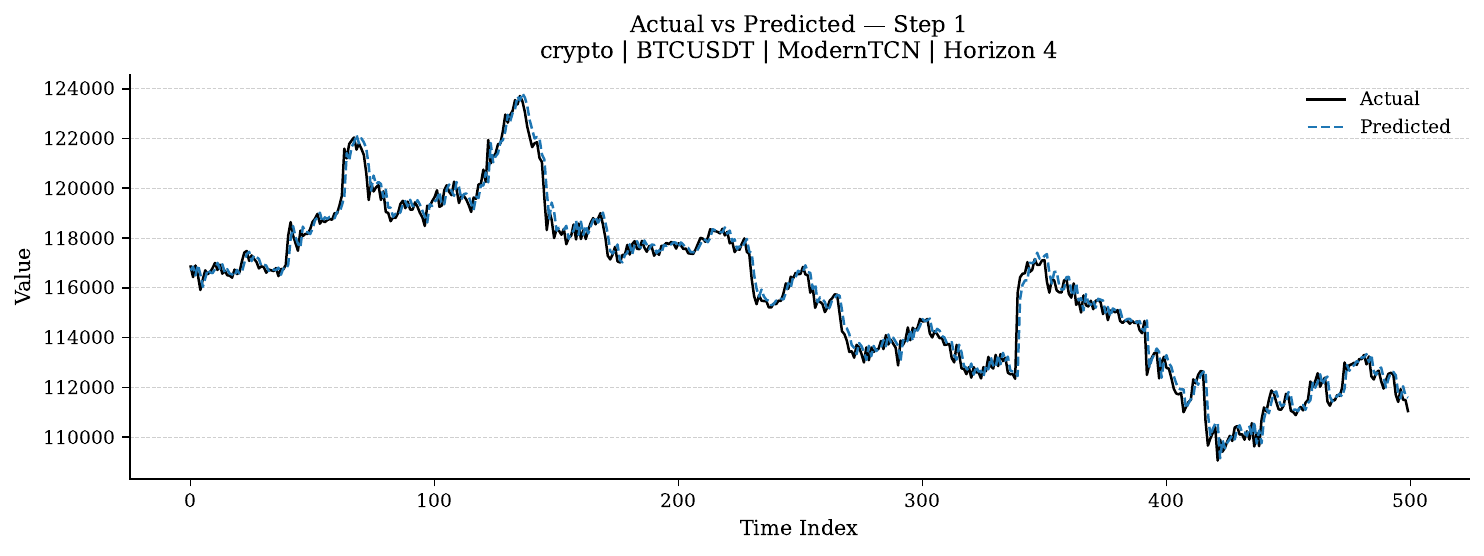}
    \caption{Step~1 ($+$1 hour ahead).}
    \label{fig:pred_btc_h4_step1}
  \end{subfigure}
  \hfill
  \begin{subfigure}[b]{0.48\textwidth}
    \includegraphics[width=\textwidth]{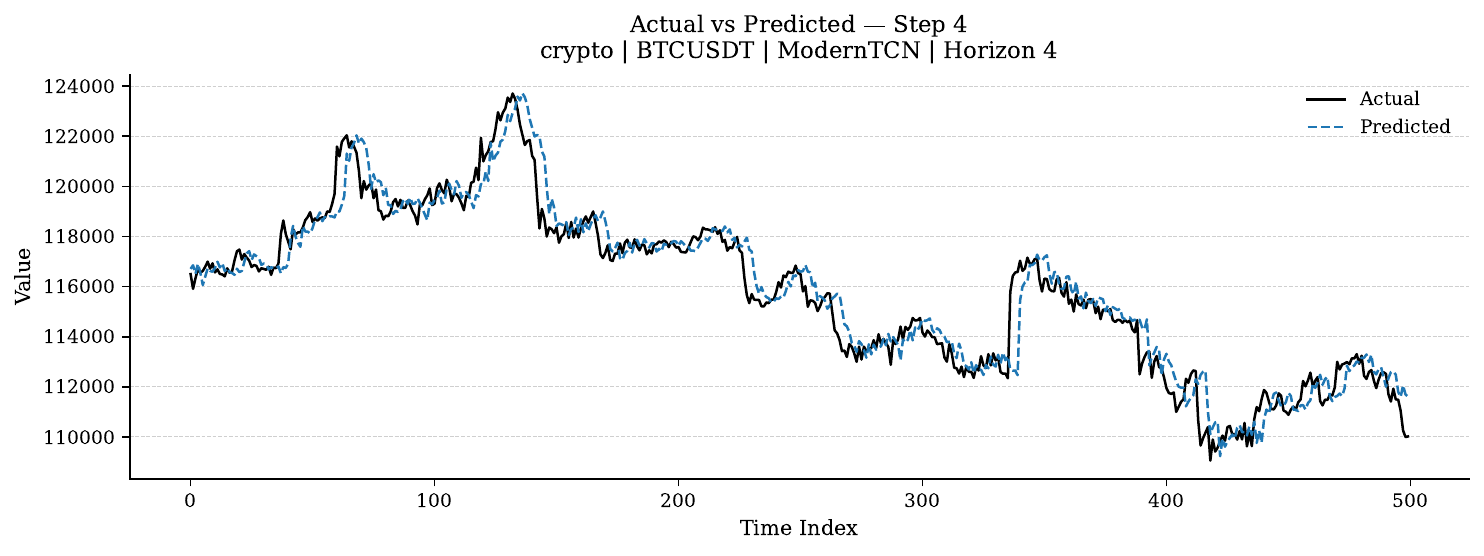}
    \caption{Step~4 ($+$4 hours ahead).}
    \label{fig:pred_btc_h4_step4}
  \end{subfigure}
  \caption{Actual versus predicted close price for \moderntcn on BTC/USDT, $h = 4$ (seed~123).  (a)~At step~1, the model tracks the actual price with high fidelity, capturing directional turns and amplitude fluctuations.  (b)~At step~4, trend structure is preserved but short-lived volatility spikes are modestly attenuated, consistent with MSE-induced shrinkage.  No systematic phase shift is observed.}
  \label{fig:pred_btc_h4_step1_step4}
\end{figure}

Figure~\ref{fig:pred_btc_h4_step1_step4} presents \moderntcn's actual-versus-predicted overlay on BTC/USDT at $\hfour$ for steps~1 and~4 of the forecast vector.  At step~1 (Figure~\ref{fig:pred_btc_h4_step1}), the predicted curve closely mirrors the actual price trajectory, capturing both direction and amplitude of hourly movements.  This near-perfect alignment is consistent with \moderntcn's lowest category-level \rmse in cryptocurrency (314.66; Table~\ref{tab:category_metrics}).  At step~4 (Figure~\ref{fig:pred_btc_h4_step4}), overall alignment is maintained, but the predicted curve shows modest amplitude attenuation during high-volatility episodes---a visual signature of the regression-to-the-mean effect inherent in MSE-optimised multi-step predictors.  No systematic phase shift or directional bias appears in either panel.

\begin{figure}[htbp]
  \centering
  \begin{subfigure}[b]{0.48\textwidth}
    \includegraphics[width=\textwidth]{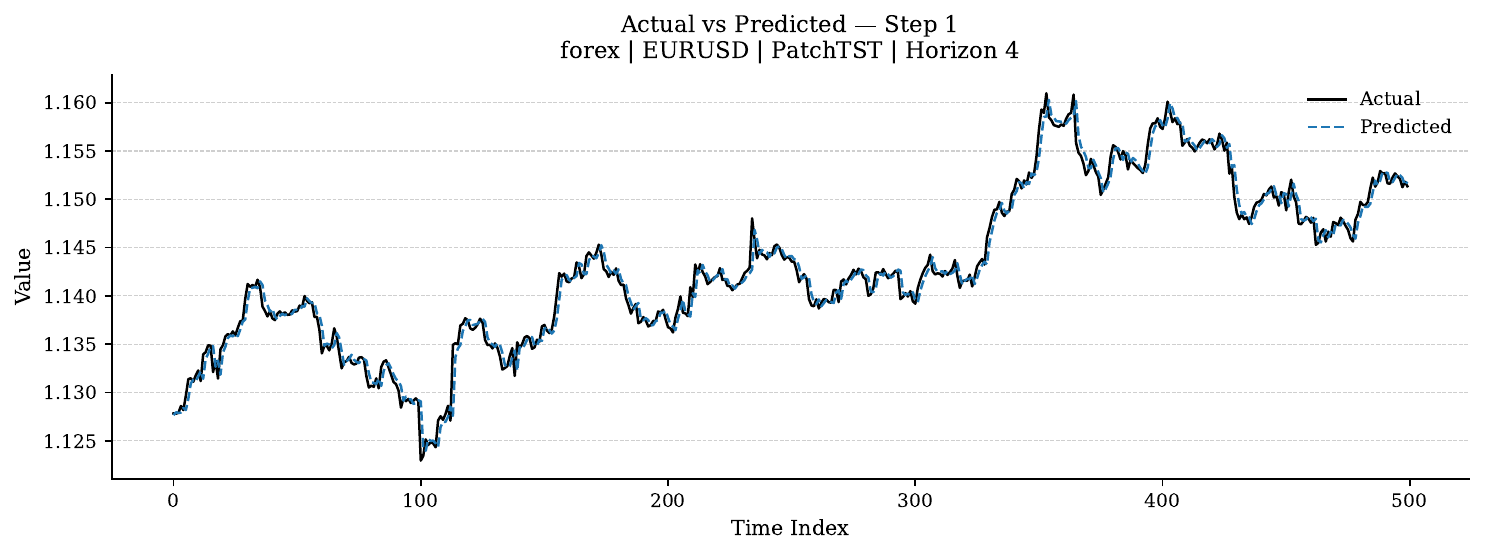}
    \caption{\patchtst on EUR/USD, step~1 ($h = 4$).}
    \label{fig:pred_eurusd_patchtst_h4_step1}
  \end{subfigure}
  \hfill
  \begin{subfigure}[b]{0.48\textwidth}
    \includegraphics[width=\textwidth]{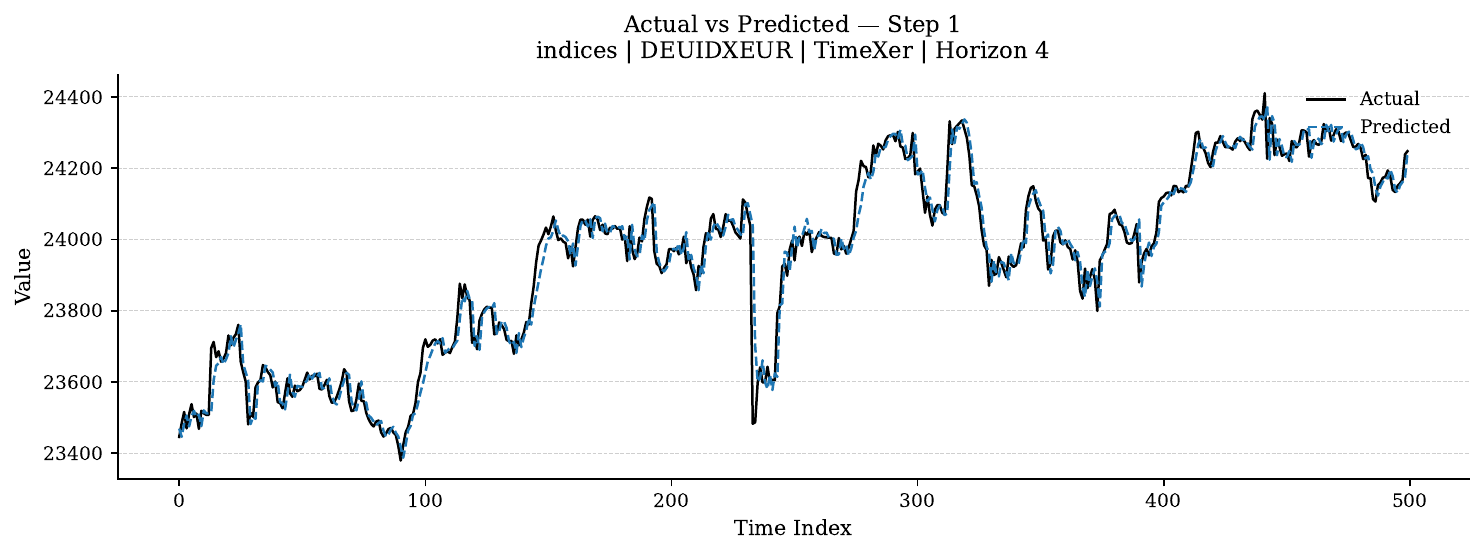}
    \caption{\timexer on Dow Jones, step~1 ($h = 4$).}
    \label{fig:pred_dji_timexer_h4_step1}
  \end{subfigure}
  \caption{Cross-architecture, cross-asset contrast at $h = 4$, step~1 (seed~123).  (a)~\patchtst on EUR/USD (rank~2) tracks the low-amplitude, mean-reverting forex dynamics with high accuracy.  (b)~\timexer on Dow Jones (rank~4) captures the directional structure but with a wider deviation band, illustrating the qualitative signature of the rank gap between top and middle tiers.}
  \label{fig:pred_short_contrast}
\end{figure}

Figure~\ref{fig:pred_short_contrast} juxtaposes \patchtst on EUR/USD and \timexer on Dow Jones, both at $\hfour$, step~1.  The EUR/USD panel (Figure~\ref{fig:pred_eurusd_patchtst_h4_step1}) shows that \patchtst's predictions follow the low-amplitude, mean-reverting dynamics of the currency pair with high precision, corroborating its category-leading \rmse in forex (0.1108; Table~\ref{tab:category_metrics}).  The Dow Jones panel (Figure~\ref{fig:pred_dji_timexer_h4_step1}) shows the middle-tier \timexer (rank~4): it tracks the general directional drift but exhibits a wider deviation band and less precise recovery of sharp reversals---a qualitative reflection of the rank gap between the top and middle tiers.

\begin{figure}[htbp]
  \centering
  \begin{subfigure}[b]{0.48\textwidth}
    \includegraphics[width=\textwidth]{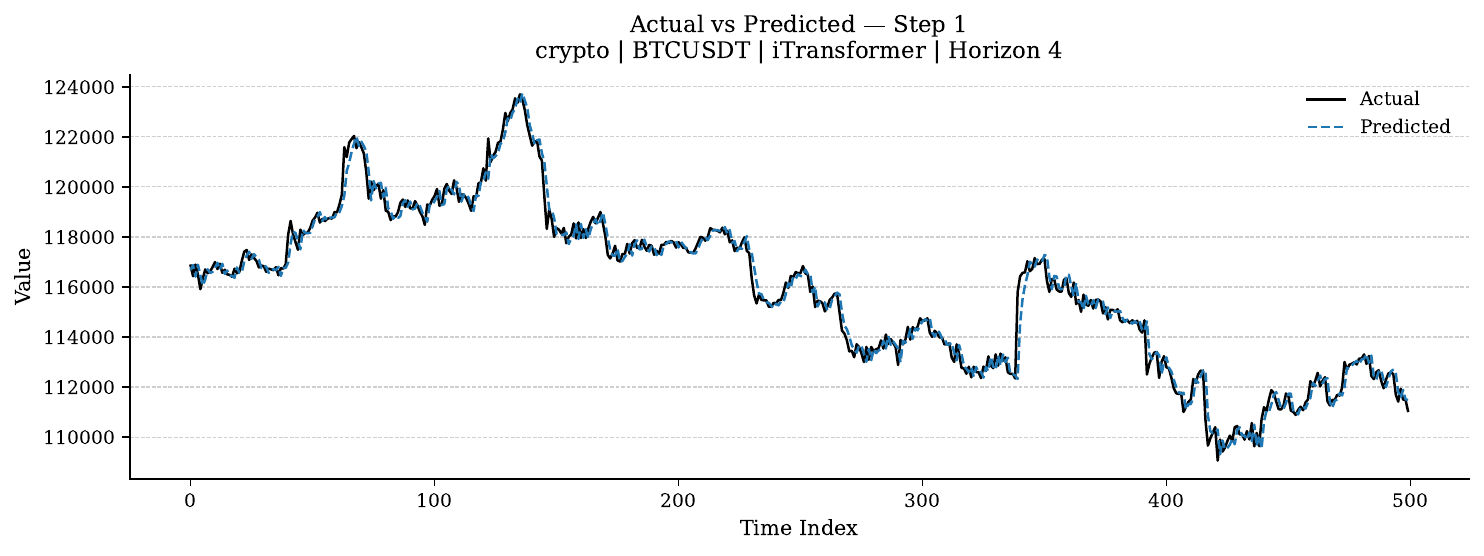}
    \caption{\itransformer on BTC/USDT, step~1 ($h = 4$).}
    \label{fig:pred_btc_itransformer_h4_step1}
  \end{subfigure}
  \hfill
  \begin{subfigure}[b]{0.48\textwidth}
    \includegraphics[width=\textwidth]{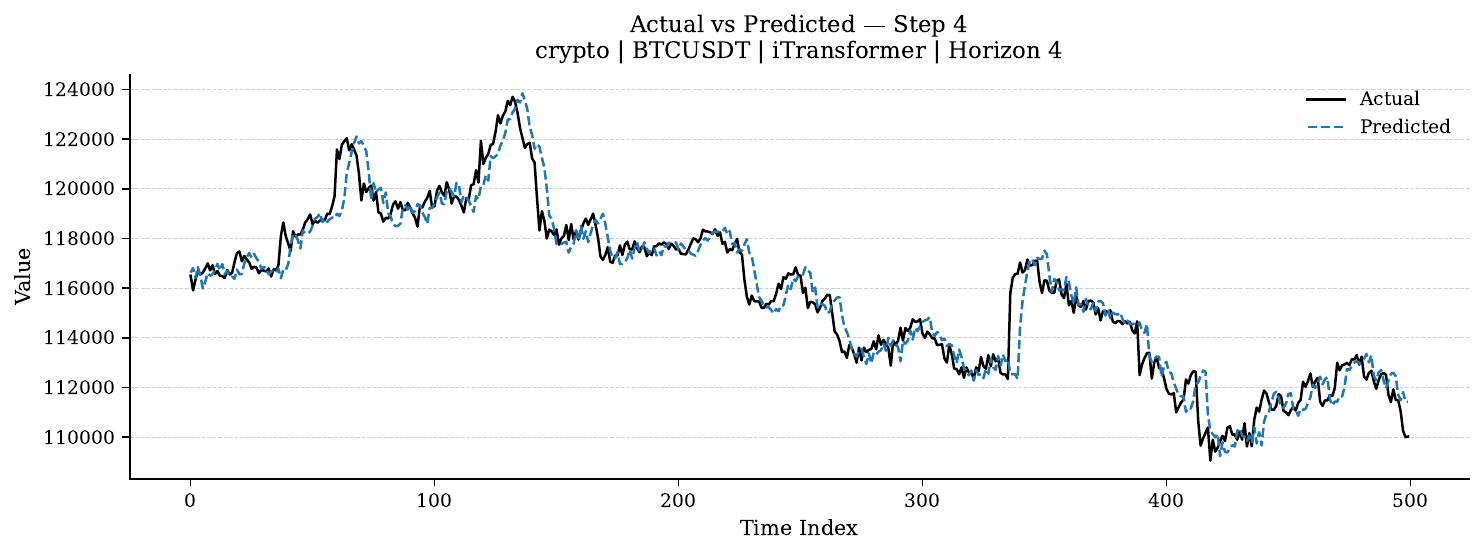}
    \caption{\itransformer on BTC/USDT, step~4 ($h = 4$).}
    \label{fig:pred_btc_itransformer_h4_step4}
  \end{subfigure}
  \caption{\itransformer (rank~3) on BTC/USDT at $\hfour$ (seed~123).  (a)~At step~1, tracking fidelity is comparable to \moderntcn (Figure~\ref{fig:pred_btc_h4_step1}).  (b)~At step~4, slightly more amplitude attenuation is visible relative to \moderntcn (Figure~\ref{fig:pred_btc_h4_step4}), consistent with the 1.6\% \rmse gap.  The inverted attention mechanism produces qualitatively similar but measurably weaker temporal representations for short-horizon cryptocurrency forecasting.}
  \label{fig:pred_itransformer_btc}
\end{figure}

Figure~\ref{fig:pred_itransformer_btc} provides a direct comparison between the third-ranked \itransformer and the top-ranked \moderntcn on BTC/USDT at $\hfour$.  Both models track the actual price trajectory closely at step~1, but at step~4 \itransformer exhibits slightly more amplitude attenuation during volatile episodes.  This visual difference is consistent with the small but consistent \rmse gap between \itransformer (743.5) and \moderntcn (731.6) on BTC/USDT at $\hfour$ (Table~\ref{tab:horizon_degradation_btc}).

\paragraph{Medium-horizon behaviour ($h = 24$, step 12).}

\begin{figure}[htbp]
  \centering
  \begin{subfigure}[b]{0.48\textwidth}
    \includegraphics[width=\textwidth]{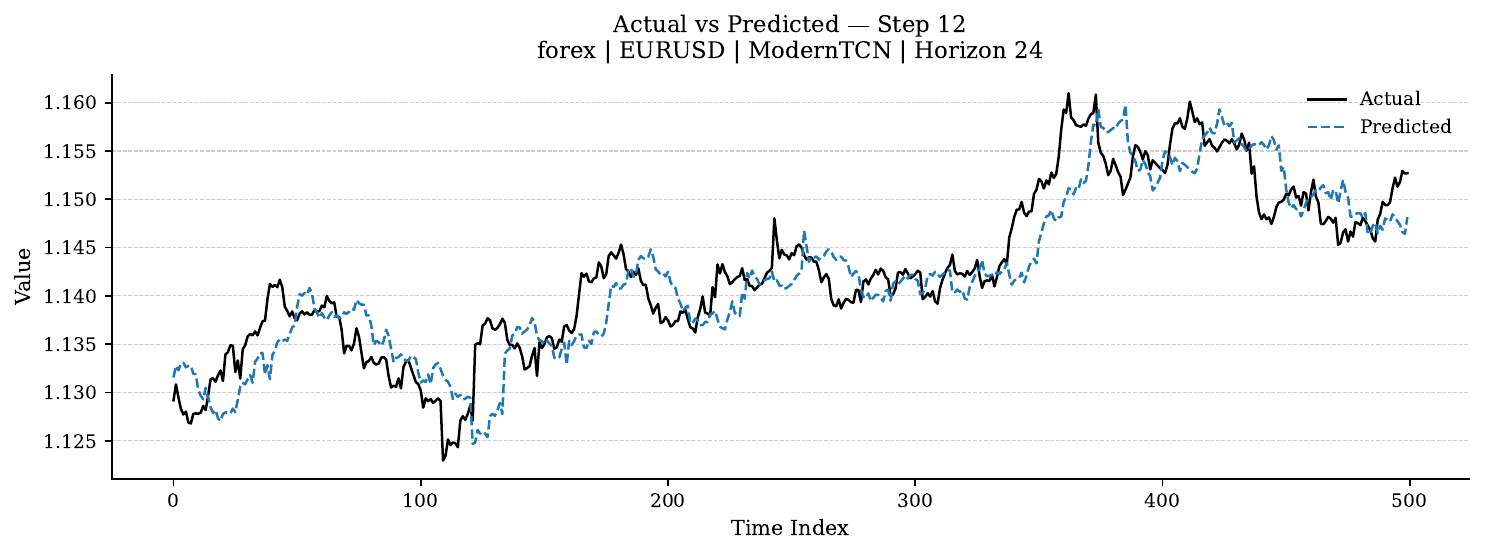}
    \caption{\moderntcn on EUR/USD, step~12 ($h = 24$).}
    \label{fig:pred_eurusd_moderntcn_h24_step12}
  \end{subfigure}
  \hfill
  \begin{subfigure}[b]{0.48\textwidth}
    \includegraphics[width=\textwidth]{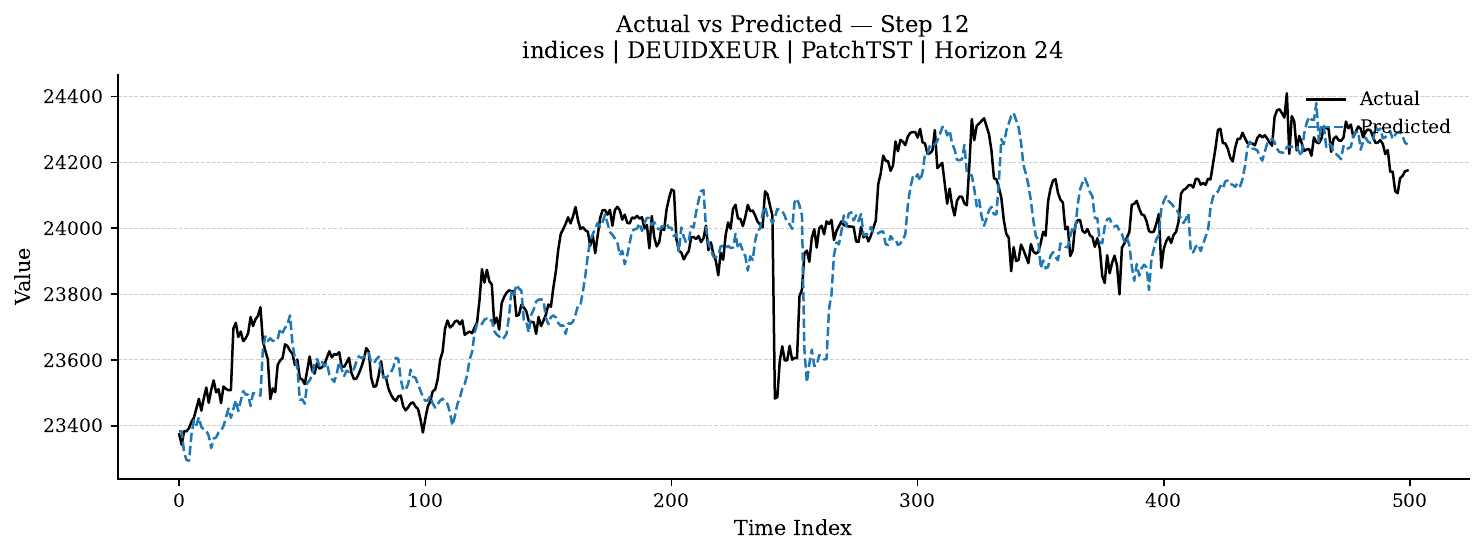}
    \caption{\patchtst on Dow Jones, step~12 ($h = 24$).}
    \label{fig:pred_dji_patchtst_h24_step12}
  \end{subfigure}
  \caption{Medium-horizon actual-versus-predicted overlays at step~12 of the $h = 24$ forecast vector (seed~123).  (a)~\moderntcn on EUR/USD: directional content is preserved 12 hours ahead while high-frequency amplitude is dampened.  (b)~\patchtst on Dow Jones: directional integrity is maintained but the forecast envelope is wider than at $\hfour$, corroborating the $2$--$2.5\times$ \rmse amplification (Table~\ref{tab:horizon_degradation_btc}).  Both top architectures exhibit amplitude attenuation as the primary degradation mode.}
  \label{fig:pred_medium_horizon}
\end{figure}

Figure~\ref{fig:pred_medium_horizon} presents step-12 overlays from two model--asset pairings at the midpoint of the $\htwentyfour$ vector.  \moderntcn on EUR/USD (Figure~\ref{fig:pred_eurusd_moderntcn_h24_step12}) shows that macro-directional structure is preserved 12 hours ahead: the predicted series follows multi-session trends while understandably missing the sharpest intra-session swings.  \patchtst on Dow Jones (Figure~\ref{fig:pred_dji_patchtst_h24_step12}) similarly maintains directional integrity at step~12, but with a visibly wider error envelope than at $\hfour$---confirming the $2$--$2.5\times$ \rmse amplification (Table~\ref{tab:horizon_degradation_btc}).  Both panels show that the dominant degradation signature is amplitude attenuation rather than phase error or directional reversal.

\paragraph{Long-horizon degradation ($h = 24$, step 24).}

\begin{figure}[htbp]
  \centering
  \includegraphics[width=0.80\textwidth]{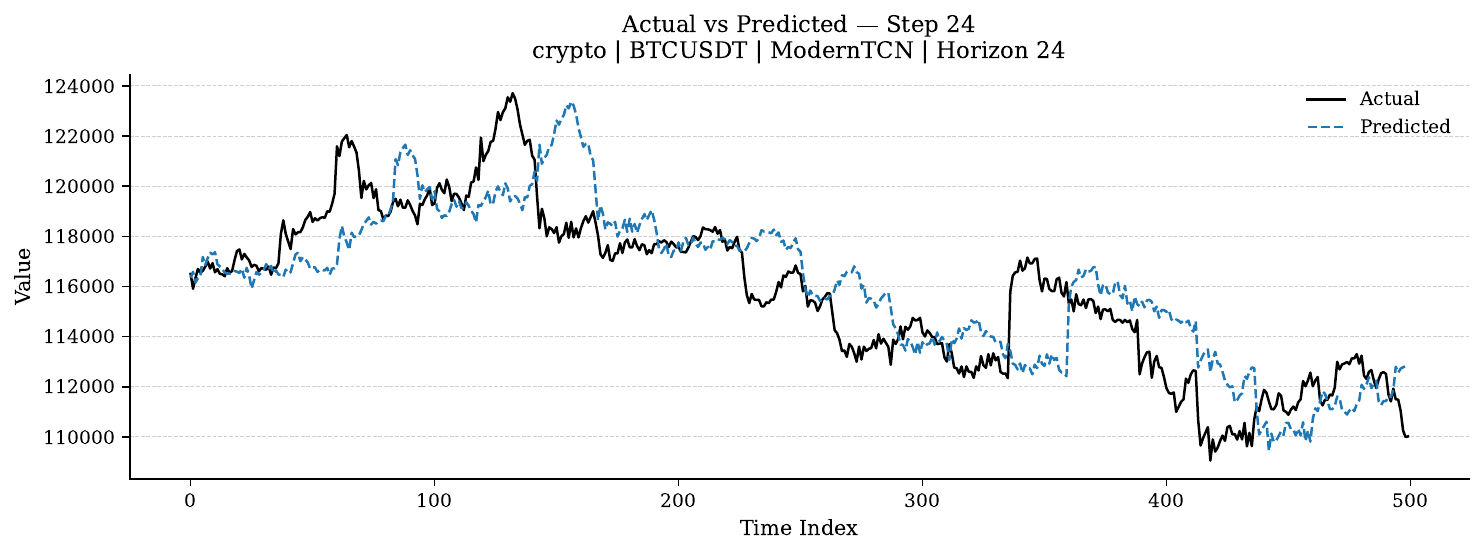}
  \caption{Long-horizon overlay: \moderntcn on BTC/USDT, $h = 24$, step~24 (seed~123).  The model retains directional integrity and captures low-frequency trend components, but high-amplitude intra-day reversals are under-predicted.  This pattern---directional fidelity without amplitude precision---is the signature of MSE-optimised direct multi-step forecasting at the maximum horizon.  The \rmse at $h = 24$ (1{,}617.4) represents a 121.1\% degradation relative to $\hfour$ (731.6; Table~\ref{tab:horizon_degradation_btc}).}
  \label{fig:pred_long_horizon}
\end{figure}

Figure~\ref{fig:pred_long_horizon} presents the 24-step-ahead overlay for \moderntcn on BTC/USDT---the most demanding combination in the benchmark, pairing the highest price volatility with the maximum forecast depth.  The predicted series retains directional drift but shows progressive amplitude compression beyond step~12, with increasingly imprecise oscillatory reversal timing.  These characteristics are consistent with \moderntcn's \rmse degradation from 731.6 ($\hfour$) to 1{,}617.4 ($\htwentyfour$; Table~\ref{tab:horizon_degradation_btc})---a 121.1\% increase that, while substantial, does not erase directional signal or introduce systematic bias.  Long-horizon predictions are \emph{trend-indicative} rather than instance-specific: architectures differ not in whether degradation occurs but in how gracefully their temporal representations transfer to the maximum horizon.

\section{Discussion}
\label{sec:discussion}

This section interprets the empirical findings from Section~\ref{sec:results}, adjudicating each hypothesis, analysing architectural mechanisms, and discussing economic implications, connections to prior work, and limitations.

\subsection{Hypothesis Adjudication}
\label{sec:hypothesis_adjudication}

\paragraph{H1: Ranking Non-Uniformity --- SUPPORTED.}
The global leaderboard (Table~\ref{tab:global_leaderboard}) reveals a clear, consistent hierarchy: \moderntcn (mean rank 1.333, 75\% win rate) and \patchtst (mean rank 2.000) lead across all 24 evaluation points, separated by more than 5.5~ranks from the bottom tier (\timesnet, \autoformer, \lstm).  The per-asset best-model matrix (Table~\ref{tab:per_asset_best_models}) shows that \moderntcn's wins span all three asset classes and both horizons, with \nhits and \patchtst achieving niche first-place finishes exclusively on cryptocurrency assets and short horizons.  The Friedman-Iman-Davenport test confirms this non-uniformity at the global level with $F(8,184) = 101.36$, $p < 10^{-15}$ (Table~\ref{tab:stat_tests_friedman}), and the same result holds within each asset class.  Post-hoc Holm-Wilcoxon tests (Section~\ref{sec:statistical_tests}) establish that 33 of 36~pairwise differences are statistically significant; the only non-significant pairs are intra-tier neighbours (\timexer~vs.~\itransformer, \autoformer~vs.~\timesnet, \dlinear~vs.~\nhits).  The \moderntcn--\patchtst gap (0.667 mean rank difference) is not individually significant by Diebold-Mariano on BTC/USDT ($p_\text{Holm} = 0.453$), reflecting near-equivalent top-tier performance rather than statistical equivalence across the full distribution.

\paragraph{H2: Cross-Horizon Ranking Stability --- SUPPORTED.}
Top-tier rankings (\moderntcn, \patchtst) are preserved at both $\hfour$ and $\htwentyfour$ across all three representative assets (Tables~\ref{tab:horizon_degradation_btc} and~\ref{tab:horizon_ranking_shift}).  EUR/USD rankings are perfectly stable; BTC/USDT and Dow Jones exhibit rank shifts confined to the middle tier (\nhits improves by 3~ranks on Dow Jones; \itransformer drops by 2~on BTC/USDT).  This stability is formally confirmed by Spearman cross-horizon rank correlations: all 12~assets yield $\rho \geq 0.683$ with $p < 0.05$ (Table~\ref{tab:stat_tests_spearman}), including $\rho = 1.000$ for EUR/USD.  The Stouffer combined $\stoufferz = 6.17$ ($p = 3.47 \times 10^{-10}$) confirms this as a systematic, not asset-specific, property.  Error amplification of $2.0$--$2.5\times$ at the longer horizon reflects increased task difficulty rather than differential model degradation.  Figure~\ref{fig:pred_long_horizon} provides qualitative confirmation: \moderntcn retains directional correlation with BTC/USDT at step~24 despite the \rmse increase.

\paragraph{H3: Variance Dominance --- STRONGLY SUPPORTED.}
The two-factor decomposition (Table~\ref{tab:variance_decomposition}) is reported in three panels to separate scale artefacts from structural effects.\footnote{We reserve \emph{strongly supported} for hypotheses where the effect holds at $>$99\% magnitude across all analysis panels; the remaining hypotheses are labelled \emph{supported}.}  On the raw price scale, architecture explains 99.90\% of variance---driven by \lstm's outlier errors ($7$--$33\times$ the category median).  After z-score normalisation, architecture accounts for 48.32\% (seed: 0.04\%; residual: 51.64\%); excluding \lstm, it rises to 68.33\% (seed: 0.02\%).  The residual reflects genuine model--slot interaction: no single architecture dominates every slot.  Across all panels, seed variance is negligible ($\leq 0.04\%$), validating three-seed replication as sufficient.  This conclusion is independently corroborated by \icc analysis: for three representative assets at $\htwentyfour$, \icc $> 0.990$ with $F > 309$ ($p < 10^{-15}$; Table~\ref{tab:stat_tests_icc}), confirming that $>99\%$ of inter-model variance is attributable to architecture rather than random seed.  Per-asset decompositions extend this result to $\hfour$, where architecture still explains 97.4\%--99.7\% of variance (Section~\ref{sec:seed_robustness}), confirming that seed-invariance holds at both forecast depths.

\paragraph{H4: Non-Monotonic Complexity--Performance --- SUPPORTED.}
The complexity--performance scatter (Figure~\ref{fig:complexity}) demonstrates a non-monotonic relationship between trainable parameter count and mean \rmse rank.  \dlinear (approximately 1{,}000 parameters) achieves rank~5, while \autoformer (approximately 438{,}000 parameters) and \lstm (approximately 172{,}000 parameters) achieve ranks 8--9.  \moderntcn (approximately 230{,}000 parameters) and \patchtst (approximately 103{,}000 parameters) occupy the top positions with moderate parameter budgets.  The Jonckheere-Terpstra test for a monotonic complexity-rank relationship finds no significant trend in any of the six category--horizon combinations (all $p > 0.35$; Table~\ref{tab:stat_tests_jt}); four of the six $z$-values are negative, suggesting an inverse tendency.  This formally confirms that \emph{how} parameters are deployed---the specific temporal inductive bias---determines forecast quality, not the raw quantity of learnable weights.

\subsection{Architecture-Specific Insights}
\label{sec:architecture_insights}

\paragraph{ModernTCN.}
\moderntcn's consistent superiority (rank~1 on 18/24 evaluation points) is consistent with complementary design features: large-kernel depthwise convolutions capture multi-range temporal dependencies without quadratic attention cost; multi-stage downsampling enables hierarchical feature extraction suited to the multi-scale dynamics of financial series; and RevIN normalisation mitigates distributional shift between training and test periods.  All of this is achieved with approximately 230{,}000 parameters---a moderate footprint.

\paragraph{PatchTST.}
\patchtst's consistent second-place ranking supports the view that patch-based tokenisation offers an effective compromise between local pattern recognition and global dependency modelling.  Segmenting the input into patches reduces the token count, enabling attention over temporally coherent segments, while channel-independent processing prevents cross-feature leakage---appropriate given the heterogeneous scales of OHLCV components.  The narrow \moderntcn--\patchtst gap (0.667 mean rank difference) suggests that both architectures capture the relevant temporal structure through distinct mechanisms.

\paragraph{iTransformer and TimeXer.}
\itransformer's inverted attention paradigm (rank~3) shows that computing attention across the five OHLCV features rather than across time steps is effective for multivariate financial forecasting.  \timexer (rank~4) further separates target and exogenous variables through cross-attention, but the marginal improvement suggests that explicit target--exogenous decomposition adds complexity without proportionate benefit when all input features are closely correlated.

\paragraph{DLinear and N-HiTS.}
\dlinear's fifth-place ranking with approximately 1{,}000 parameters and no nonlinear activations corroborates the finding by \parencite{Zeng2023} that simple linear mappings can be surprisingly effective.  Its trend--seasonal decomposition captures the dominant low-frequency structure of financial series at minimal cost.  \nhits (rank~6) benefits from multi-rate pooling across temporal resolutions but processes only the target channel, potentially limiting cross-feature exploitation.  Its notably lower cross-horizon degradation on BTC/USDT (85.3\% vs.\ 121\% for \moderntcn; Table~\ref{tab:horizon_degradation_btc}) suggests that hierarchical temporal decomposition transfers well across horizons for trending series.  Notably, \nhits achieves the lowest \rmse on ETH/USDT (both horizons) and ADA/USDT ($\hfour$)---both lower-capitalisation, higher-volatility cryptocurrency assets---accounting for all three of its first-place finishes (Table~\ref{tab:per_asset_best_models}).  This suggests that its multi-rate pooling is particularly suited to the superimposed multi-scale oscillatory patterns characteristic of altcoin markets, where multiple speculative timescales dominate the price dynamics.

\paragraph{TimesNet, Autoformer, and LSTM.}
\timesnet (rank~7) and \autoformer (rank~8) both employ frequency-domain inductive biases---FFT-based 2D reshaping and auto-correlation, respectively---that presuppose periodic structure largely absent in hourly financial data.  Their designs are better suited to domains with strong seasonality such as electricity demand or weather forecasting.

\lstm's consistently poor performance (worst-ranked across all conditions; errors $7$--$33\times$ higher than the best model) confirms that the recurrent architecture is not competitive for multi-step financial forecasting.  Compressing all temporal context into a fixed-dimensional hidden state severely limits representation capacity for direct multi-step prediction.

\subsection{Asset-Class Dynamics}
\label{sec:asset_dynamics}

The three asset classes present distinct forecasting challenges shaped by different market microstructures (Section~\ref{sec:asset_universe}).  Cryptocurrency markets exhibit the highest absolute errors (mean \rmse 314--2{,}399; Table~\ref{tab:category_metrics}) due to elevated price levels, 24/7 trading, and speculative volatility.  Forex markets produce the smallest errors (0.110--3.668), reflecting low price magnitudes and high liquidity, while equity indices occupy an intermediate position (111--1{,}549).

Despite these scale differences, \emph{relative rankings are largely preserved}: \moderntcn and \patchtst consistently occupy the top two positions in every category (Table~\ref{tab:category_metrics}, Figure~\ref{fig:category_boxplot}), indicating general-purpose temporal modelling capabilities rather than class-specific advantages.

Mid-tier variation is more pronounced.  \timexer ranks 3rd in cryptocurrency but 4th in forex and indices.  \nhits (rank~6 overall) performs relatively better on cryptocurrency, where multi-rate pooling may better capture multi-scale volatility.  \dlinear performs better on forex (rank~5), where simpler, mean-reverting dynamics may be well-approximated by linear projections.

\paragraph{Asset-specific niche advantages.}
The per-asset best-model matrix (Table~\ref{tab:per_asset_best_models}) reveals a finer-grained picture than category-level rankings.  \nhits achieves the lowest \rmse on ETH/USDT at both horizons and on ADA/USDT at $\hfour$---all lower-capitalisation cryptocurrency assets characterised by higher relative volatility and more pronounced multi-scale dynamics.  This suggests that \nhits's hierarchical multi-rate pooling, which decomposes the input at several temporal resolutions, captures the superimposed short- and medium-term oscillation patterns that dominate altcoin price series.  \patchtst wins on BTC/USDT at $\hfour$ (with a margin of only 0.08\% over \moderntcn), on ADA/USDT at $\htwentyfour$, and on GBP/USD at $\hfour$, indicating that patch-based self-attention is competitive for short-horizon prediction on assets with diverse temporal structure.

Critically, no model other than \moderntcn wins on any forex or equity index evaluation point at $\htwentyfour$.  This long-horizon, cross-category dominance---16 out of 16~possible wins outside the cryptocurrency domain at $\htwentyfour$---suggests that the combination of large-kernel depthwise convolutions and multi-stage downsampling provides the most transferable temporal representations when the forecast window extends.

\paragraph{Statistical separability varies by market microstructure.}
Diebold-Mariano tests reveal that the top-tier gap is market-dependent.  On EUR/USD ($\htwentyfour$), \moderntcn is statistically separable from every other architecture, including \patchtst ($p_\text{Holm} = 2.77 \times 10^{-24}$; Section~\ref{sec:statistical_tests}).  On BTC/USDT at the same horizon, \moderntcn~vs.~\patchtst is not significant ($p_\text{Holm} = 0.453$).  This disparity reflects the differing signal-to-noise ratios: EUR/USD's lower intrinsic noise amplifies small but consistent performance differences into statistical separability, whereas BTC/USDT's high volatility masks the same differences.  Practitioners operating in low-noise markets can thus have higher confidence in \moderntcn's superiority; in high-noise crypto markets, the top-tier models should be treated as near-equivalent.

\subsection{Economic Interpretation}
\label{sec:economic_interpretation}

While this study evaluates forecasting \emph{accuracy} rather than trading \emph{profitability}, several economically relevant observations emerge.

\paragraph{Cost of model selection error.}
The \rmse gap between \moderntcn and the worst modern alternative (\autoformer) ranges from 60\%--70\% across categories (Table~\ref{tab:category_metrics}); the gap to \lstm is an order of magnitude larger.  Where forecast error translates into sizing or timing errors, systematic architecture evaluation yields substantial returns relative to default model selection.

\paragraph{Directional accuracy.}
The mean \da across all 54 model--category--horizon combinations is 50.08\%, with no combination deviating meaningfully from the 50\% baseline (Table~\ref{tab:directional_accuracy}).  MSE-optimised architectures do not exhibit established directional skill at hourly resolution.  This observation has two compatible interpretations: weak-form efficiency may leave little predictable sign information, and MSE minimisation may contract forecasts toward the conditional mean for a noisy, quasi-martingale target, producing DA near 50\% even without a strict efficiency conclusion.  The current point-forecast design cannot identify which mechanism dominates.

This distinction matters for loss-function design.  Point MSE targets the conditional mean and is appropriate for level-error comparisons, but it does not directly reward correct signs, calibrated uncertainty, or asymmetric trading payoffs.  RNAD addresses this broader design space by combining regime normalisation with tail-robust and direction-aware loss components \parencite{Saidd2026RNAD}; its contribution is a specialised training objective, not an architecture comparison under a shared loss.  Probabilistic forecasting could instead report a predictive distribution or calibrated intervals, while quantile regression could represent asymmetric tail risk without committing to a single mean path.  Directional or utility-aware differentiable surrogates, possibly combined with MSE in a multi-objective loss, could explicitly supervise sign information; however, such objectives can trade level accuracy for directional performance and require their own calibration and economic evaluation.  None of these alternatives, including RNAD, was trained in the present benchmark, so they are methodological extensions rather than claims about the evaluated models.

\paragraph{Variance decomposition implication.}
Architecture choice explains the overwhelming majority of forecast variance while seed explains $\leq 0.04\%$.  The practical corollary: effort invested in architecture selection yields far higher returns than effort spent on seed selection or initialisation-based ensembles.

\paragraph{Caveat.}
These economic interpretations are preliminary.  \rmse and \mae measure statistical accuracy, not economic value.  Trading performance requires consideration of transaction costs, market impact, slippage, position sizing, and risk-adjusted returns (Sharpe ratio, maximum drawdown).  This study provides the statistical foundation for economic evaluations but does not assess trading profitability.

\subsection{Comparison with Prior Literature}
\label{sec:prior_comparison}

\paragraph{Linear models are competitive.}
\dlinear's fifth-place ranking with approximately 1{,}000 parameters is consistent with the finding by \parencite{Zeng2023} that linear temporal mappings can match or exceed more complex alternatives.  This benchmark extends that finding from standard time-series benchmarks (ETTh, Weather, Electricity) to financial data across three asset classes under controlled HPO.

\paragraph{Patch-based attention is effective.}
\patchtst's consistent second-place ranking supports the view of \parencite{Nie2023} that patch tokenisation is an effective strategy for time-series Transformers, and this effectiveness extends to financial data across asset classes and horizons.

\paragraph{Seed variance is small in financial forecasting.}
The 99.90\% raw model vs.\ 0.01\% seed decomposition (z-normalised: 48.3\% vs.\ 0.04\%) extends prior work by \parencite{Bouthillier2021} documenting the importance of separating implementation variance from algorithmic performance.  In financial forecasting with fixed splits and deterministic preprocessing, seed variance is even smaller than in the general settings examined there, justifying the three-seed protocol.

\paragraph{Recurrent models are not competitive.}
The order-of-magnitude inferiority of \lstm corroborates observations by \parencite{Lim2021} and \parencite{Hewamalage2021} on the declining competitiveness of recurrent architectures.  This study provides the most controlled evidence to date for this conclusion in the financial domain.

\paragraph{Positioning relative to compound gap.}
Prior studies (Table~\ref{tab:prior_studies}) address at most two or three of the five gaps.  This study simultaneously addresses: controlled HPO~(G1), multi-seed evaluation~(G2), multi-horizon analysis~(G3), formal pairwise statistical correction with Holm-Wilcoxon and Diebold-Mariano tests~(G4), and multi-asset-class coverage~(G5).  All five methodological gaps are therefore addressed, placing this study at the intersection of rigorous experimental design and financial time-series benchmarking.

\subsection{Limitations and Threats to Validity}
\label{sec:limitations}

The following limitations are organised by severity, from most impactful to least:

\begin{enumerate}[label=\textbf{L\arabic*.}]
  \item \textbf{HPO budget.}  Five Optuna trials represent a limited search budget.  Models with larger search spaces may be disadvantaged, and the post hoc HPO-versus-non-HPO rank comparison (Table~\ref{tab:hpo_nonhpo_ranks}) does not test whether five trials found each model's optimum.  Increasing the budget may shift mid-tier rankings, though the top-tier ordering is stable in the available evaluation outputs.  The equal budget is important for internal fairness; a larger symmetric search is future work.

  \item \textbf{Statistical validation.}  A comprehensive battery of formal statistical tests is reported (Section~\ref{sec:statistical_tests} and Tables~\ref{tab:stat_tests_friedman}--\ref{tab:stat_tests_jt}), including Friedman-Iman-Davenport omnibus tests, Holm-Wilcoxon pairwise comparisons, Diebold-Mariano tests, Spearman/Stouffer cross-horizon correlations, \icc, and Jonckheere-Terpstra complexity tests.  The main remaining limitation is that per-category pairwise Wilcoxon tests are underpowered due to small block size ($n = 8$); a categorical-level study with more assets per class would overcome this constraint.

  \item \textbf{Feature set.}  The OHLCV-only restriction, while useful for isolating architectural effects and maintaining identical information across models, may underestimate models designed to exploit technical indicators, order-book data, sentiment features, calendar variables, or other covariates.  Rankings could differ under richer input configurations; the present results should not be extrapolated to those settings.

  \item \textbf{Temporal and regime scope.}  Each asset is evaluated on one fixed historical sample of 30{,}000 windows with a single chronological 70/15/15 split.  Although the assets span several years in the source ranges, the study does not perform validation on multiple non-overlapping periods or explicitly identify separate market regimes.  Consequently, cross-regime generalisability is unestablished.  All experiments also use H1 frequency; transfer to tick-level, M1, H4, or daily data is not established, and the relative importance of inductive biases may change with the time scale.

  \item \textbf{Asset universe.}  Twelve instruments across three asset classes provide a representative but not exhaustive sample.  Commodities, fixed income, and emerging-market instruments are absent.

  \item \textbf{Horizon granularity.}  Degradation is characterised at only two points ($\hfour$ and $\htwentyfour$).  Intermediate horizons ($h = 8, 12, 16$) would yield smoother degradation curves and more precise characterisation of architecture-specific scaling behaviour.

  \item \textbf{Effect-size reporting.}  A critical difference diagram (Figure~\ref{fig:cd_diagram}) visualises the Holm-Wilcoxon significance structure, and Diebold--Mariano pairwise tests are reported for representative assets (Section~\ref{sec:statistical_tests}).  However, Cohen's~$d$ standardised effect sizes---which would facilitate cross-study comparison---are not computed.  While the Holm-adjusted \pvalue{}s and rank differences provide implicit effect scaling, explicit $d$~values remain a useful complement and are identified as future work.

  \item \textbf{Out-of-sample recency.}  The test set comprises the final 15\% of historical data, evaluated in batch.  A live forward-walk evaluation with rolling retraining would provide stronger evidence for real-time deployment.
\end{enumerate}

\section{Conclusion}
\label{sec:conclusion}

\subsection{Summary of Contributions}
\label{sec:contribution_summary}

This paper presented a controlled comparison of \nmodels deep learning architectures---spanning Transformer, MLP, convolutional, and recurrent families---for multi-horizon financial time-series forecasting across 918 experimental runs.  Five contributions were made: (C1)~protocol-controlled fair comparison, (C2)~multi-seed robustness quantification, (C3)~cross-horizon generalisation analysis, (C4)~asset-class-specific deployment guidance, and (C5)~release of a fully open benchmarking framework.

\subsection{Principal Findings}
\label{sec:principal_findings}
All four hypotheses from Section~\ref{sec:hypotheses} are supported (see Section~\ref{sec:hypothesis_adjudication} for detailed adjudication):

\begin{description}[leftmargin=0.5cm, font=\bfseries]
  \item[H1 (Ranking non-uniformity): SUPPORTED.]  A clear three-tier hierarchy emerges, with \moderntcn and \patchtst separated from the bottom tier by over 5.5 mean-rank positions.

  \item[H2 (Cross-horizon stability): SUPPORTED.]  Top-tier rankings are preserved across horizons despite $2$--$2.5\times$ error amplification; rank shifts are confined to mid-tier models.

  \item[H3 (Variance dominance): STRONGLY SUPPORTED.]  Architecture explains $\geq 99.9\%$ of raw variance; seed variance is negligible ($\leq 0.04\%$), confirming three-seed replication suffices.

  \item[H4 (Non-monotonic complexity): SUPPORTED.]  \dlinear (approximately 1{,}000 parameters) outranks \autoformer (approximately 438K) and \lstm (approximately 172K); architectural inductive bias dominates raw capacity.
\end{description}

\subsection{Practical Recommendations}
\label{sec:recommendations}

Building on the empirical evidence from 648 final training runs and the hypothesis adjudication (Section~\ref{sec:hypothesis_adjudication}):

\begin{itemize}[nosep]
  \item \textbf{Default recommendation:} \moderntcn.  Rank~1 on 75\% of evaluation points with moderate cost.  Large-kernel depthwise convolutions and multi-stage downsampling provide effective general-purpose temporal modelling across all asset classes and horizons.  On low-noise markets (EUR/USD), its superiority is statistically unambiguous ($p_\text{Holm} < 10^{-20}$ vs.\ all competitors; Section~\ref{sec:statistical_tests}).

  \item \textbf{Transformer alternative:} \patchtst.  Consistently rank~2, with a narrow gap from \moderntcn (0.667).  Patch-based tokenisation with channel independence suits settings where Transformer-family models are preferred.

  \item \textbf{Altcoin specialist:} \nhits.  Achieves the lowest \rmse on ETH/USDT and ADA/USDT (Table~\ref{tab:per_asset_best_models}), suggesting niche advantage on lower-capitalisation cryptocurrency assets due to its multi-rate pooling design.

  \item \textbf{Resource-constrained environments:} \dlinear.  Rank~5 with approximately 1{,}000 parameters and no nonlinear activations, suitable when inference latency or memory is binding.

  \item \textbf{Not recommended:} \lstm.  Ranks last across all conditions with errors $7$--$33\times$ higher than the best model.
\end{itemize}

\paragraph{Directional accuracy.}  MSE-optimised architectures produce no established directional advantage in this benchmark (mean \da~$= 50.08\%$; Table~\ref{tab:directional_accuracy}).  This is compatible with weak-form efficiency and/or regression-to-the-mean under MSE; the present design cannot distinguish the two.  Applications requiring directional skill should evaluate explicit directional, quantile, or probabilistic objectives rather than infer it from point-level RMSE.

These recommendations are conditioned on the experimental scope described in Section~\ref{sec:experimental_design} (OHLCV features, H1 frequency, 12~assets, 2~horizons) and the limitations acknowledged in Section~\ref{sec:limitations}.

\subsection{Future Work}
\label{sec:future_work}

The following extensions are prioritised by expected impact:

\begin{enumerate}

  \item \textbf{Cohen's $d$ effect-size matrices} to complement the Holm-Wilcoxon and Diebold-Mariano pairwise analyses already reported (Section~\ref{sec:statistical_tests}), facilitating standardised cross-study comparison.

  \item \textbf{Increased HPO budget and seed count} to strengthen ranking evidence, particularly for mid-tier models where the current 5-trial budget may be limiting; this should be applied symmetrically to all architectures.

  \item \textbf{Directional and probabilistic objectives} via differentiable sign-aware losses, quantile regression, and calibrated distributional forecasting, investigating whether MSE-accurate architectures also achieve superior directional or uncertainty-aware performance under explicit supervision.

  \item \textbf{Multi-regime validation} using new, non-overlapping historical periods and symmetric retraining across all models to test whether the present ranking hierarchy transfers across market regimes.

  \item \textbf{Extended horizon coverage} ($h \in \{1, 8, 12, 48, 96\}$) for smoother degradation curves and more precise characterisation of architecture-specific scaling behaviour.

  \item \textbf{Asset-specific model selection} exploring whether the niche advantages observed for \nhits on lower-capitalisation cryptocurrency (Table~\ref{tab:per_asset_best_models}) extend to a broader altcoin universe.

  \item \textbf{Richer feature sets} (technical indicators, sentiment, order-book data) to assess ranking sensitivity to input dimensionality.

  \item \textbf{Heterogeneous ensembles} of top architectures for potential accuracy gains, particularly combining \moderntcn's convolutional inductive bias with \patchtst's patch-based attention.

  \item \textbf{Alternative frequencies} (H4, daily, tick-level) to test cross-frequency generalisation.

  \item \textbf{Expanded asset universe} (commodities, fixed income, emerging markets).

  \item \textbf{Forward-walk evaluation} with rolling retraining for real-time deployment evidence.
\end{enumerate}

\section*{Conflict of Interest}
\label{sec:conflict_of_interest}

The authors declare that they have no known competing financial interests or personal relationships that could have appeared to influence the work reported in this paper.

\section{Reproducibility Statement}
\label{sec:reproducibility}

The accompanying repository snapshot contains the manuscript source, the
frozen configurations and configuration-hash references used after HPO, and
the generated benchmark outputs needed to audit the reported aggregates.  The
scope of the available artefacts is stated explicitly below.

\subsection{Code and Data Availability}
\label{sec:code_data}

The current snapshot contains the LaTeX manuscript source, the tabulated
benchmark outputs, per-seed metrics, aggregated statistics, benchmark
summaries, statistical test results, figures, and frozen configuration
references.  It does not contain trial-level Optuna logs or a complete
training-source tree; accordingly, this snapshot supports audit and
reproduction of the reported aggregations, but does not by itself claim a
clean end-to-end retraining of all 918 runs.  No trial-stability statistic is
reported in this revision because the underlying trial-level records are not
present.
\subsection{Determinism Controls}
\label{sec:determinism}

Randomness is managed through a centralised seed-management module
(\texttt{src/utils/seed.py}) invoked at the start of every experimental run.
Seeds are applied consistently to the Python standard library, NumPy,
PyTorch CPU and CUDA random number generators, and the
\texttt{PYTHONHASHSEED} environment variable.  The cuDNN backend is
configured with \texttt{deterministic=True} and
\texttt{benchmark=False}.  DataLoader workers derive their seeds
deterministically from the primary seed, preserving consistent
data-loading order across runs.

PyTorch's fully deterministic algorithm mode
(\texttt{use\_deterministic\_algorithms})
was not enabled, so minor floating-point variation remains possible
across different GPU architectures.  This limitation is acknowledged
in Section~\ref{sec:limitations} and mitigated by the three-seed
replication protocol.

\subsection{Configuration Versioning}
\label{sec:config_versioning}

The configuration hierarchy captures all experimental degrees of freedom: HPO sampler settings and trial budgets; final training schedules, seed lists, and early-stopping criteria; per-model hyperparameter search spaces; dataset window sizes, split ratios, and horizon definitions; and asset category assignments.  After HPO, the best hyperparameter set for each (model, category, horizon) triple is serialised as a frozen configuration file (Table~\ref{tab:best_hyperparameters}) and held fixed for all subsequent stages.

\subsection{Execution Overview}
\label{sec:execution}

The pipeline proceeds through three sequential stages---hyperparameter optimisation, multi-seed final training, and benchmarking with statistical validation---each accessible via a unified command-line interface.  Stages may be executed independently with optional filtering by model, seed, horizon, or configuration path.  Training supports automatic checkpoint resumption, restoring model weights, optimiser and scheduler states, and all random-number-generator states from the most recent checkpoint.

\subsection{Environment Specification}
\label{sec:environment}

The software environment is fully specified by the repository's \texttt{environment.yml} (Conda; Python version, deep learning framework, CUDA toolkit) and \texttt{requirements.txt} (pinned library versions).  All experiments were run using PyTorch, as described by \parencite{Paszke2019}, on CUDA-enabled hardware.  Results are reproducible within the same software environment, subject to standard floating-point and hardware-level numerical variation.

\clearpage
\printbibliography

\clearpage
\appendix

\section{Additional Empirical Results}
\label{sec:app_empirical}

\subsection{Representative Per-Asset Results}
\label{sec:app_per_asset}

One representative instrument from each asset class---BTC/USDT (cryptocurrency), EUR/USD (forex),
and Dow Jones/USA30IDXUSD (equity indices)---is presented below.  These are the HPO representative
assets and therefore provide the most controlled inter-model comparison.  \rmse bar charts for all
remaining instruments (ETH/USDT, BNB/USDT, ADA/USDT, USD/JPY, GBP/USD, AUD/USD, S\&P~500,
NASDAQ~100, and DAX), together with corresponding \mae and rank variants, are available in the Online
Supplementary Materials.  Results are reported as mean~$\pm$~standard deviation across seeds 123,
456, and 789.

\subsubsection{Cryptocurrency}
\label{sec:app_crypto}

\begin{figure}[htbp]
  \centering
  \includegraphics[width=0.45\textwidth]{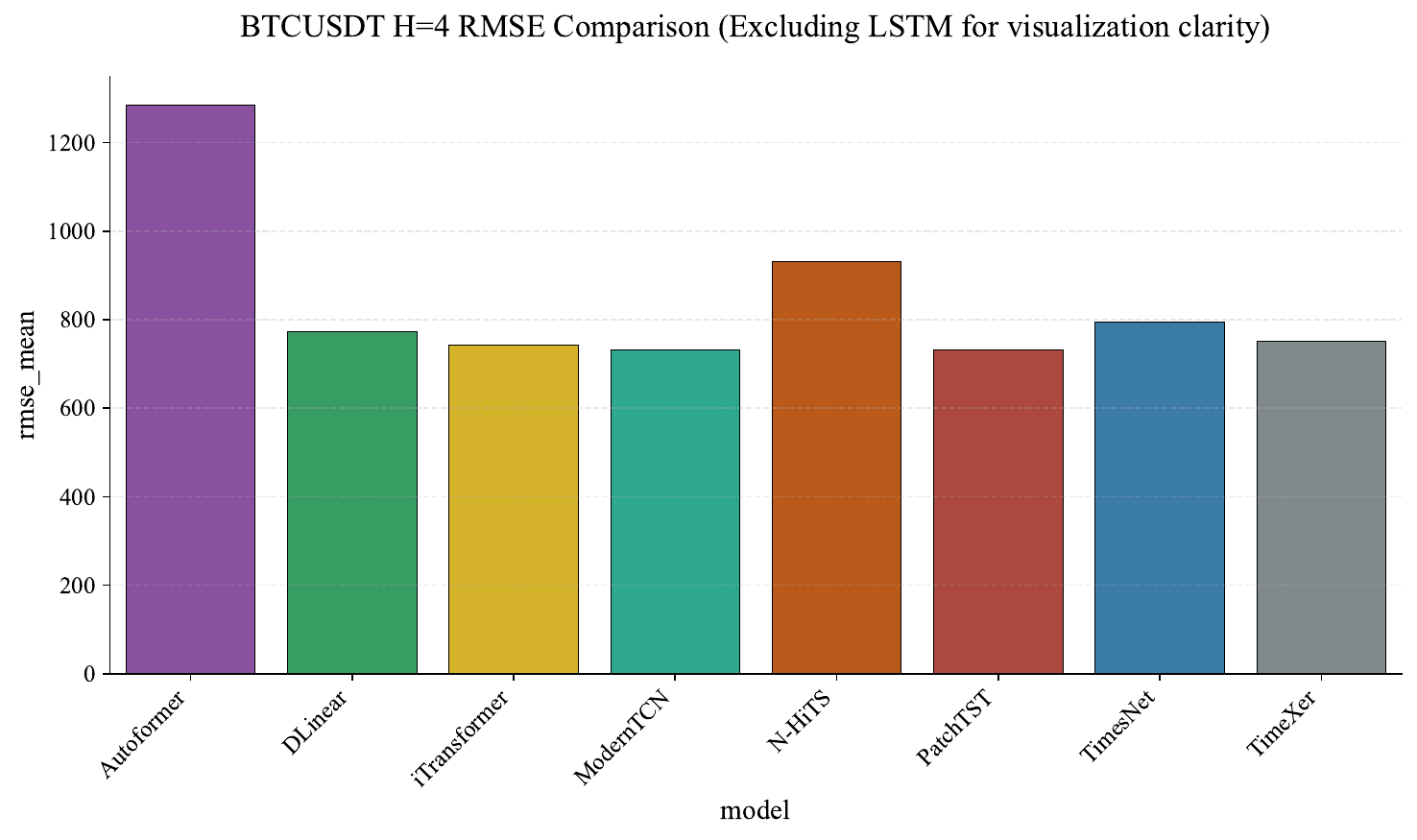}
  \hfill
  \includegraphics[width=0.45\textwidth]{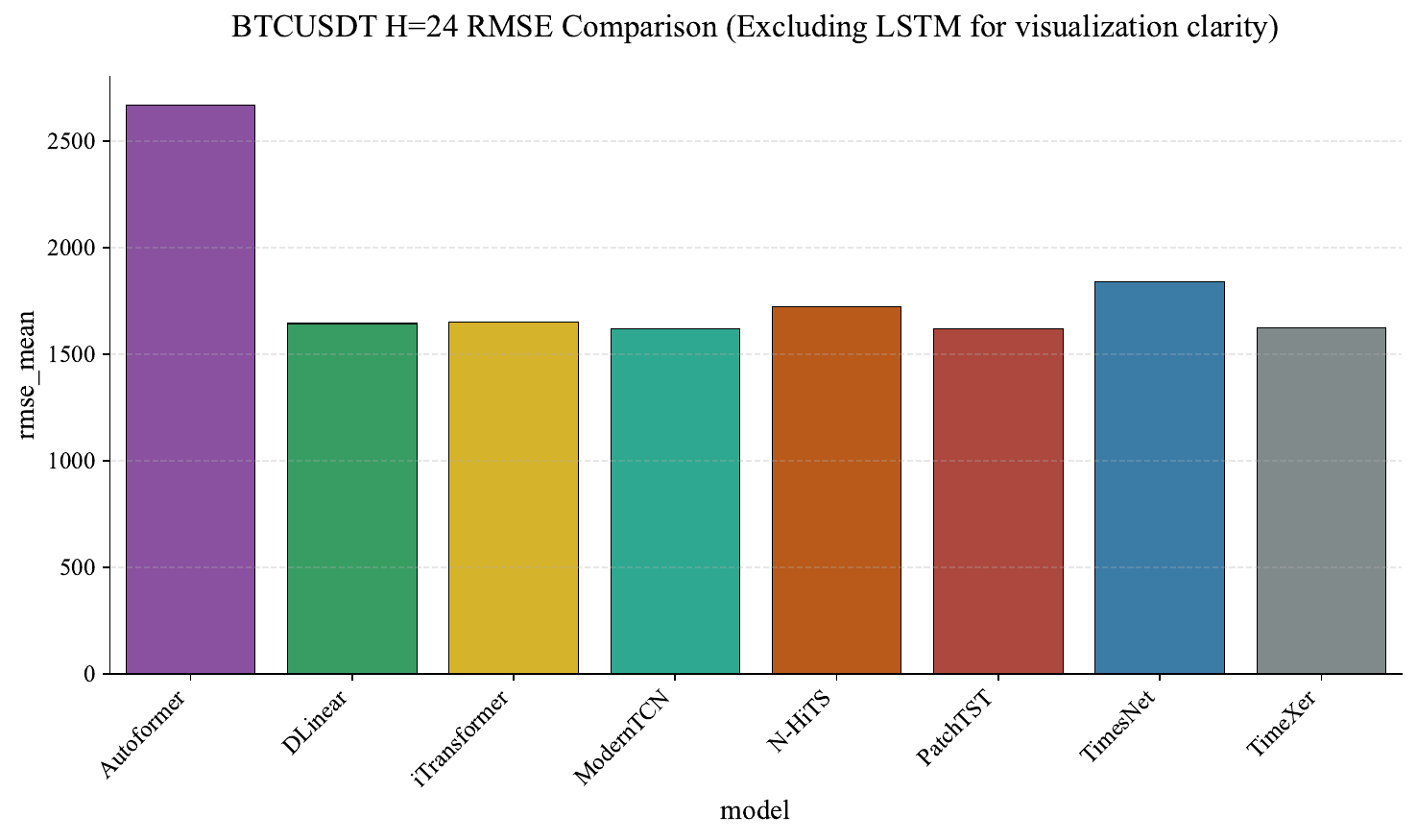}
  \caption{\rmse comparison for BTC/USDT at $\hfour$ (left) and $\htwentyfour$ (right), excluding \lstm
           for visual clarity.  \patchtst and \moderntcn achieve the two lowest errors at $\hfour$;
           \moderntcn leads marginally at $\htwentyfour$.  Mean~$\pm$~std across three seeds.}
  \label{fig:btcusdt_rmse_h4_h24}
\end{figure}

\subsubsection{Forex}
\label{sec:app_forex}

\begin{figure}[htbp]
  \centering
  \includegraphics[width=0.45\textwidth]{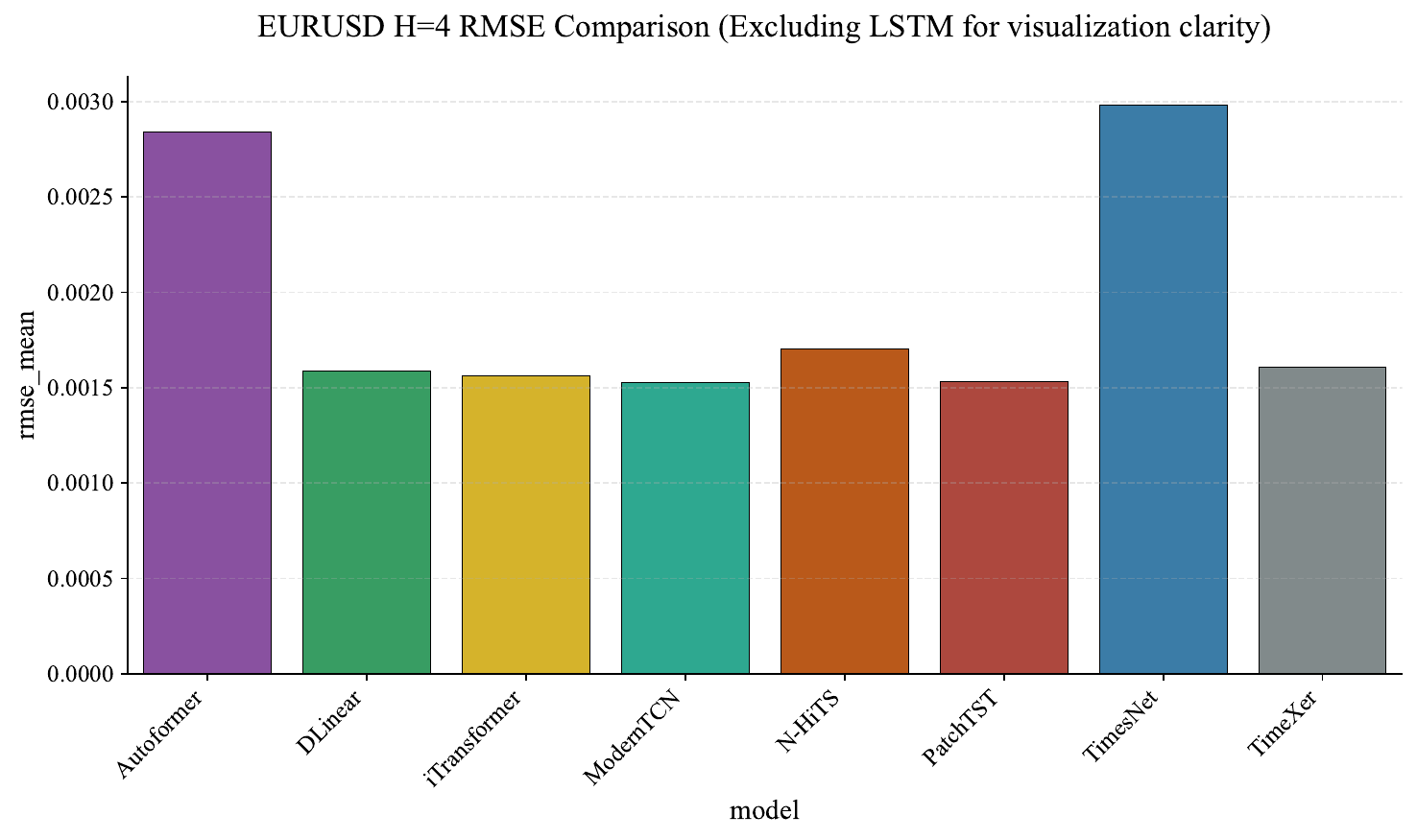}
  \hfill
  \includegraphics[width=0.45\textwidth]{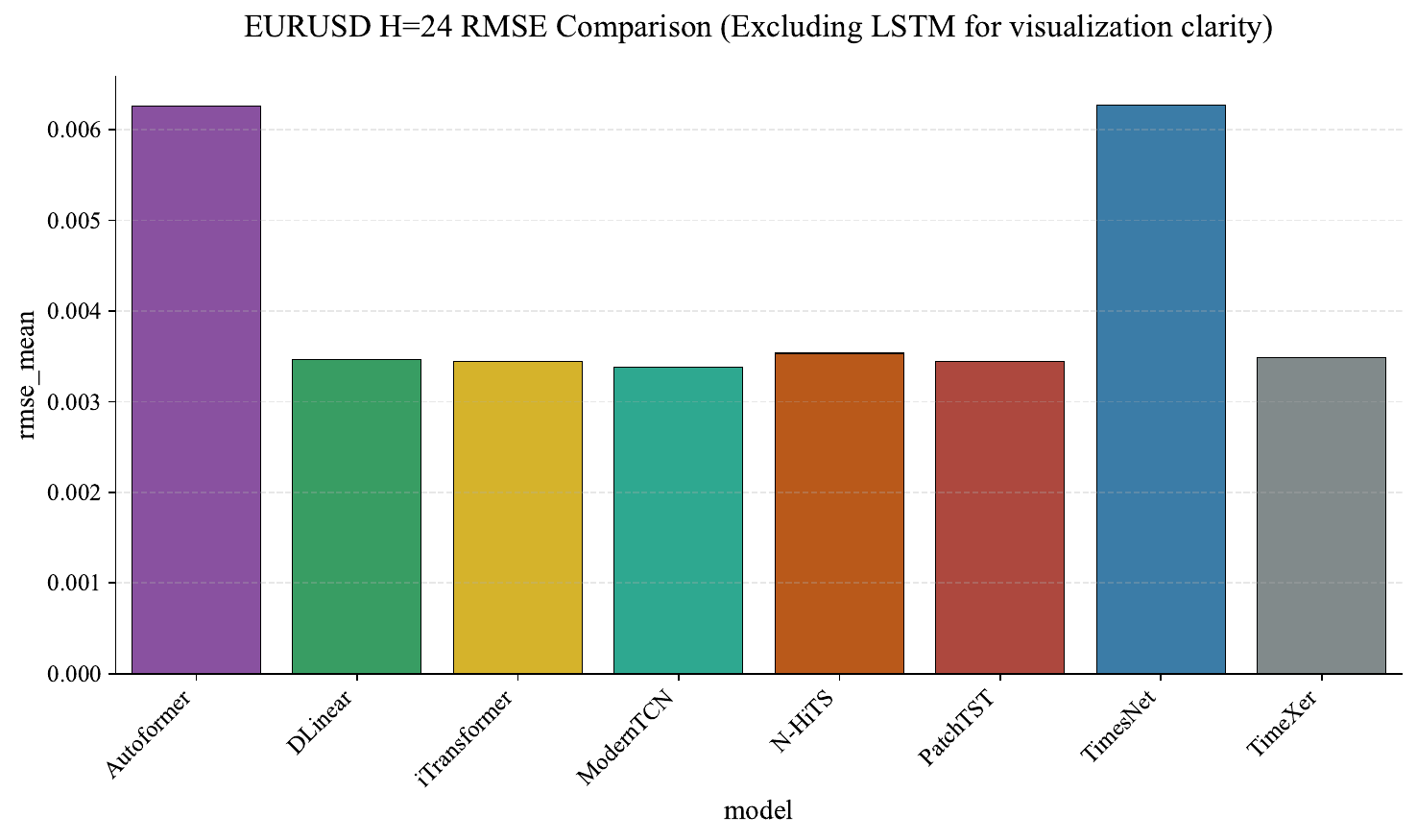}
  \caption{\rmse comparison for EUR/USD at $\hfour$ (left) and $\htwentyfour$ (right), excluding \lstm.
           \moderntcn and \patchtst rank highest across both horizons; the remaining modern architectures
           cluster within a narrow error band.  Mean~$\pm$~std across three seeds.}
  \label{fig:eurusd_rmse_h4_h24}
\end{figure}

\subsubsection{Equity Indices}
\label{sec:app_indices}

\begin{figure}[htbp]
  \centering
  \includegraphics[width=0.45\textwidth]{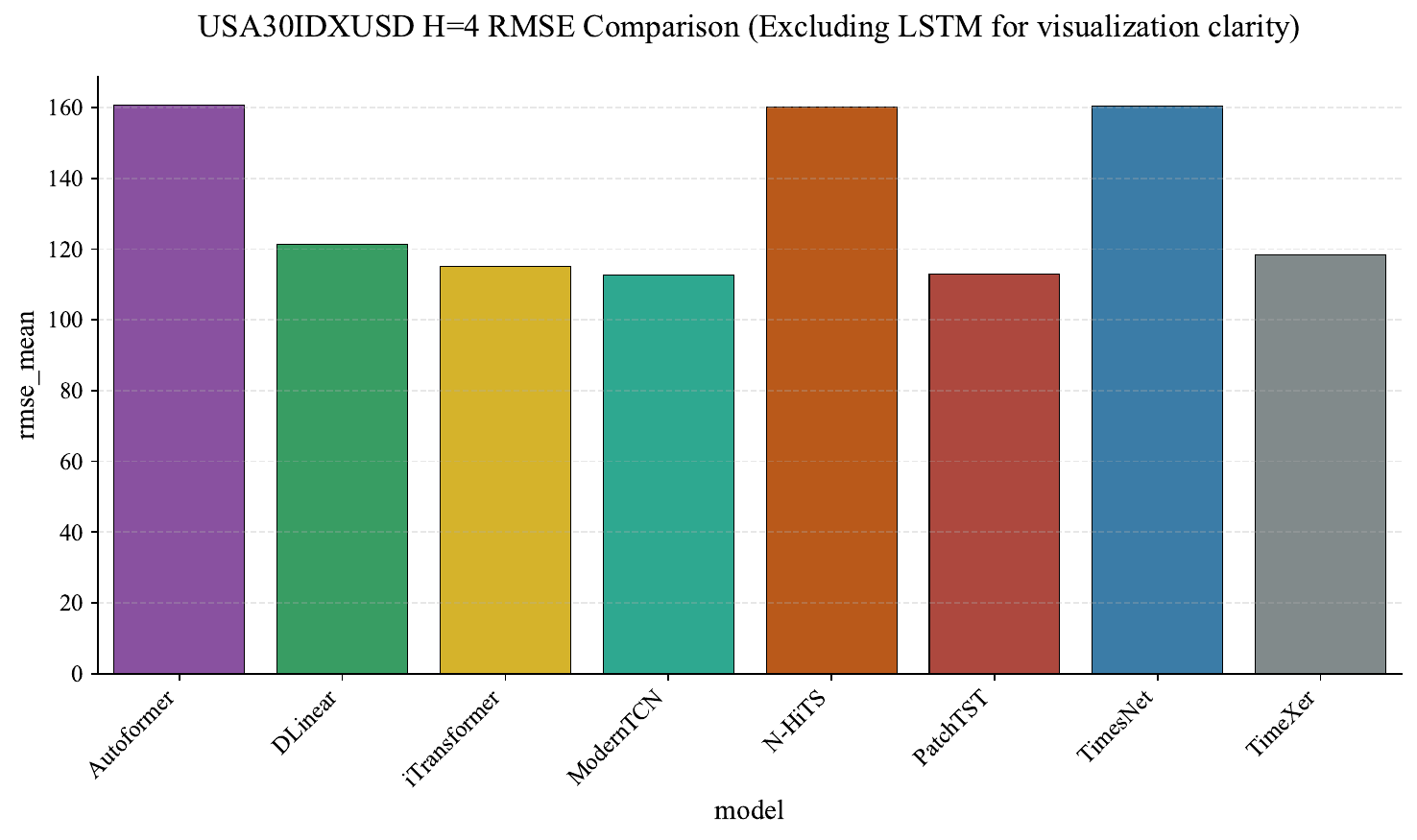}
  \hfill
  \includegraphics[width=0.45\textwidth]{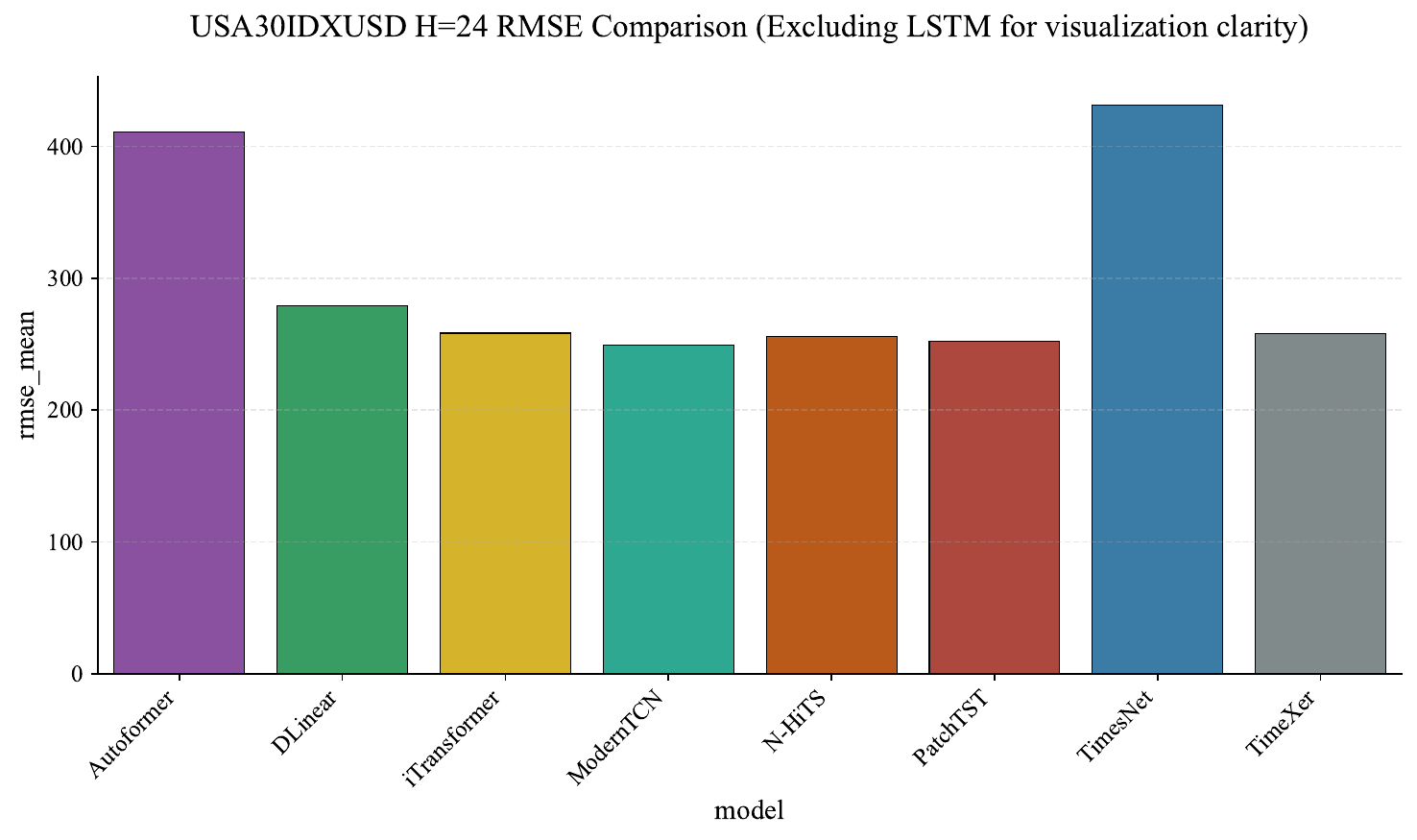}
  \caption{\rmse comparison for Dow Jones (USA30IDXUSD) at $\hfour$ (left) and $\htwentyfour$ (right),
           excluding \lstm.  \moderntcn ranks first at both horizons; \patchtst follows closely.
           Mean~$\pm$~std across three seeds.}
  \label{fig:usa30idxusd_rmse_h4_h24}
\end{figure}

\subsection{Remaining Per-Asset Results}
\label{sec:app_remaining_assets}

\begin{figure}[htbp]
  \centering
  \includegraphics[width=0.45\textwidth]{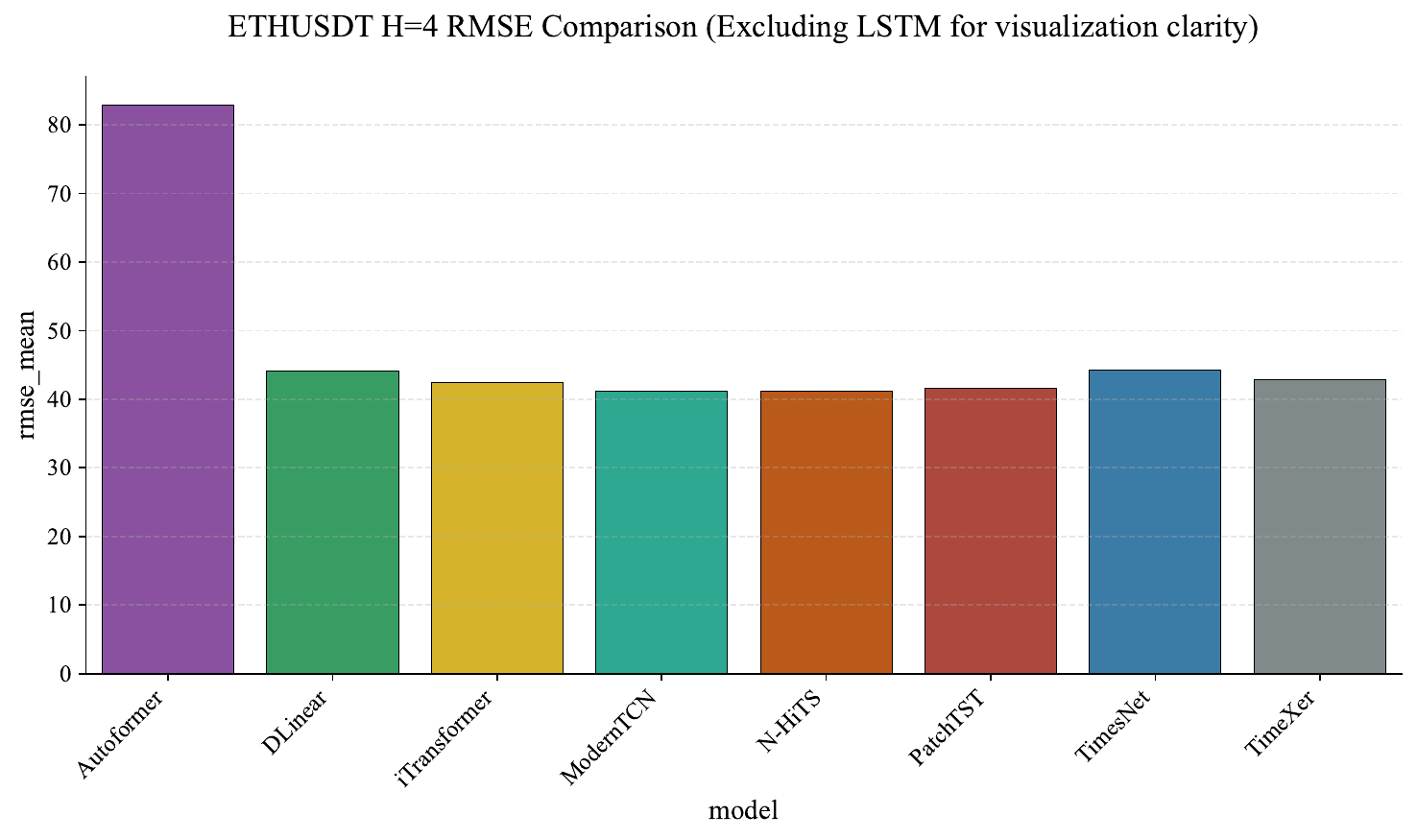}
  \hfill
  \includegraphics[width=0.45\textwidth]{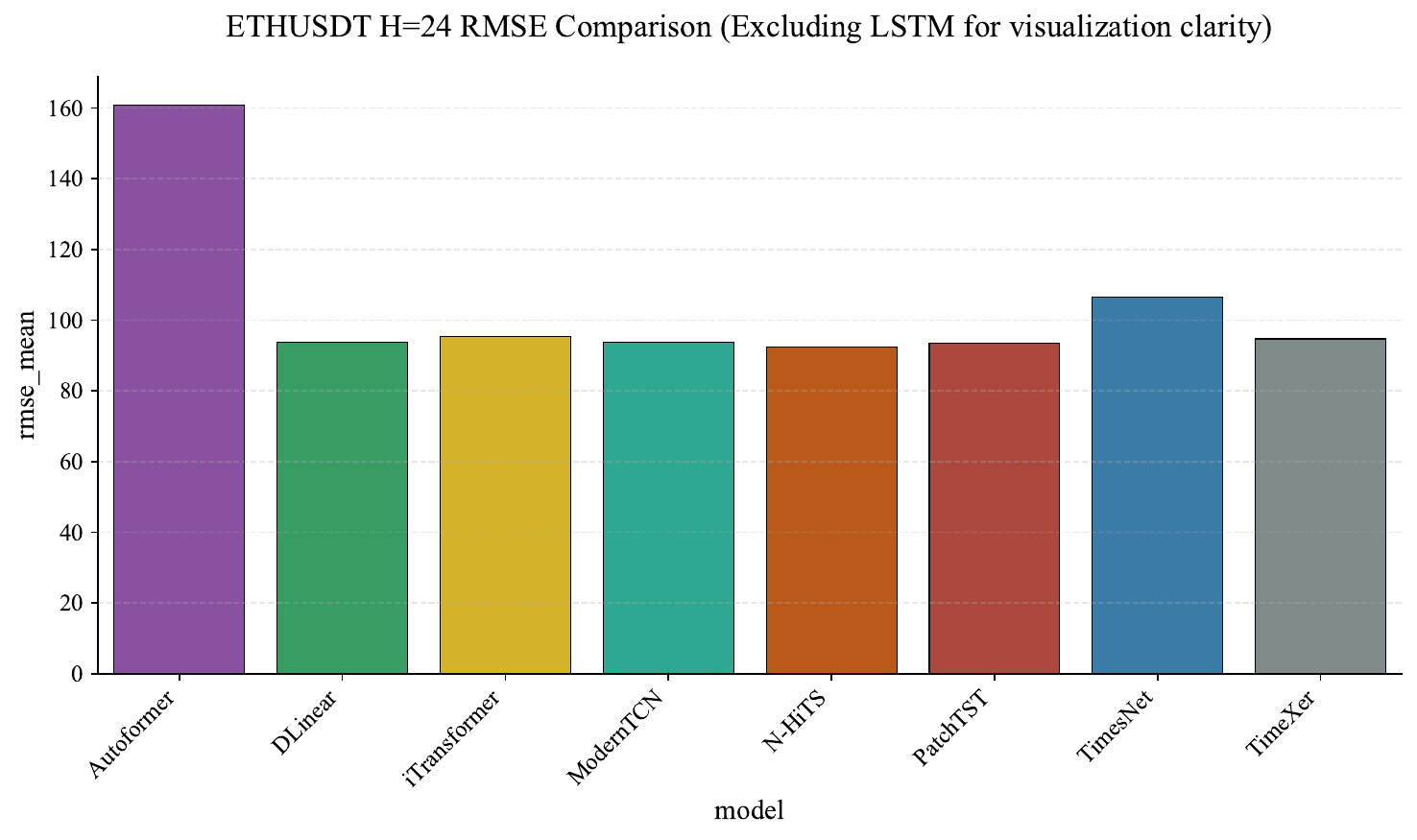}
  \caption{\rmse comparison for ETH/USDT at $\hfour$ (left) and $\htwentyfour$ (right),
           excluding \lstm.  \nhits achieves the lowest \rmse at both horizons,
           one of only two assets where \moderntcn does not rank first.
           Mean~$\pm$~std across three seeds.}
  \label{fig:ethusdt_rmse_h4_h24}
\end{figure}

\begin{figure}[htbp]
  \centering
  \includegraphics[width=0.45\textwidth]{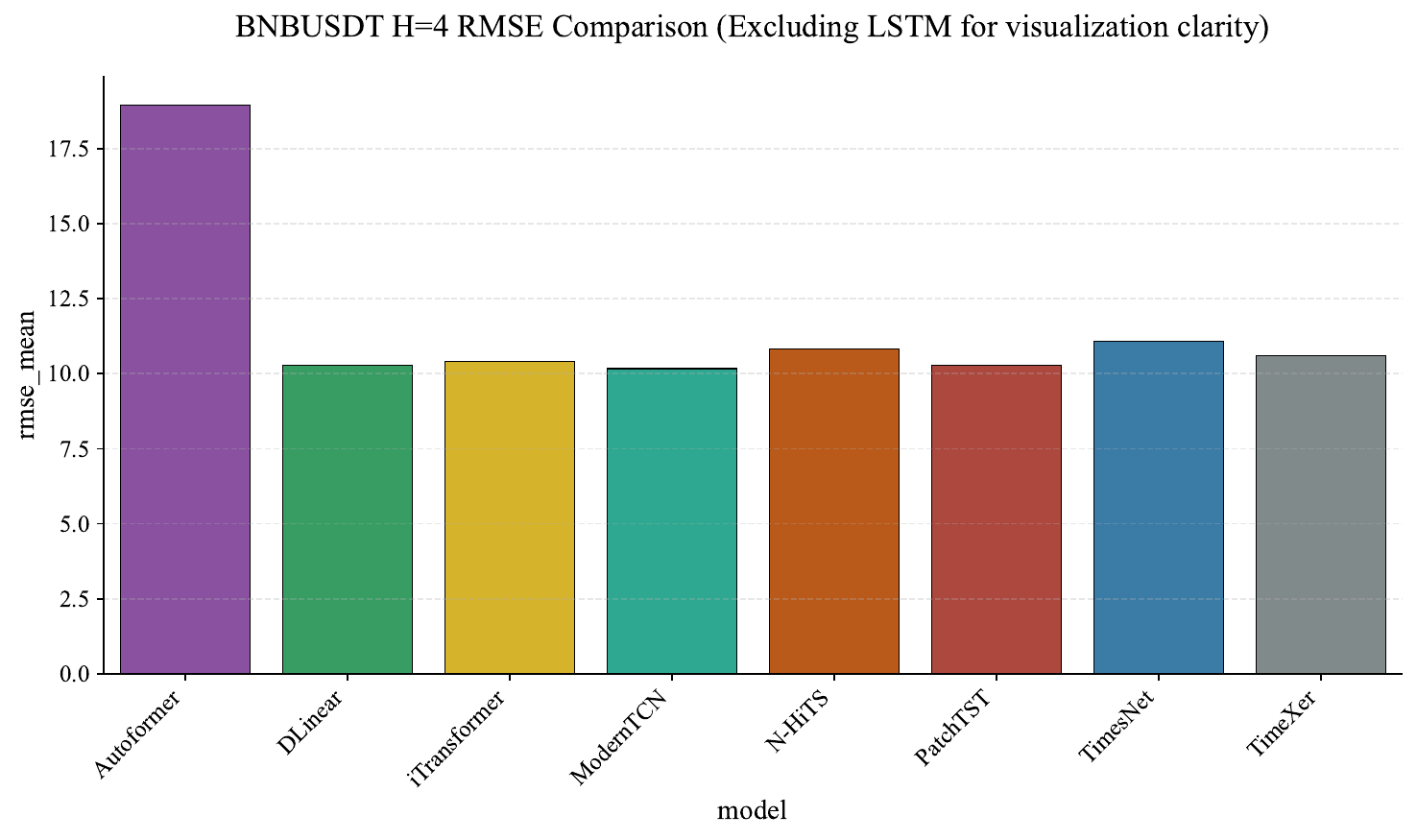}
  \hfill
  \includegraphics[width=0.45\textwidth]{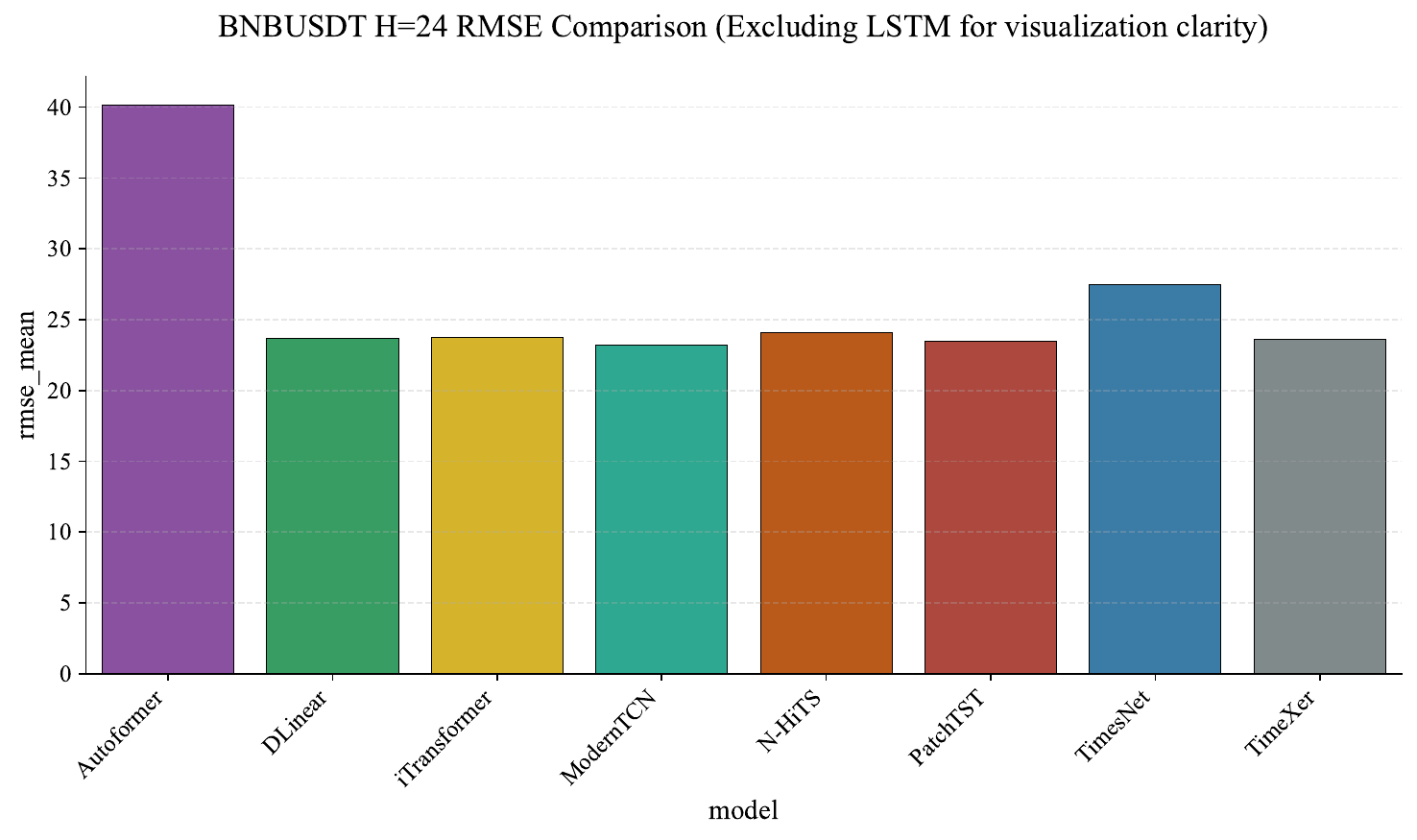}
  \caption{\rmse comparison for BNB/USDT at $\hfour$ (left) and $\htwentyfour$ (right),
           excluding \lstm.  \moderntcn leads at both horizons with a clear margin.
           Mean~$\pm$~std across three seeds.}
  \label{fig:bnbusdt_rmse_h4_h24}
\end{figure}

\begin{figure}[htbp]
  \centering
  \includegraphics[width=0.45\textwidth]{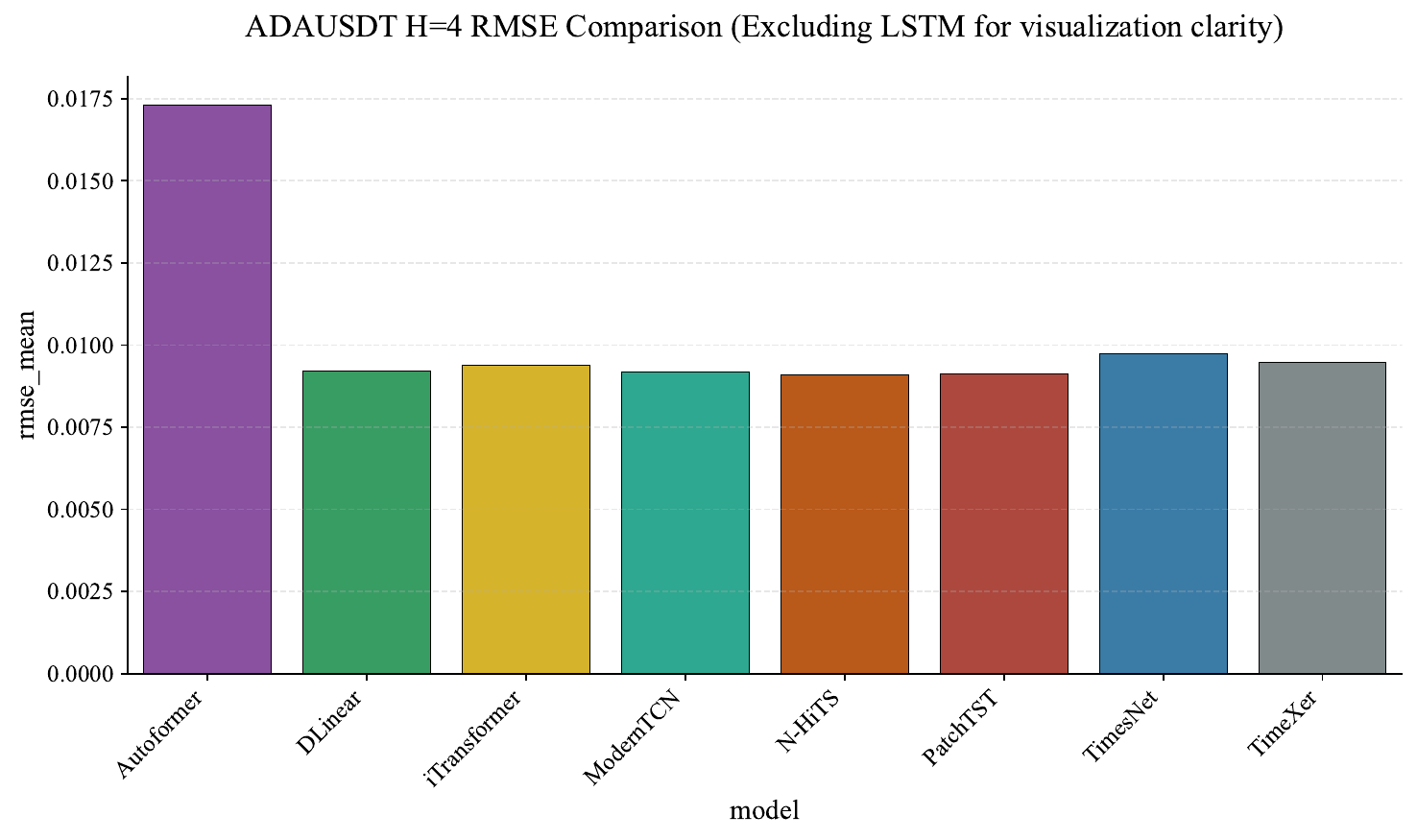}
  \hfill
  \includegraphics[width=0.45\textwidth]{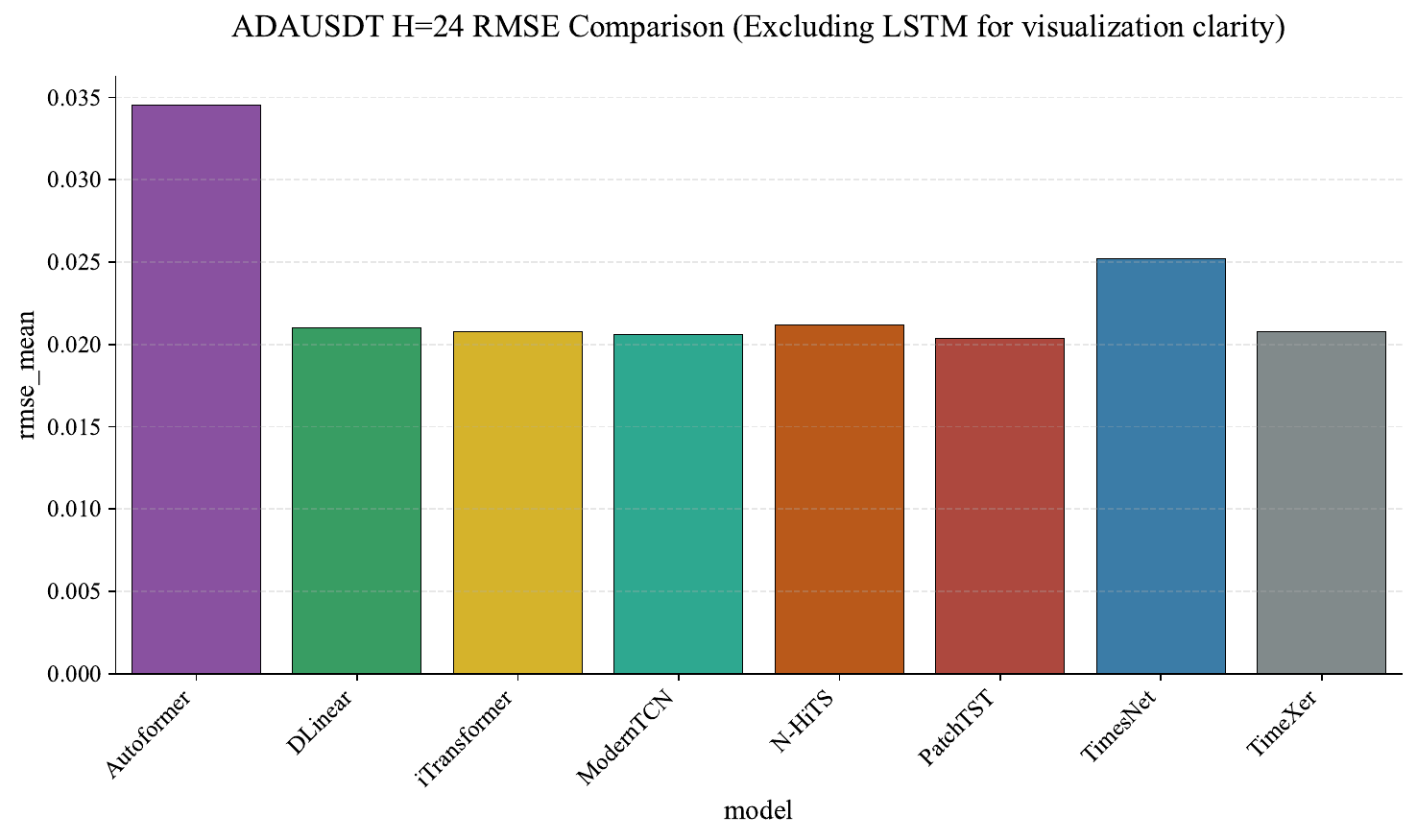}
  \caption{\rmse comparison for ADA/USDT at $\hfour$ (left) and $\htwentyfour$ (right),
           excluding \lstm.  \nhits leads at $\hfour$; \patchtst leads at $\htwentyfour$.
           This asset exhibits the lowest cross-horizon Spearman $\rho$ ($0.683$;
           Table~\ref{tab:stat_tests_spearman}), consistent with more variable
           rankings between horizons.
           Mean~$\pm$~std across three seeds.}
  \label{fig:adausdt_rmse_h4_h24}
\end{figure}

\begin{figure}[htbp]
  \centering
  \includegraphics[width=0.45\textwidth]{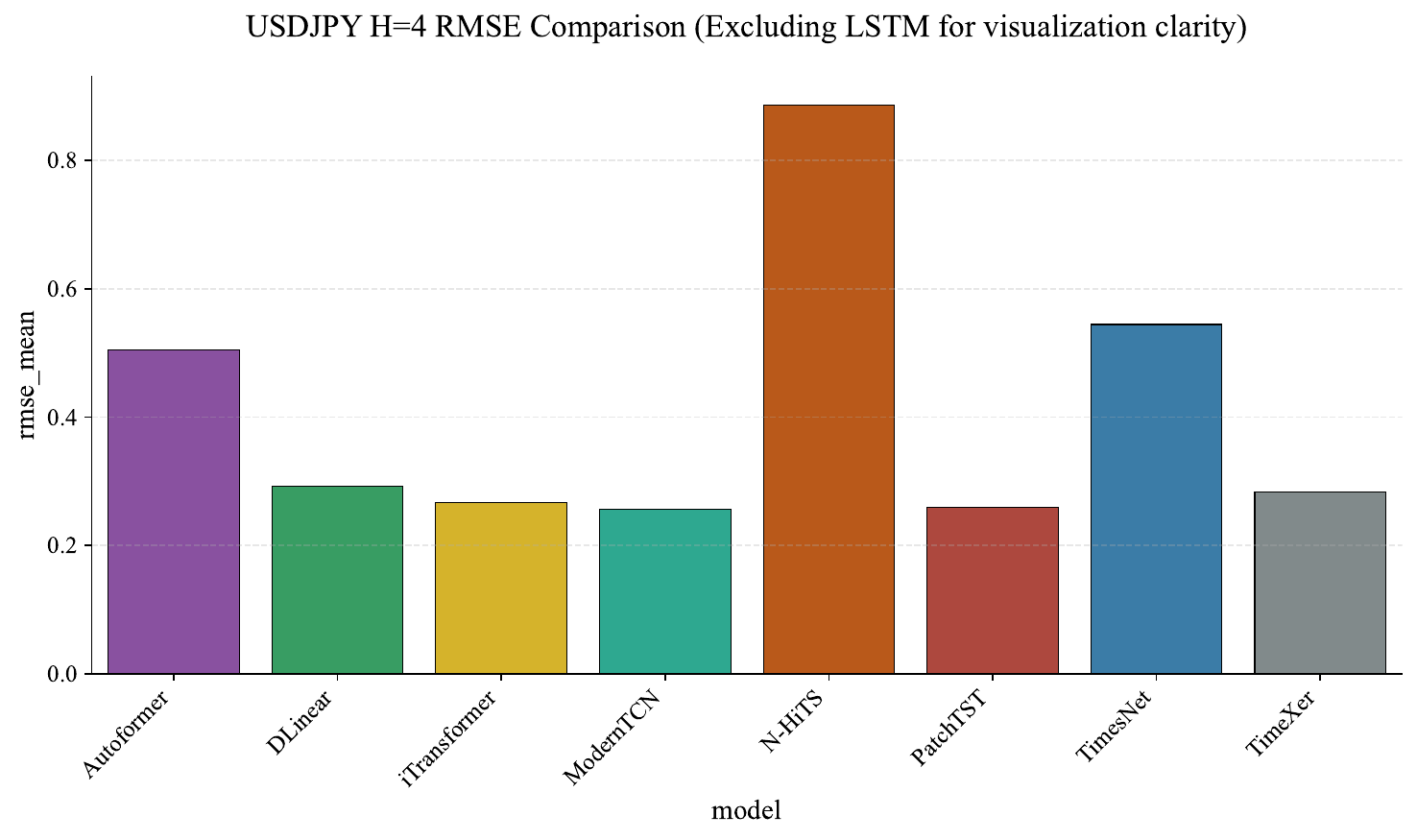}
  \hfill
  \includegraphics[width=0.45\textwidth]{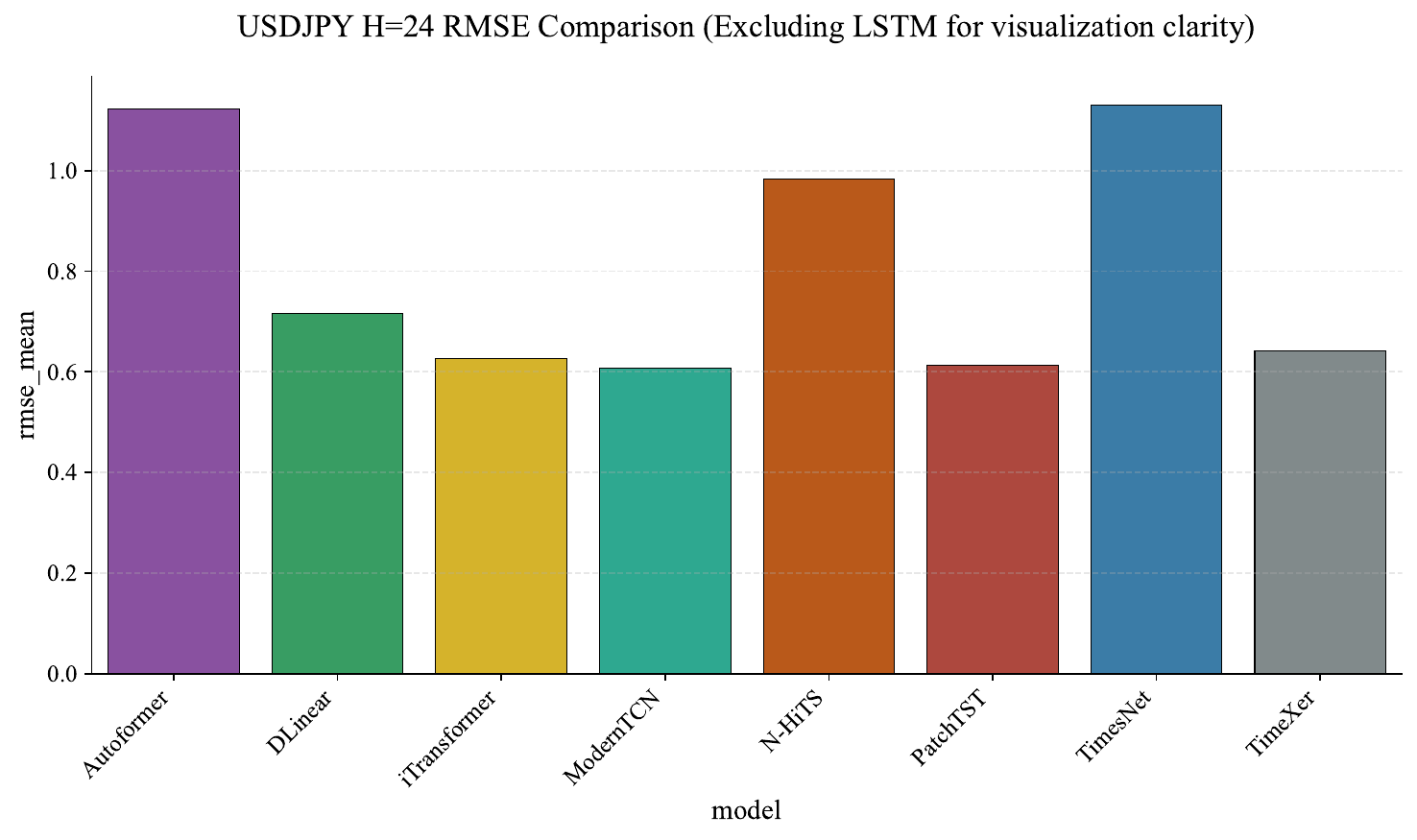}
  \caption{\rmse comparison for USD/JPY at $\hfour$ (left) and $\htwentyfour$ (right),
           excluding \lstm.  \moderntcn ranks first; the modern architecture cluster
           is tightly packed.
           Mean~$\pm$~std across three seeds.}
  \label{fig:usdjpy_rmse_h4_h24}
\end{figure}

\begin{figure}[htbp]
  \centering
  \includegraphics[width=0.45\textwidth]{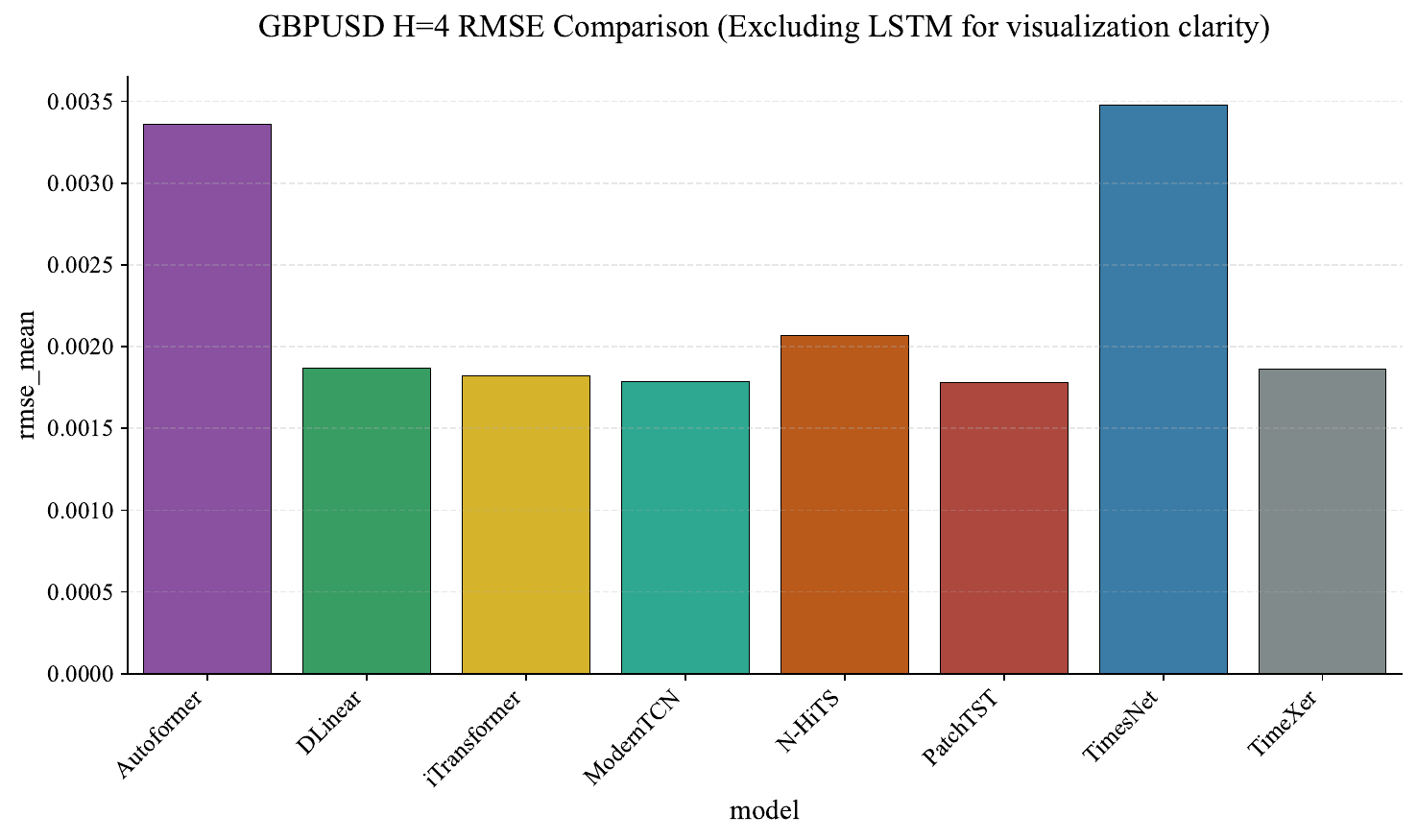}
  \hfill
  \includegraphics[width=0.45\textwidth]{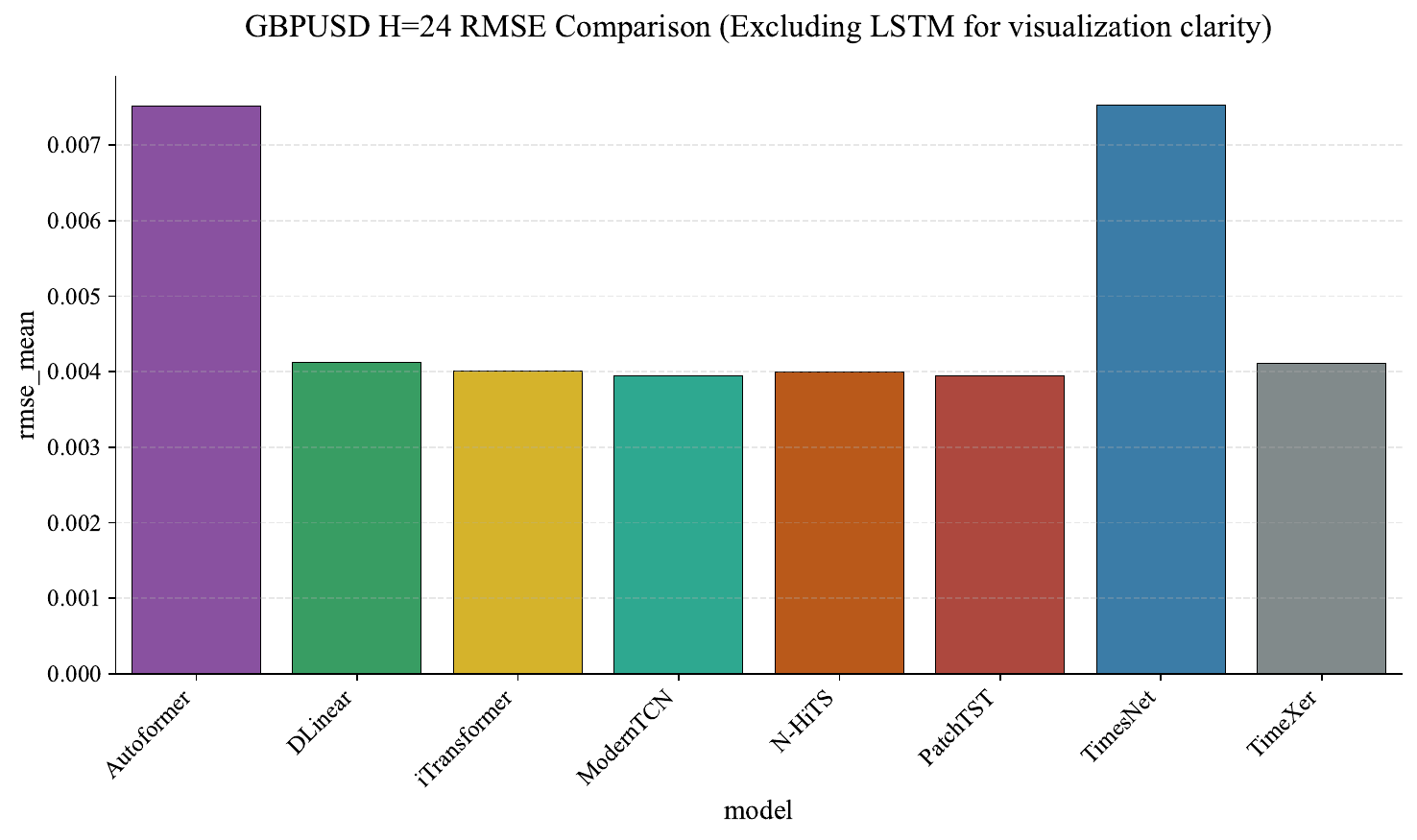}
  \caption{\rmse comparison for GBP/USD at $\hfour$ (left) and $\htwentyfour$ (right),
           excluding \lstm.  \patchtst leads at $\hfour$; \moderntcn leads
           at $\htwentyfour$.  Mean~$\pm$~std across three seeds.}
  \label{fig:gbpusd_rmse_h4_h24}
\end{figure}

\begin{figure}[htbp]
  \centering
  \includegraphics[width=0.45\textwidth]{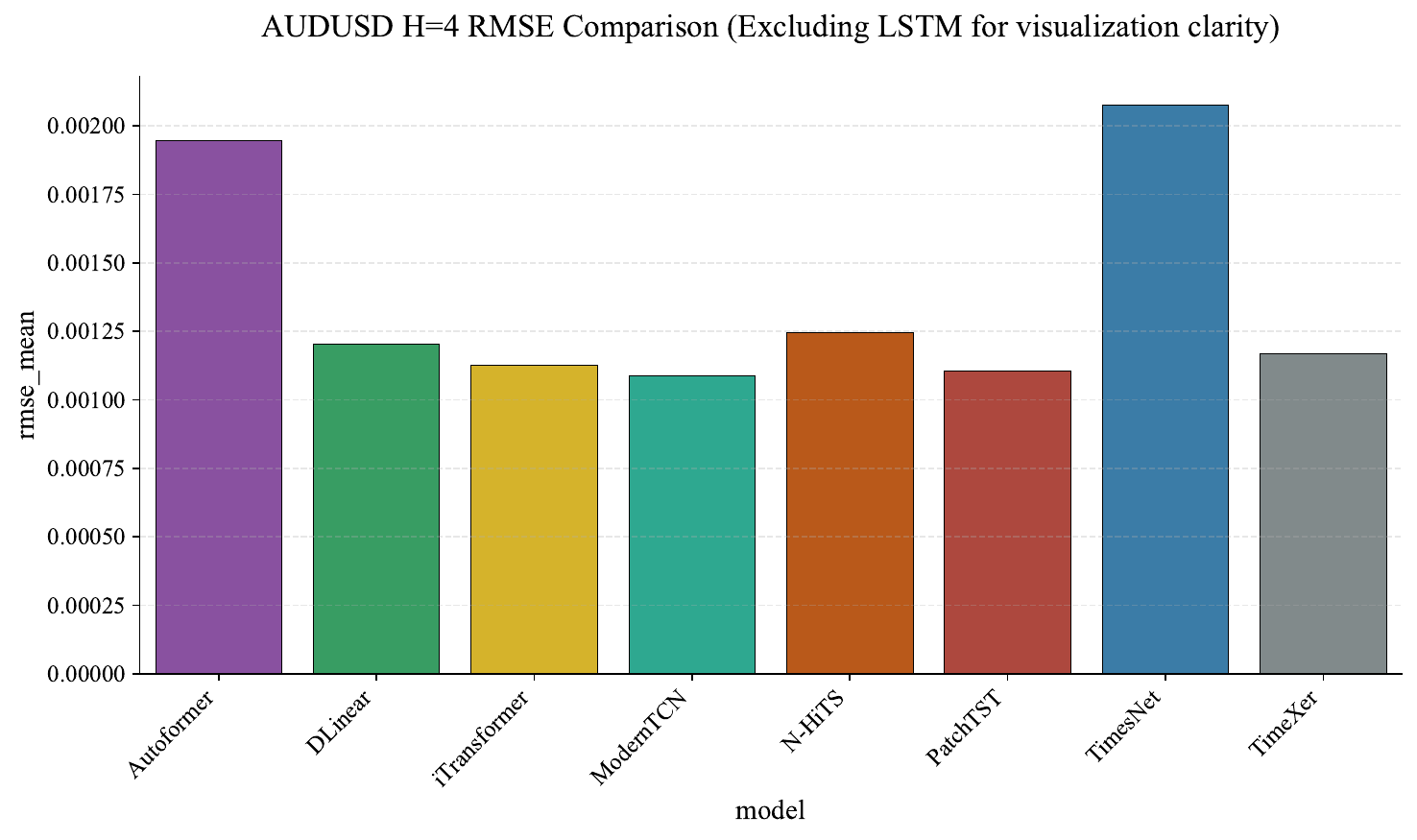}
  \hfill
  \includegraphics[width=0.45\textwidth]{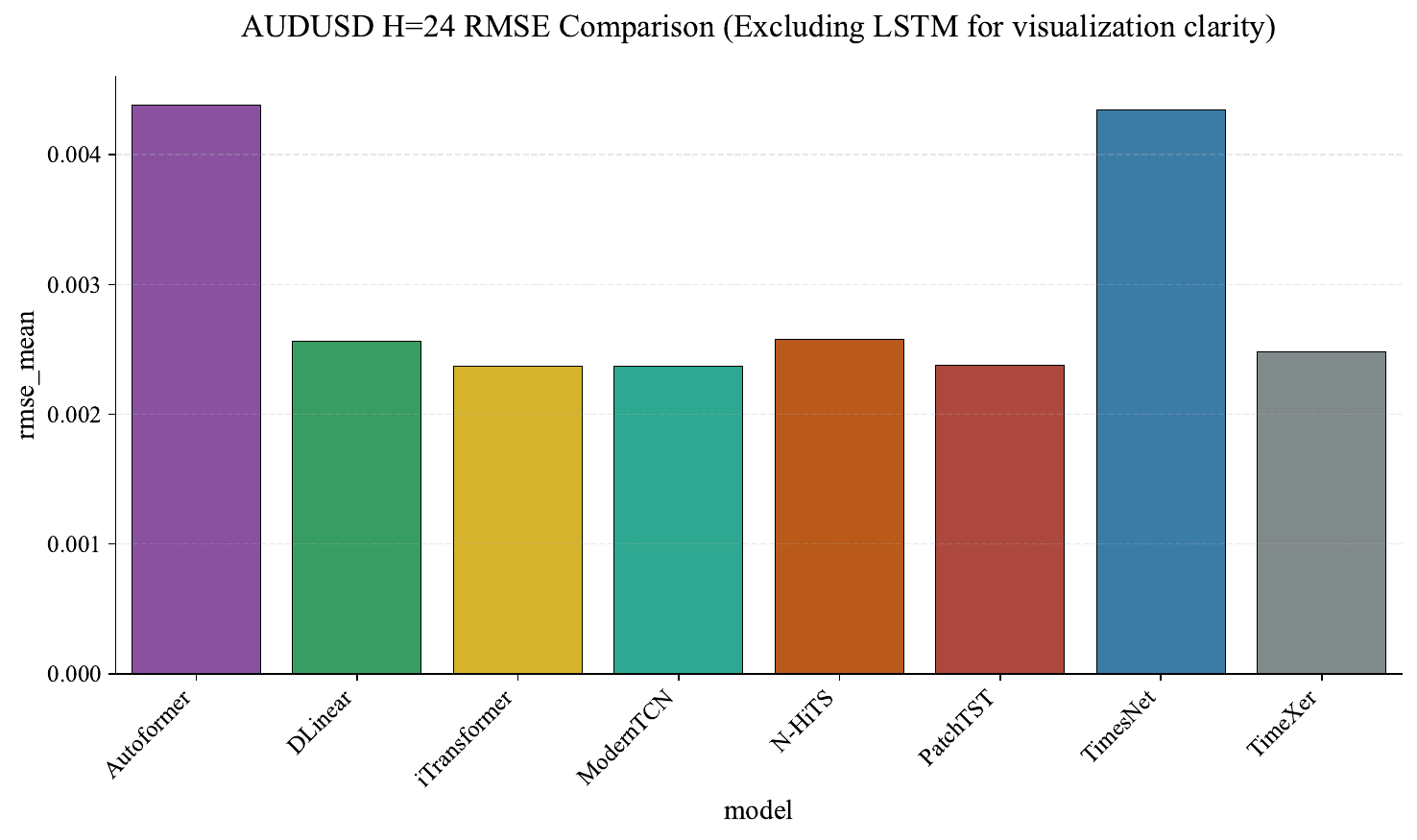}
  \caption{\rmse comparison for AUD/USD at $\hfour$ (left) and $\htwentyfour$ (right),
           excluding \lstm.  \moderntcn ranks first at both horizons.
           Mean~$\pm$~std across three seeds.}
  \label{fig:audusd_rmse_h4_h24}
\end{figure}

\begin{figure}[htbp]
  \centering
  \includegraphics[width=0.45\textwidth]{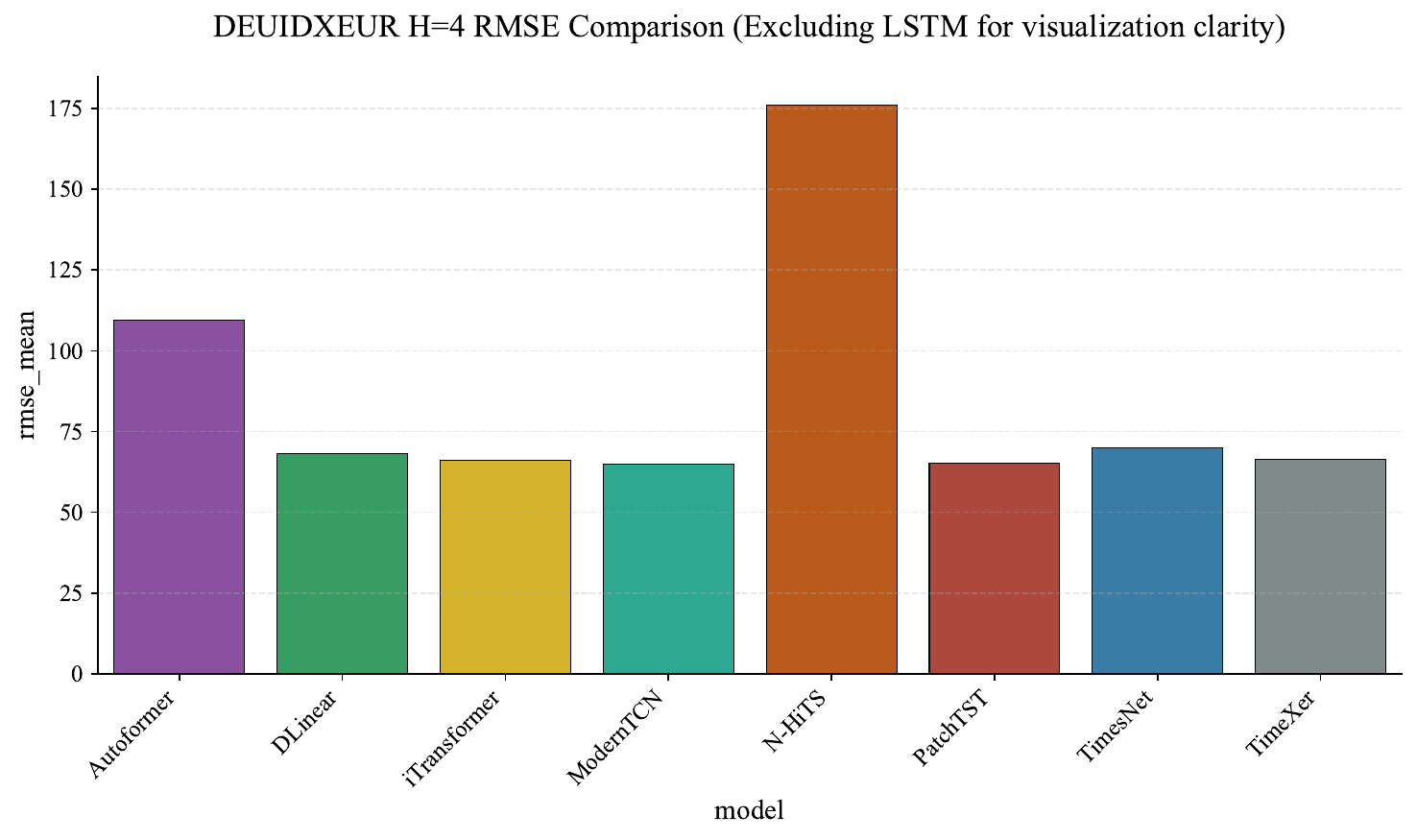}
  \hfill
  \includegraphics[width=0.45\textwidth]{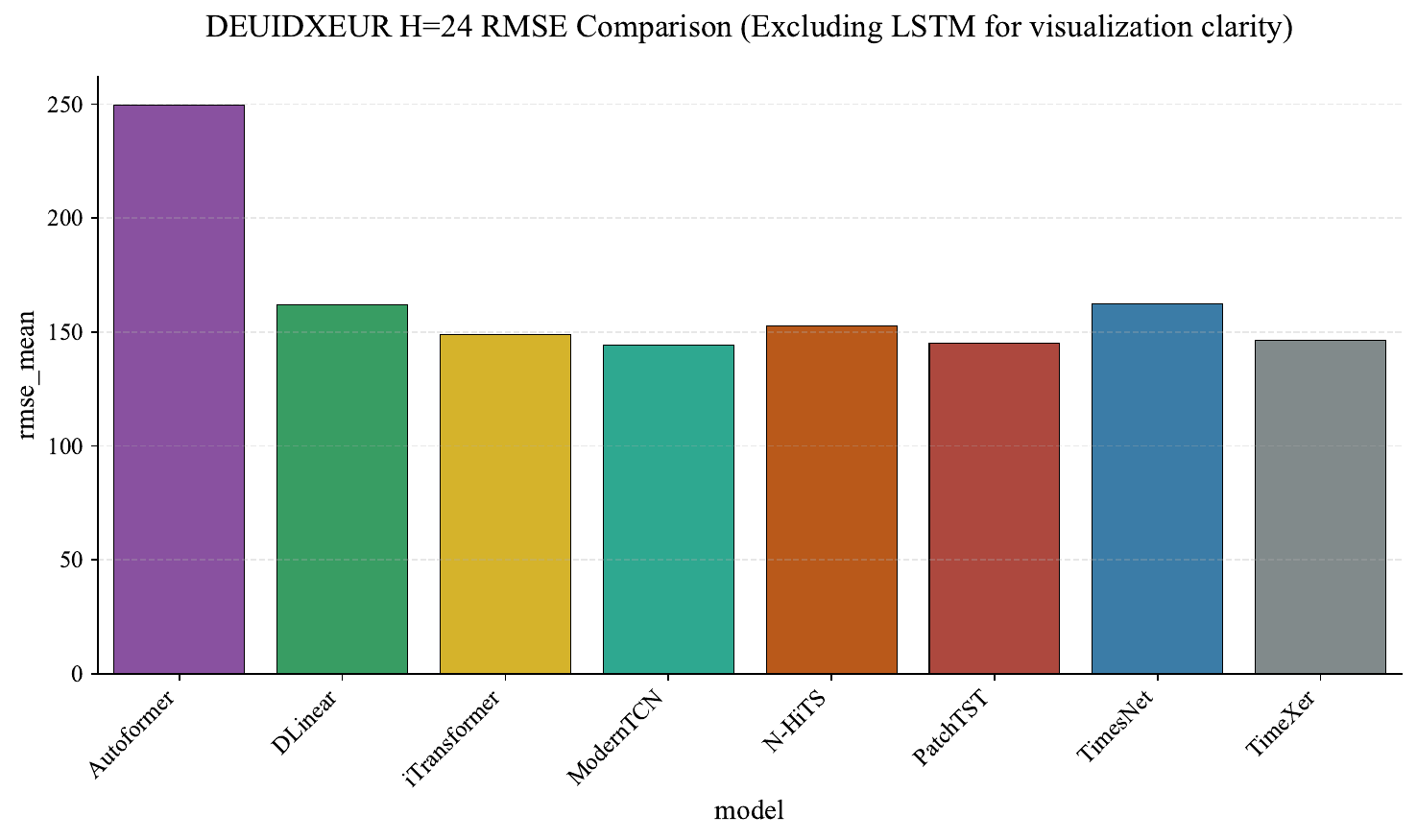}
  \caption{\rmse comparison for DAX (DEUIDXEUR) at $\hfour$ (left) and $\htwentyfour$ (right),
           excluding \lstm.  \moderntcn leads at both horizons; the top-four
           cluster (\moderntcn, \patchtst, \itransformer, \timexer) is tightly grouped.
           Mean~$\pm$~std across three seeds.}
  \label{fig:deuidxeur_rmse_h4_h24}
\end{figure}

\begin{figure}[htbp]
  \centering
  \includegraphics[width=0.45\textwidth]{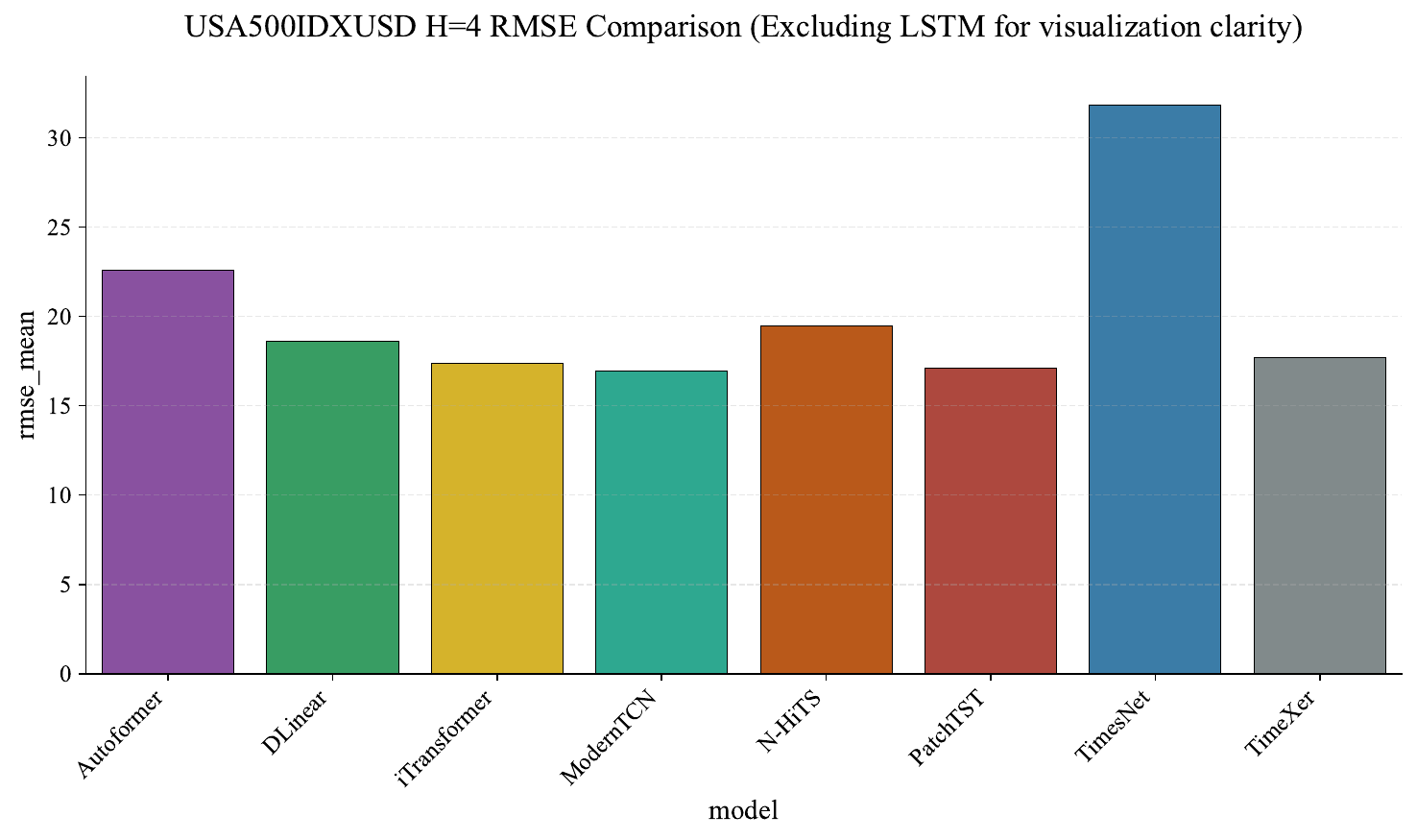}
  \hfill
  \includegraphics[width=0.45\textwidth]{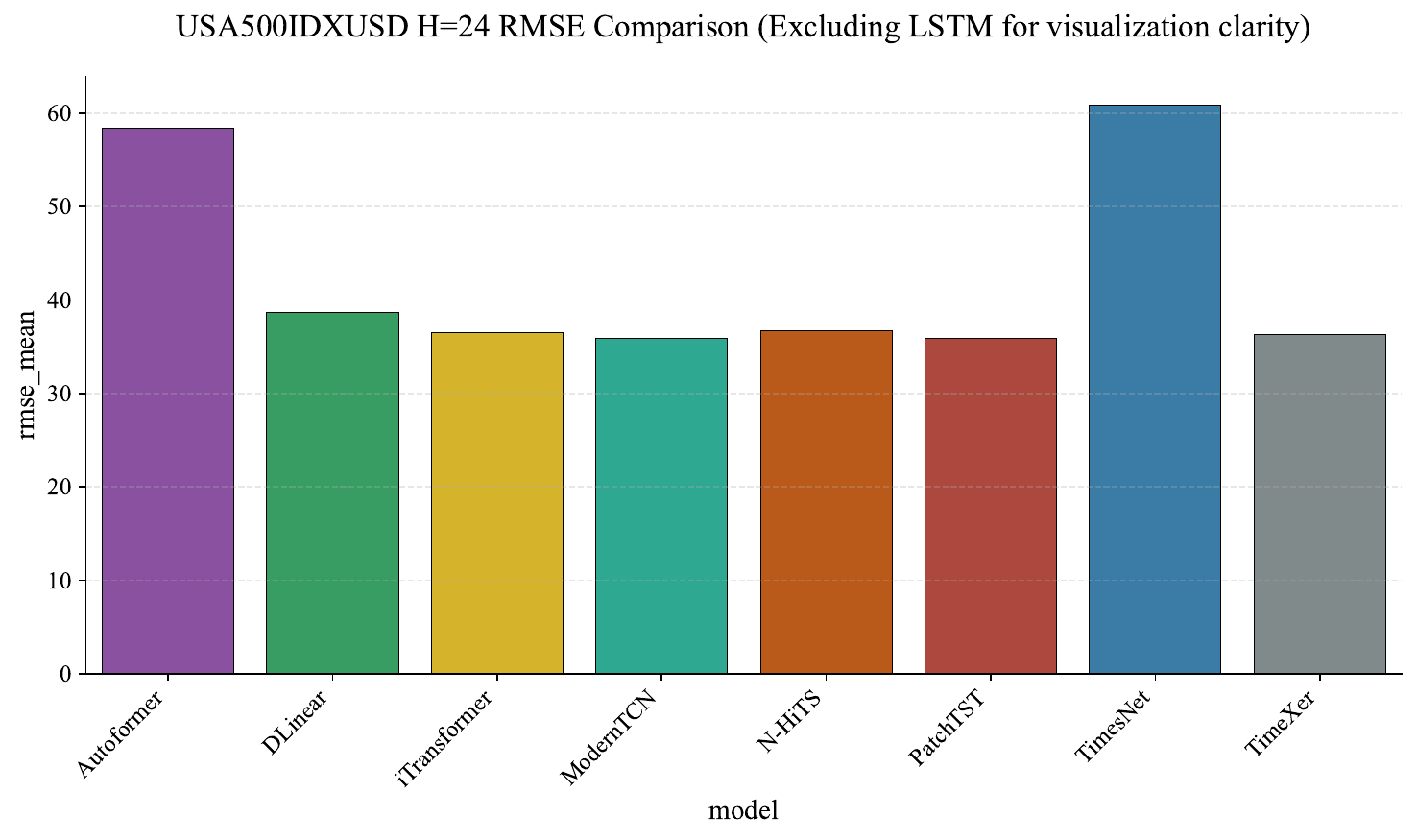}
  \caption{\rmse comparison for S\&P~500 (USA500IDXUSD) at $\hfour$ (left) and $\htwentyfour$ (right),
           excluding \lstm.  \moderntcn leads; the cross-horizon Spearman
           $\rho = 0.967$ (Table~\ref{tab:stat_tests_spearman}) indicates near-perfect
           rank preservation.  Mean~$\pm$~std across three seeds.}
  \label{fig:usa500idxusd_rmse_h4_h24}
\end{figure}

\begin{figure}[htbp]
  \centering
  \includegraphics[width=0.45\textwidth]{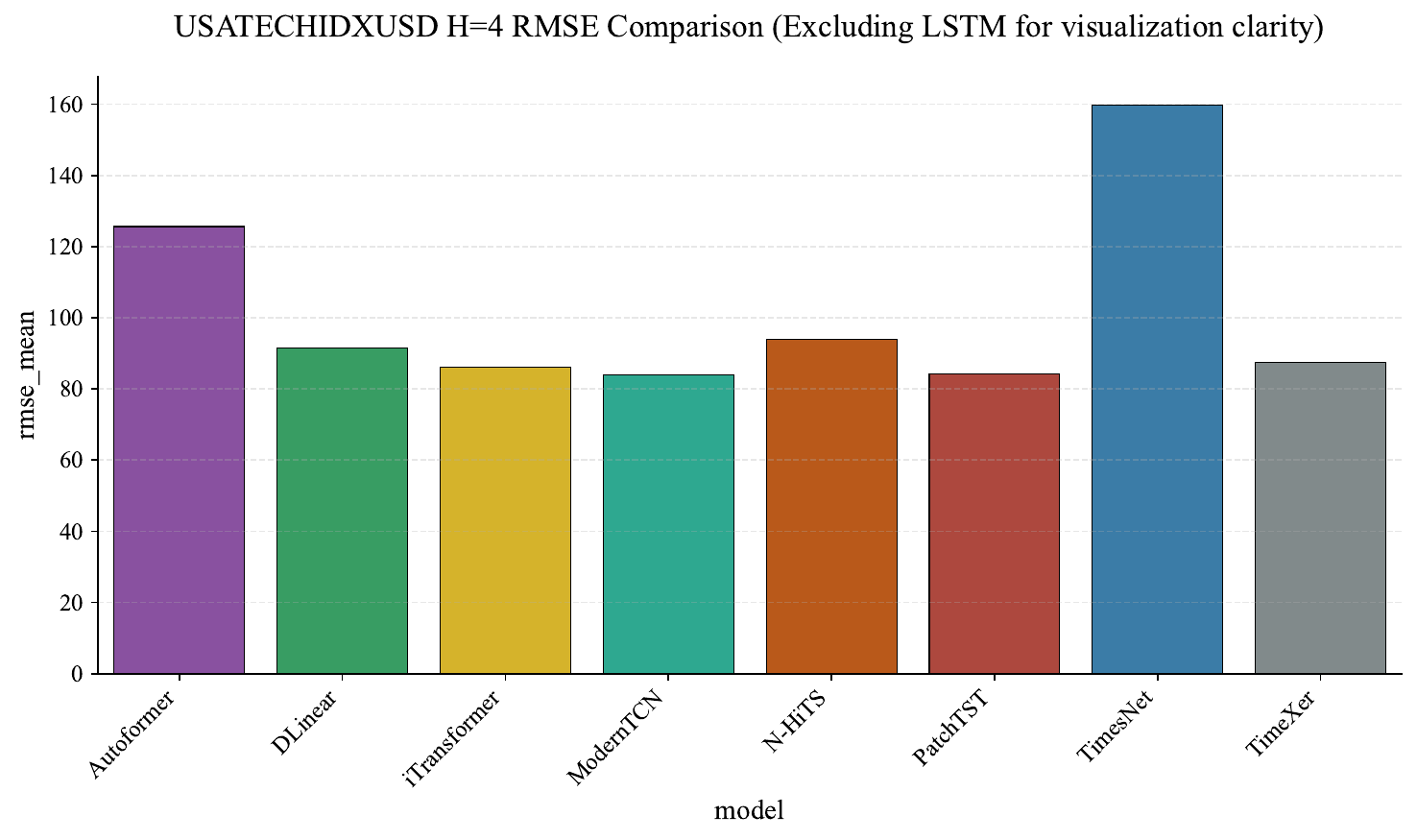}
  \hfill
  \includegraphics[width=0.45\textwidth]{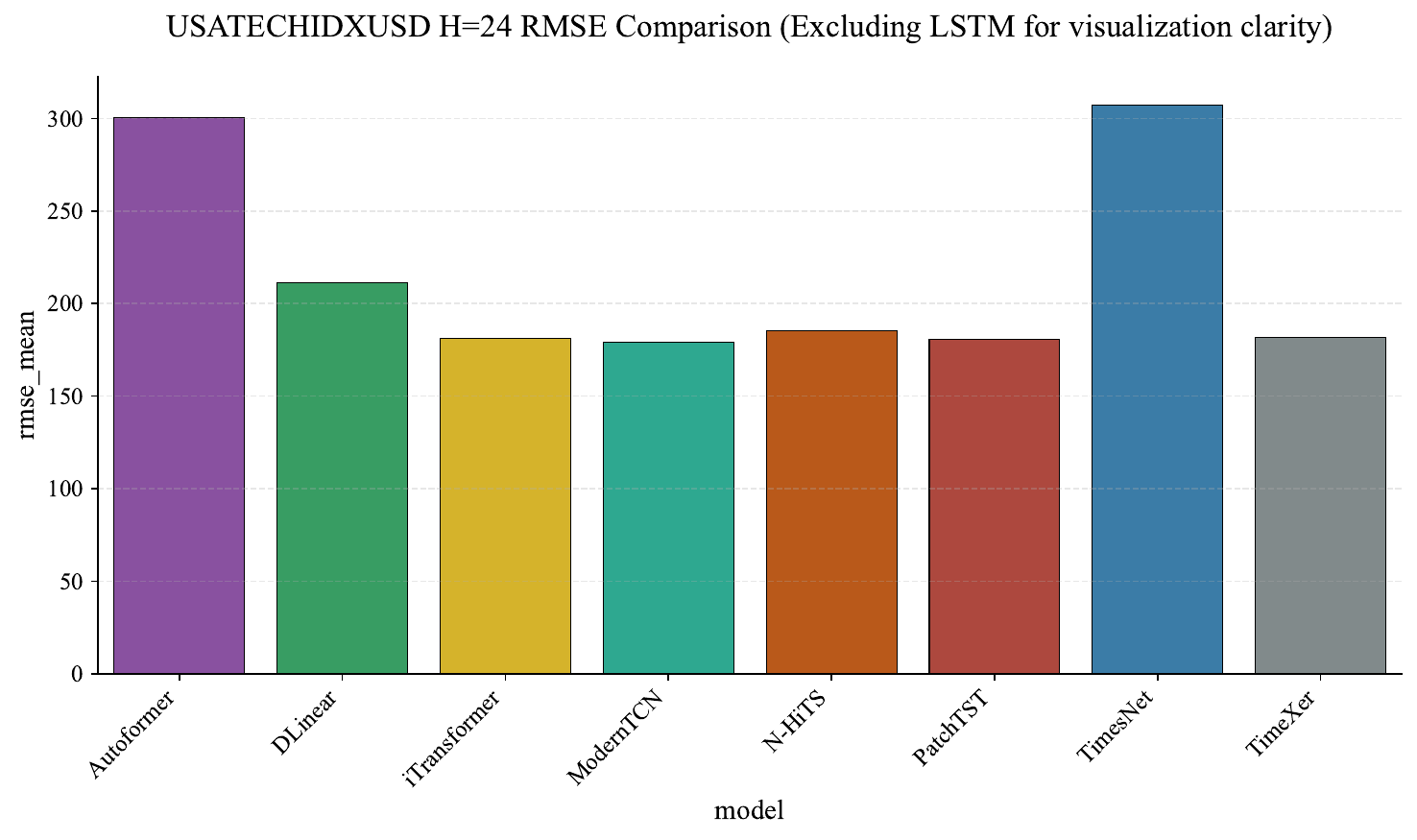}
  \caption{\rmse comparison for NASDAQ~100 (USATECHIDXUSD) at $\hfour$ (left) and $\htwentyfour$ (right),
           excluding \lstm.  \moderntcn leads; the cross-horizon Spearman
           $\rho = 0.983$ (Table~\ref{tab:stat_tests_spearman}) is the highest
           among all assets.  Mean~$\pm$~std across three seeds.}
  \label{fig:usatechidxusd_rmse_h4_h24}
\end{figure}

Figures~\ref{fig:ethusdt_rmse_h4_h24}--\ref{fig:usatechidxusd_rmse_h4_h24} present \rmse
bar charts for the nine assets not shown in Section~\ref{sec:app_per_asset}, completing the
full 12-asset comparison.  Across all assets, the three-tier structure
(\moderntcn/\patchtst at top; \itransformer/\timexer/\dlinear/\nhits in the middle;
\timesnet/\autoformer at the bottom) is maintained, with the notable exception of
ETH/USDT and ADA/USDT where \nhits achieves the lowest \rmse
(Table~\ref{tab:per_asset_best_models}).

\subsubsection{Cryptocurrency --- Remaining Assets}

\subsubsection{Forex --- Remaining Assets}

\subsubsection{Equity Indices --- Remaining Assets}

\subsection{Cross-Horizon Analysis}
\label{sec:app_horizon}

\begin{figure}[htbp]
  \centering
  \includegraphics[width=0.75\textwidth]{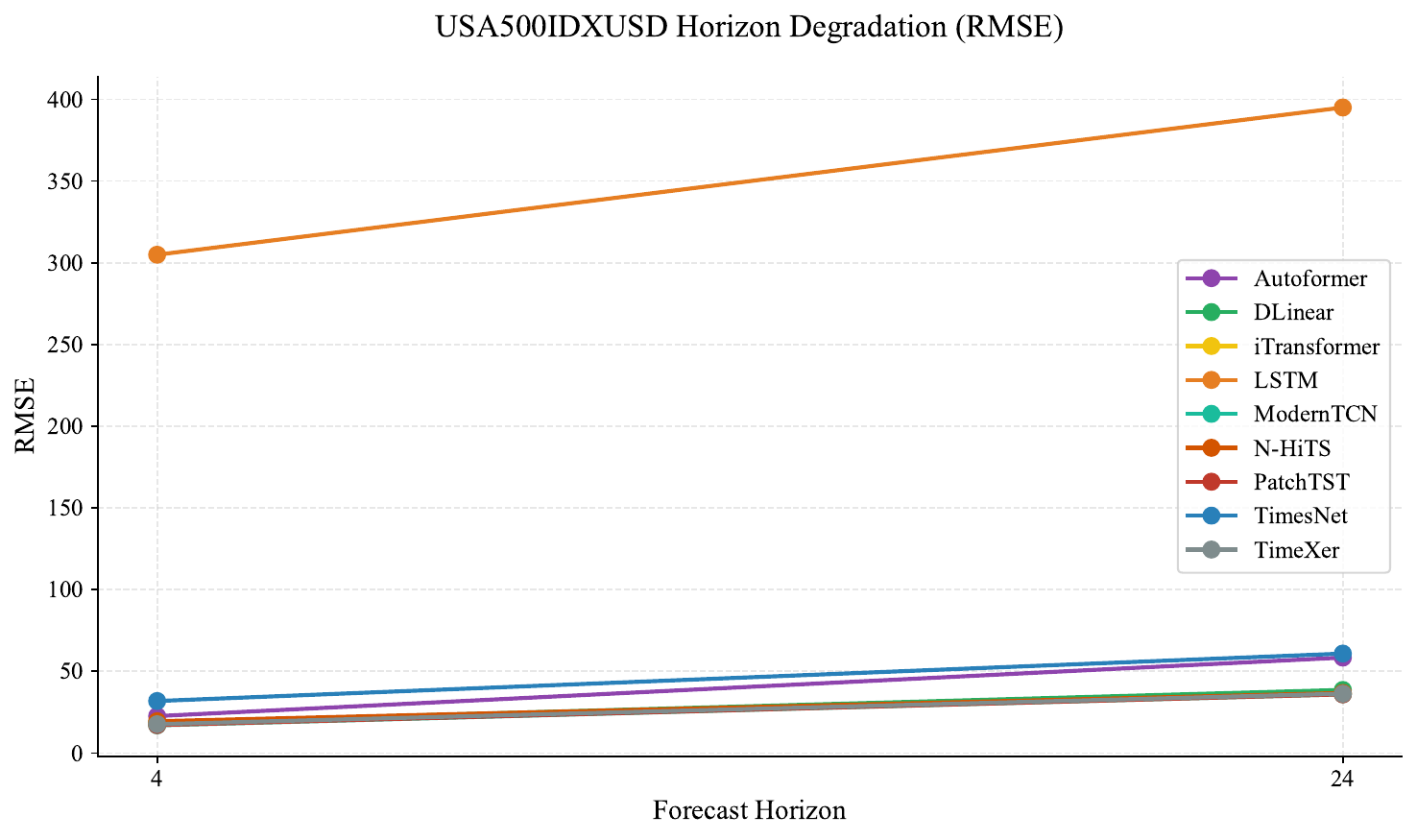}
  \caption{Cross-horizon \rmse degradation for all \nmodels models.
           Compare with Figure~\ref{fig:horizon_degradation} (no-\lstm variant) for finer inter-model
           discrimination.}
  \label{fig:app_horizon_degradation_all}
\end{figure}

Figure~\ref{fig:app_horizon_degradation_all} reproduces the horizon-degradation line plot from the
main text (Figure~\ref{fig:horizon_degradation}) with all \nmodels models included.  The no-\lstm
variant is used as the primary reference because \lstm's elevated baseline
compresses the vertical scale, obscuring differences among the eight modern architectures.

\subsection{Horizon Sensitivity Heatmap}
\label{sec:app_horizon_sensitivity}

\begin{figure}[htbp]
  \centering
  \includegraphics[width=0.85\textwidth]{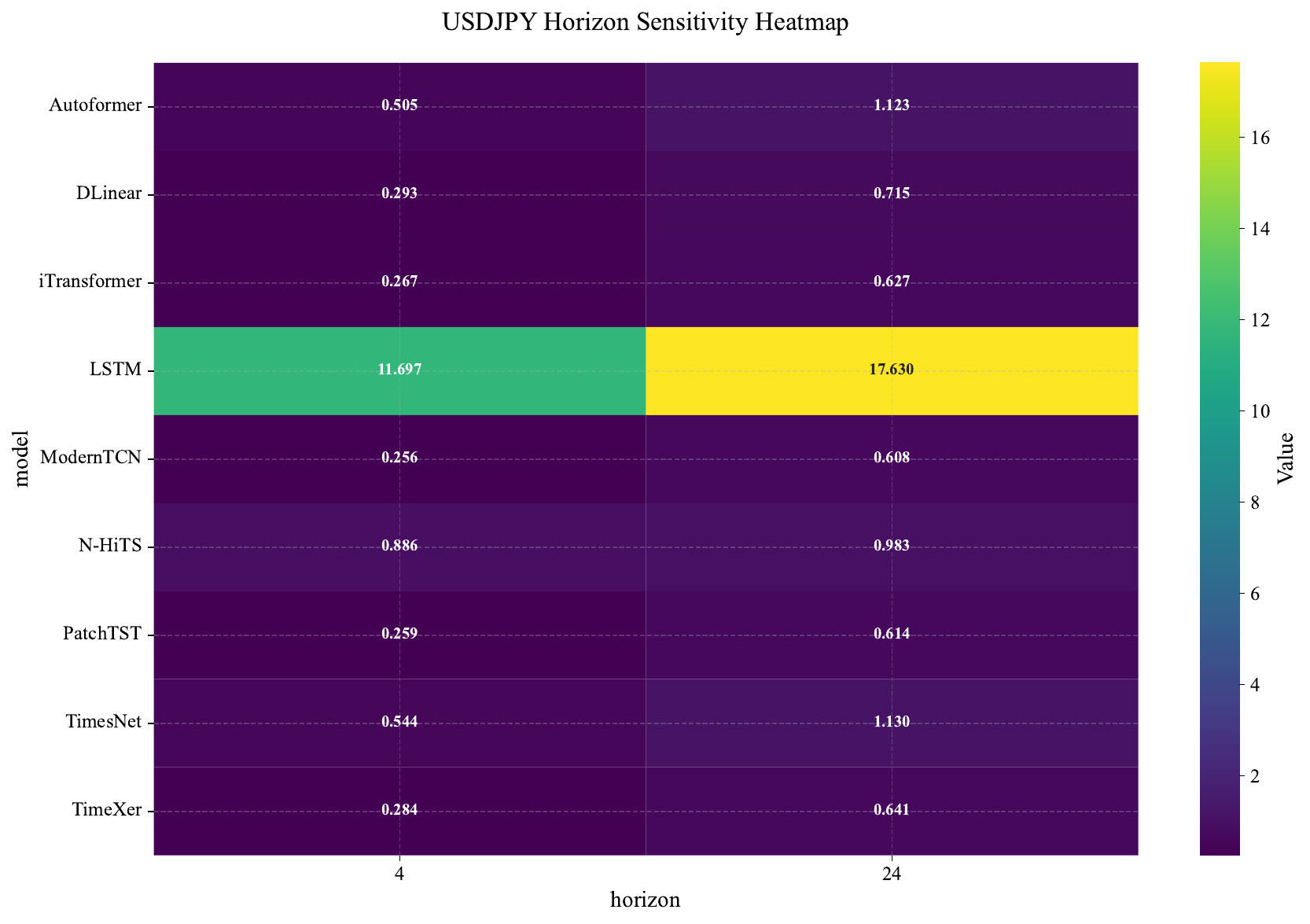}
  \caption{Horizon sensitivity heatmap: percentage \rmse degradation
           ($\Delta\% = 100\times(\text{RMSE}_{24} - \text{RMSE}_4)/\text{RMSE}_4$)
           for eight modern architectures across all twelve assets.
           \lstm is excluded for visual clarity.
           Values below 90\% (low degradation) appear in cooler colours;
           values above 150\% appear in warmer shades.
           \nhits on Dow Jones achieves the lowest degradation (59.7\%);
           \timesnet on Dow Jones the highest (169.1\%).}
  \label{fig:app_horizon_sensitivity_heatmap}
\end{figure}

Figure~\ref{fig:app_horizon_sensitivity_heatmap} displays the percentage \rmse
degradation from $\hfour$ to $\htwentyfour$ for each model--asset combination (excluding \lstm).
Cool cells indicate architectures whose representations transfer well
across horizons; warm cells highlight combinations where temporal structure
deteriorates rapidly with forecast depth.

\subsection{Dual-Plot Variants (All Models Including \lstm)}
\label{sec:app_dual_plots}

\begin{figure}[htbp]
  \centering
  \includegraphics[width=0.85\textwidth]{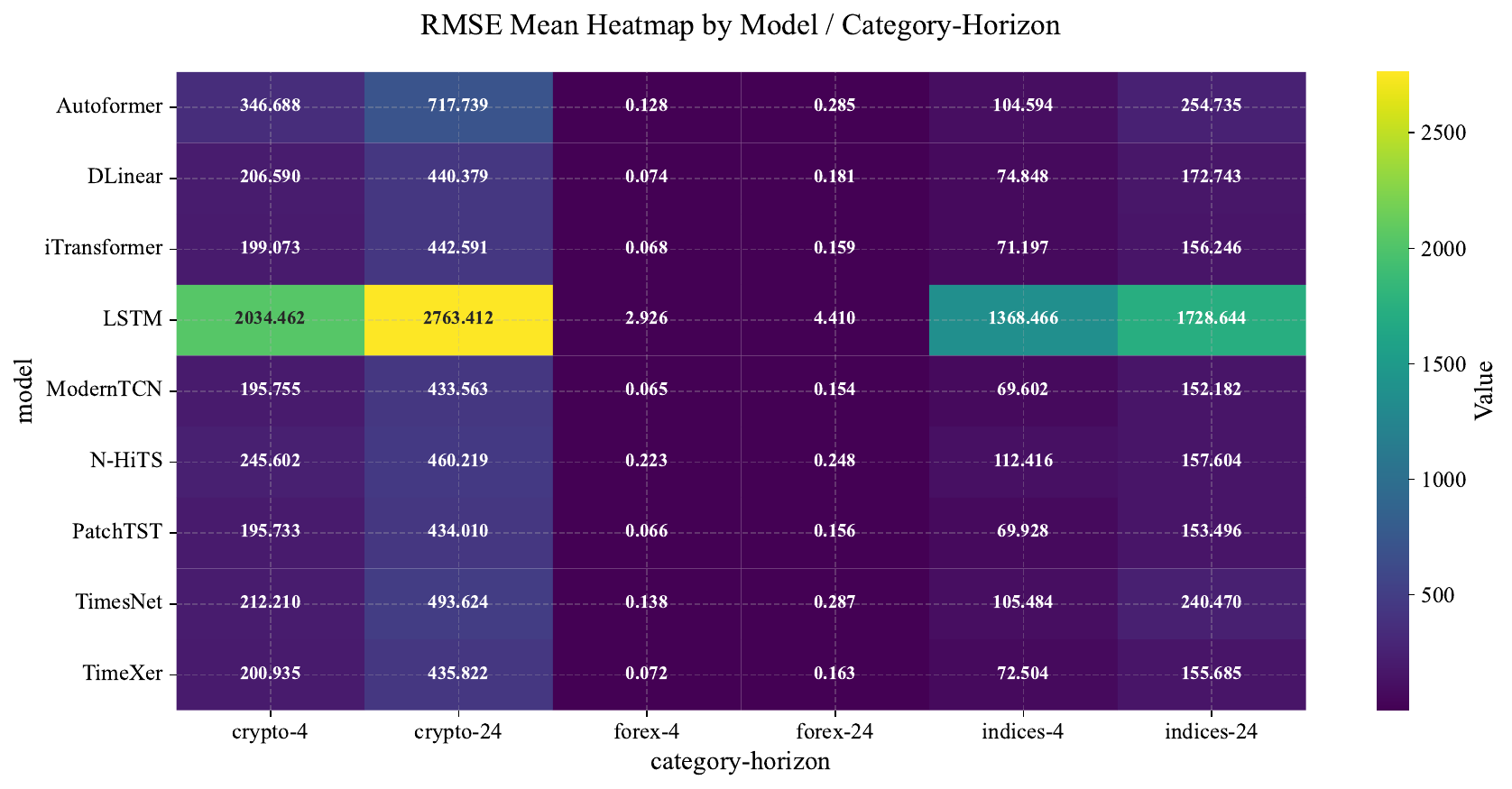}
  \caption{Global \rmse heatmap for all \nmodels architectures (including \lstm) across 24 evaluation points.  \lstm's extreme errors dominate the colour scale, which is why the no-\lstm variant (Figure~\ref{fig:global_heatmap}) is used as the primary figure in the main body.}
  \label{fig:app_global_heatmap_all}
\end{figure}

\begin{figure}[htbp]
  \centering
  \includegraphics[width=\textwidth]{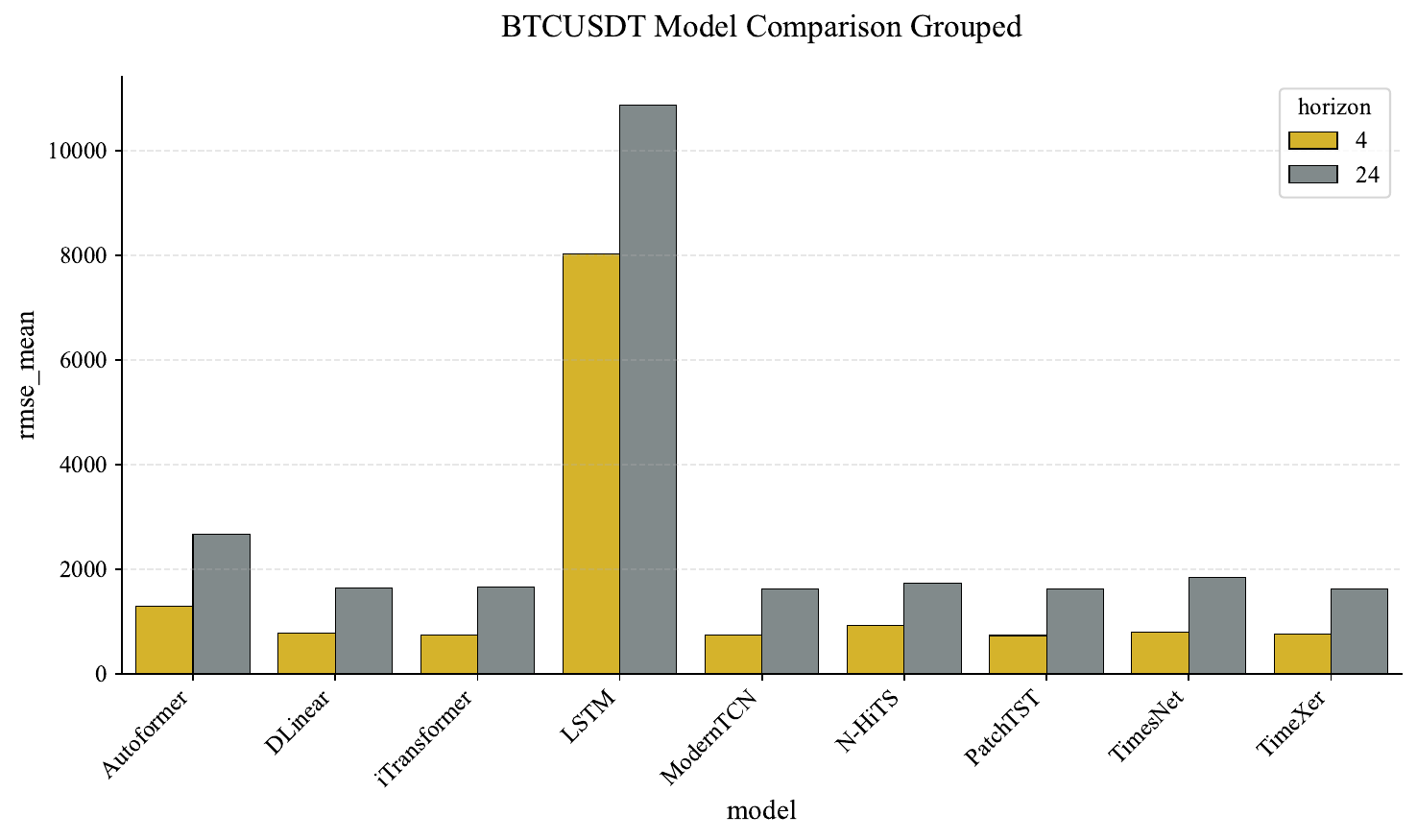}
  \caption{Cross-horizon \rmse comparison for all nine architectures across twelve assets, including \lstm.  The recurrent baseline illustrates the generational performance gap relative to modern time-series architectures.}
  \label{fig:app_asset_model_comparison_grouped_all}
\end{figure}

Figure~\ref{fig:app_global_heatmap_all} shows the global \rmse heatmap including all nine models; compare with Figure~\ref{fig:global_heatmap} for finer discrimination among modern architectures.  Figure~\ref{fig:app_asset_model_comparison_grouped_all} presents the cross-horizon comparison for the full nine-model set.  \lstm produces the highest errors across all assets and horizons, with its error floor exceeding even the lowest-performing modern architectures.

\subsection{Appendix: Efficiency Including LSTM Models}
\label{sec:app_efficiency_lstm}

\begin{figure}[htbp]
  \centering
  \begin{subfigure}[b]{0.48\textwidth}
    \includegraphics[width=\textwidth]{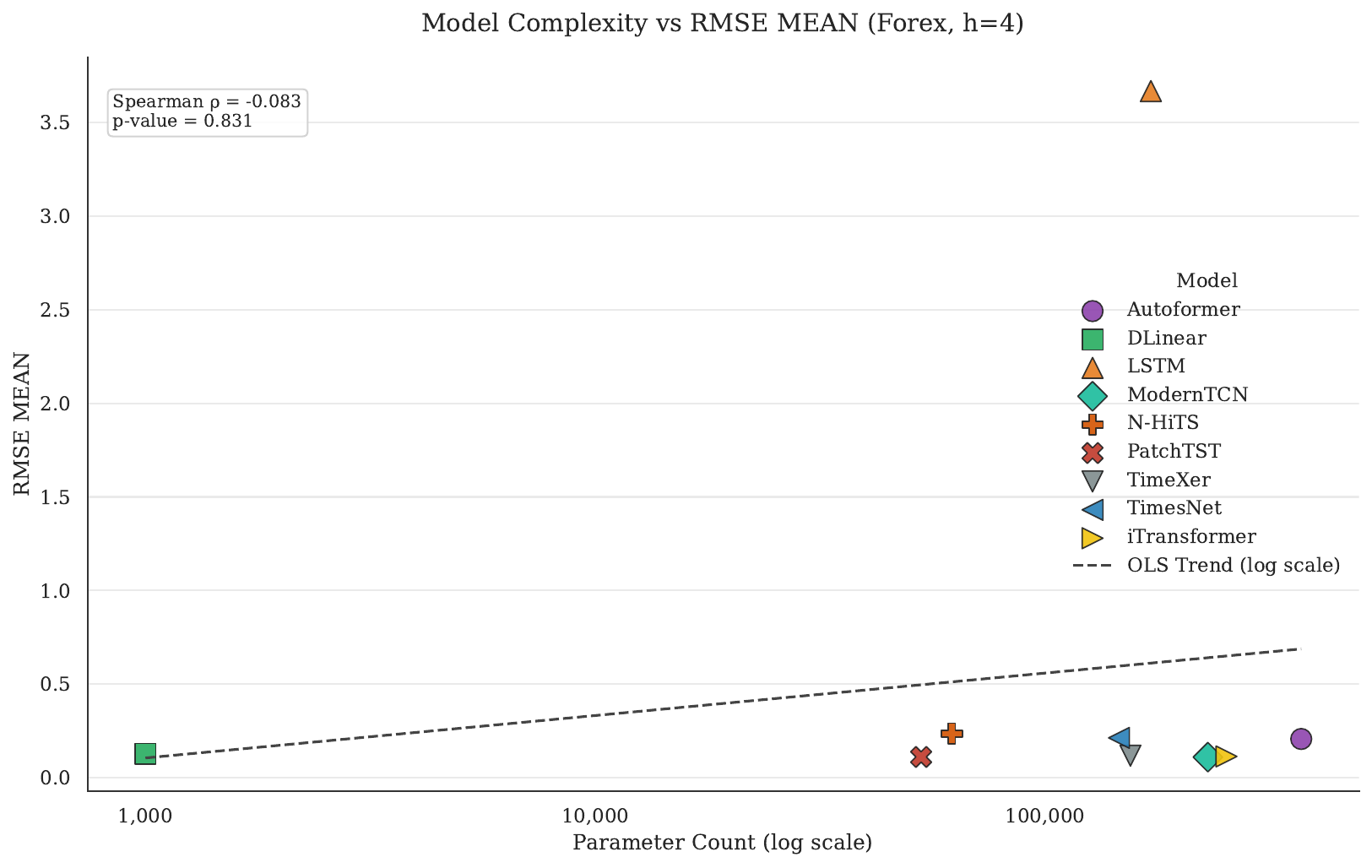}
    \caption{Horizon $h=4$.}
    \label{fig:app_complexity_h4}
  \end{subfigure}
  \hfill
  \begin{subfigure}[b]{0.48\textwidth}
    \includegraphics[width=\textwidth]{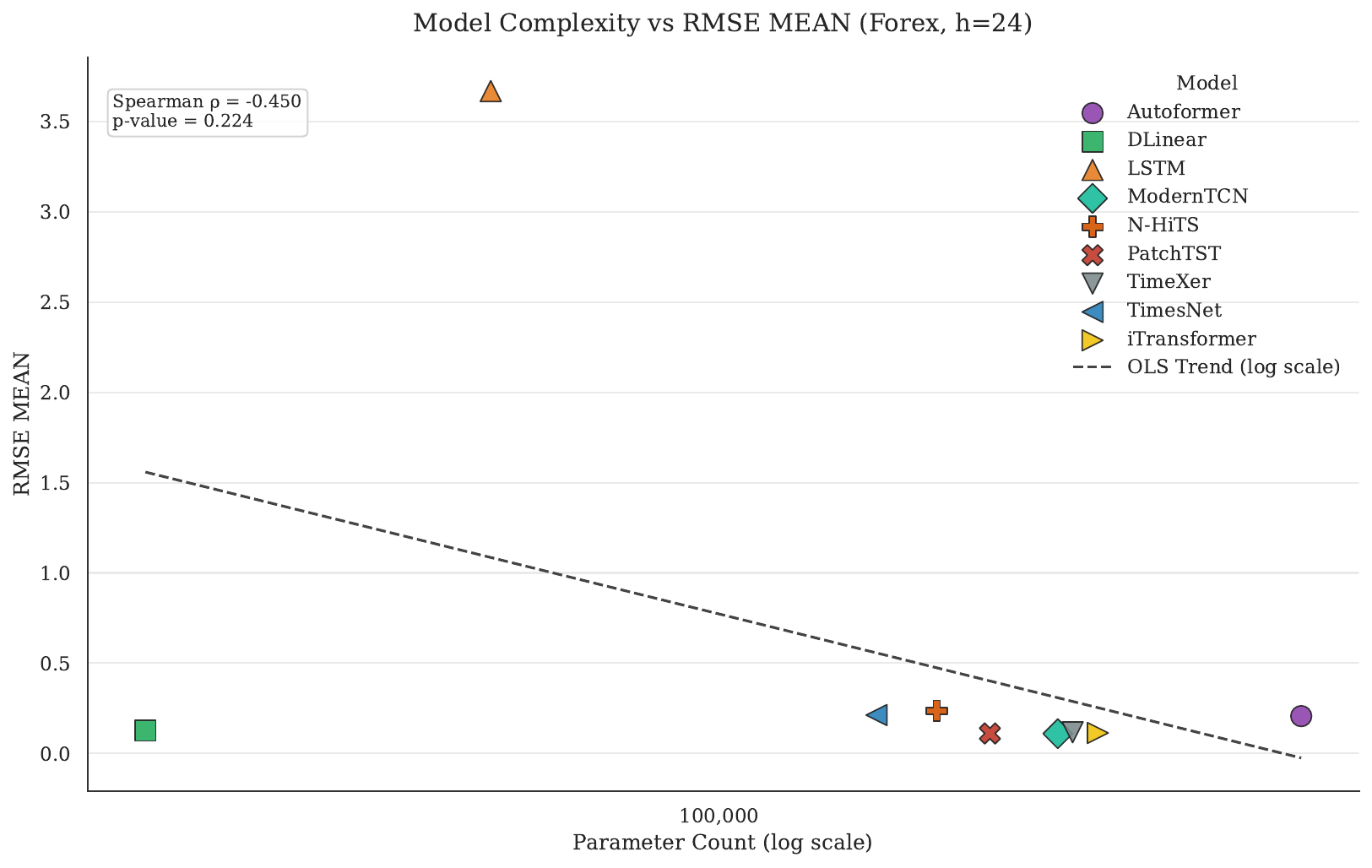}
    \caption{Horizon $h=24$.}
    \label{fig:app_complexity_h24}
  \end{subfigure}
  \caption{Extended complexity--performance relationship including \lstm.  The recurrent baseline lies far from the Pareto frontier defined by modern architectures.}
  \label{fig:app_complexity_full}
\end{figure}

While the main-body analysis (Section~\ref{sec:complexity}) focuses on modern architectures, Figure~\ref{fig:app_complexity_full} includes \lstm.  Despite a parameter budget ($\approx 172{,}000$) that is moderate in absolute terms, \lstm achieves the worst ranking at both horizons, reinforcing that modern architectural components provide far higher return on parameter investment than recurrent inductive biases for this task.

\subsection{Category Hierarchical Rankings}
\label{sec:app_category_rankings}

\begin{figure}[htbp]
  \centering
  \begin{subfigure}[b]{0.48\textwidth}
    \includegraphics[width=\textwidth]{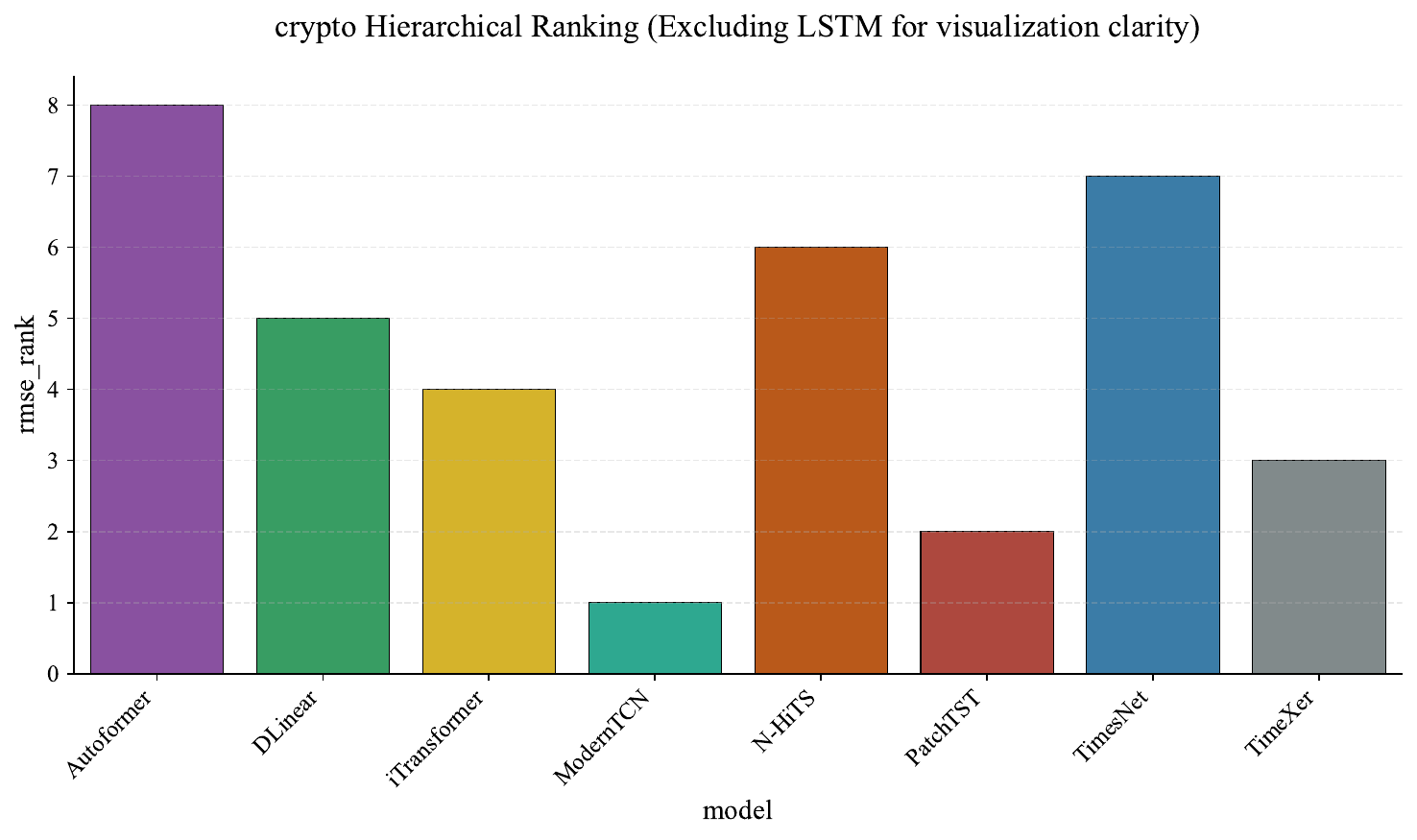}
    \caption{Cryptocurrency.}
    \label{fig:app_crypto_hier_ranking}
  \end{subfigure}
  \hfill
  \begin{subfigure}[b]{0.48\textwidth}
    \includegraphics[width=\textwidth]{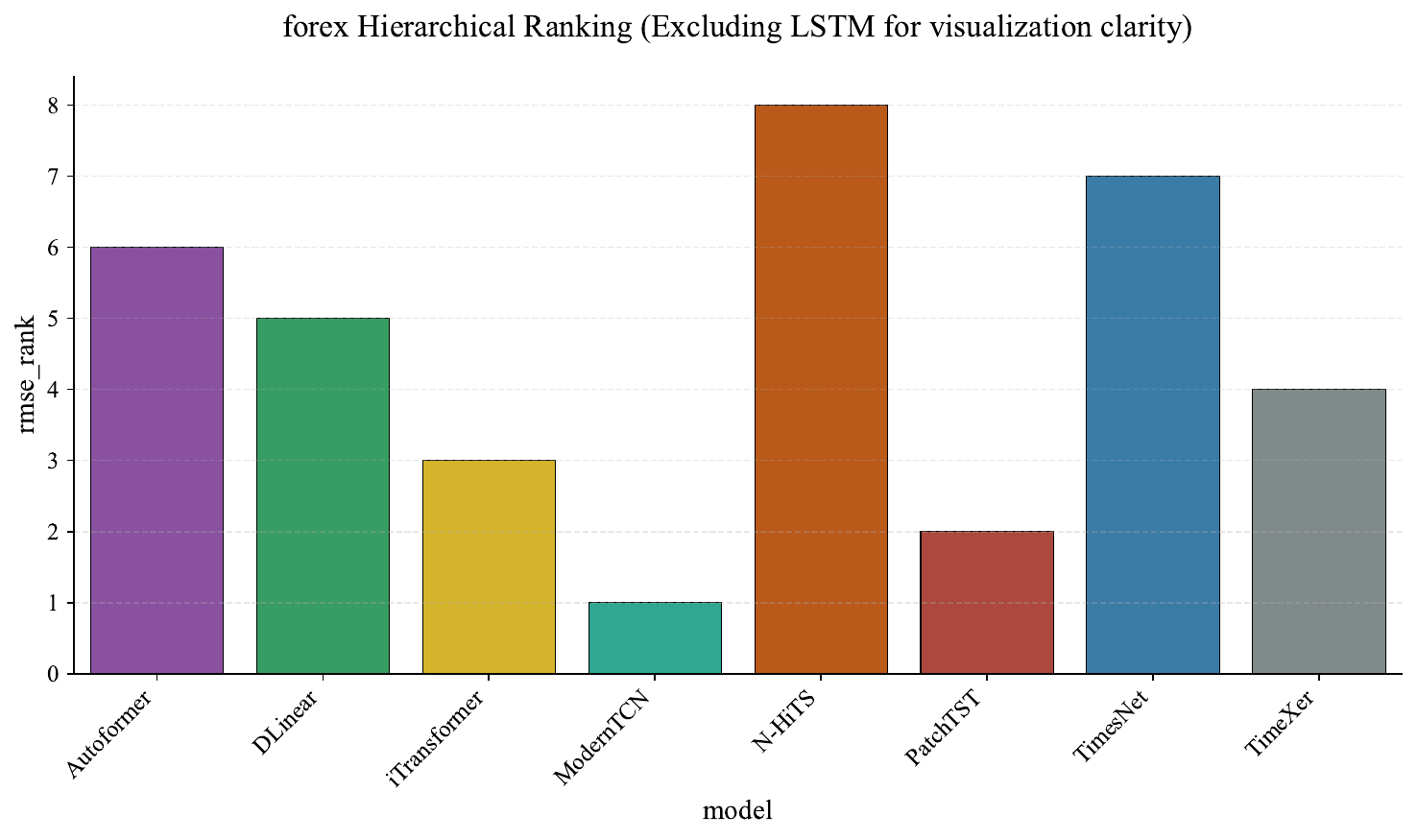}
    \caption{Forex.}
    \label{fig:app_forex_hier_ranking}
  \end{subfigure}
  \par\medskip
  \hfill
  \begin{subfigure}[b]{0.48\textwidth}
    \includegraphics[width=\textwidth]{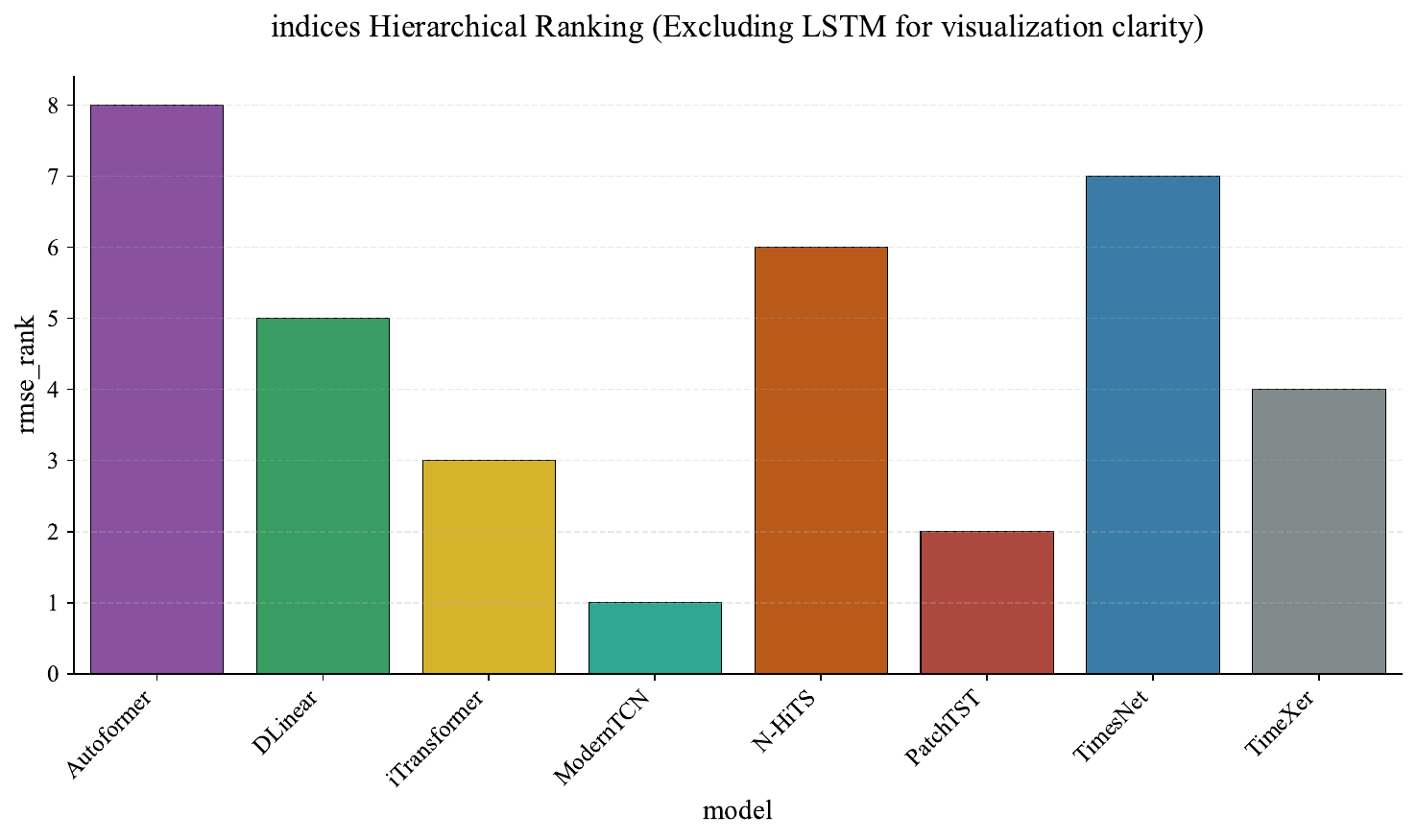}
    \caption{Equity Indices.}
    \label{fig:app_indices_hier_ranking}
  \end{subfigure}
  \hfill
  \caption{Category hierarchical ranking dendrograms for the eight modern architectures
           (\lstm excluded).  Branch lengths encode performance dissimilarity within
           each asset class.  \moderntcn and \patchtst are consistently co-clustered at
           the top of each dendrogram, confirming their joint dominance is stable
           across all three asset classes.}
  \label{fig:app_category_hierarchical_rankings}
\end{figure}

Figures~\ref{fig:app_crypto_hier_ranking}--\ref{fig:app_indices_hier_ranking} display
hierarchical ranking dendrograms for each asset class, grouping architectures by
performance similarity.  These complement the leaderboard tables by revealing clustering
structure: closely branched architectures perform similarly; widely separated branches
indicate consistent performance gaps.  The no-\lstm variant is shown.

\subsection{Category Performance Matrices}
\label{sec:app_category_matrices}

\begin{figure}[htbp]
  \centering
  \begin{subfigure}[b]{0.48\textwidth}
    \includegraphics[width=\textwidth]{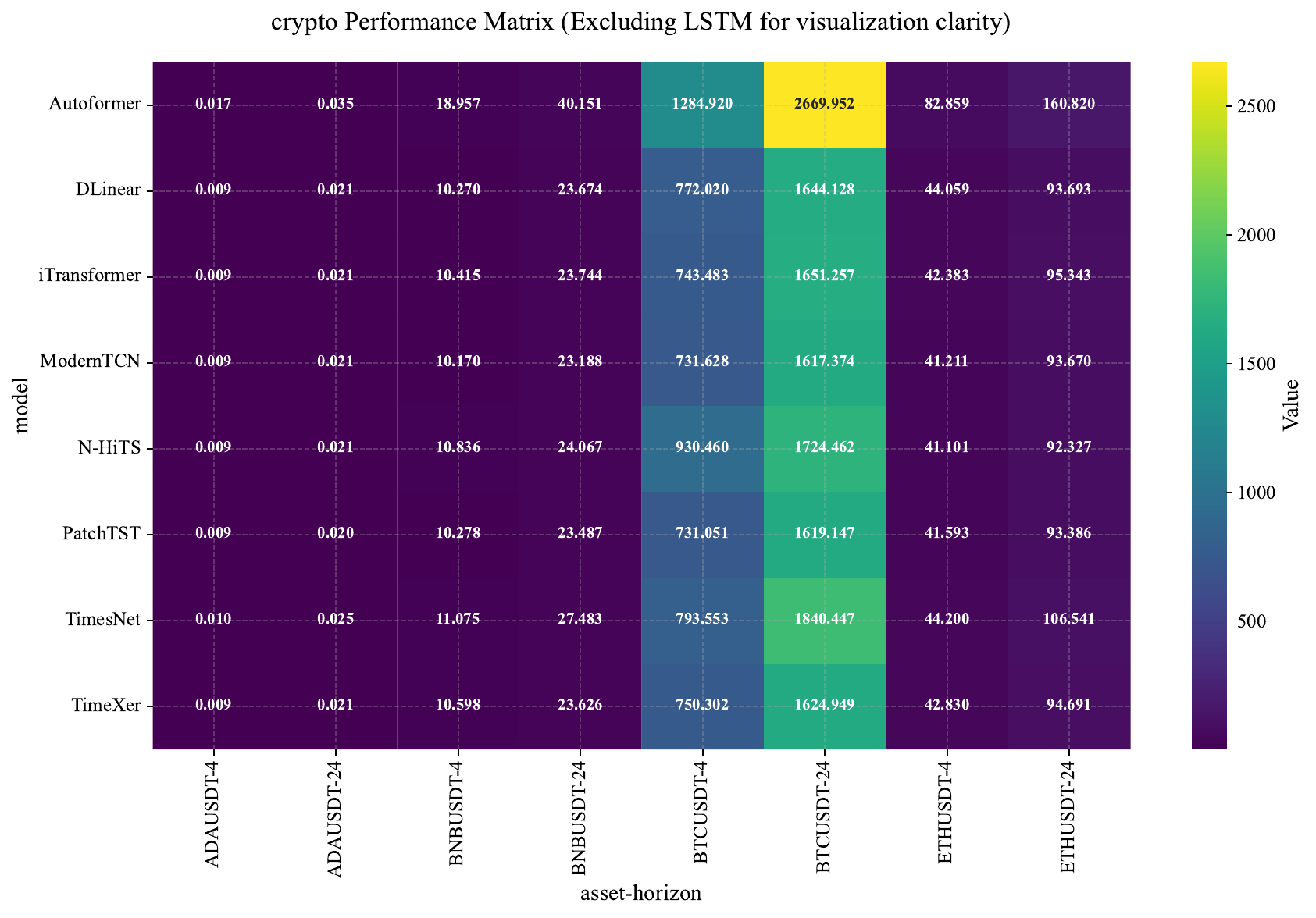}
    \caption{Cryptocurrency.}
    \label{fig:app_crypto_perf_matrix}
  \end{subfigure}
  \hfill
  \begin{subfigure}[b]{0.48\textwidth}
    \includegraphics[width=\textwidth]{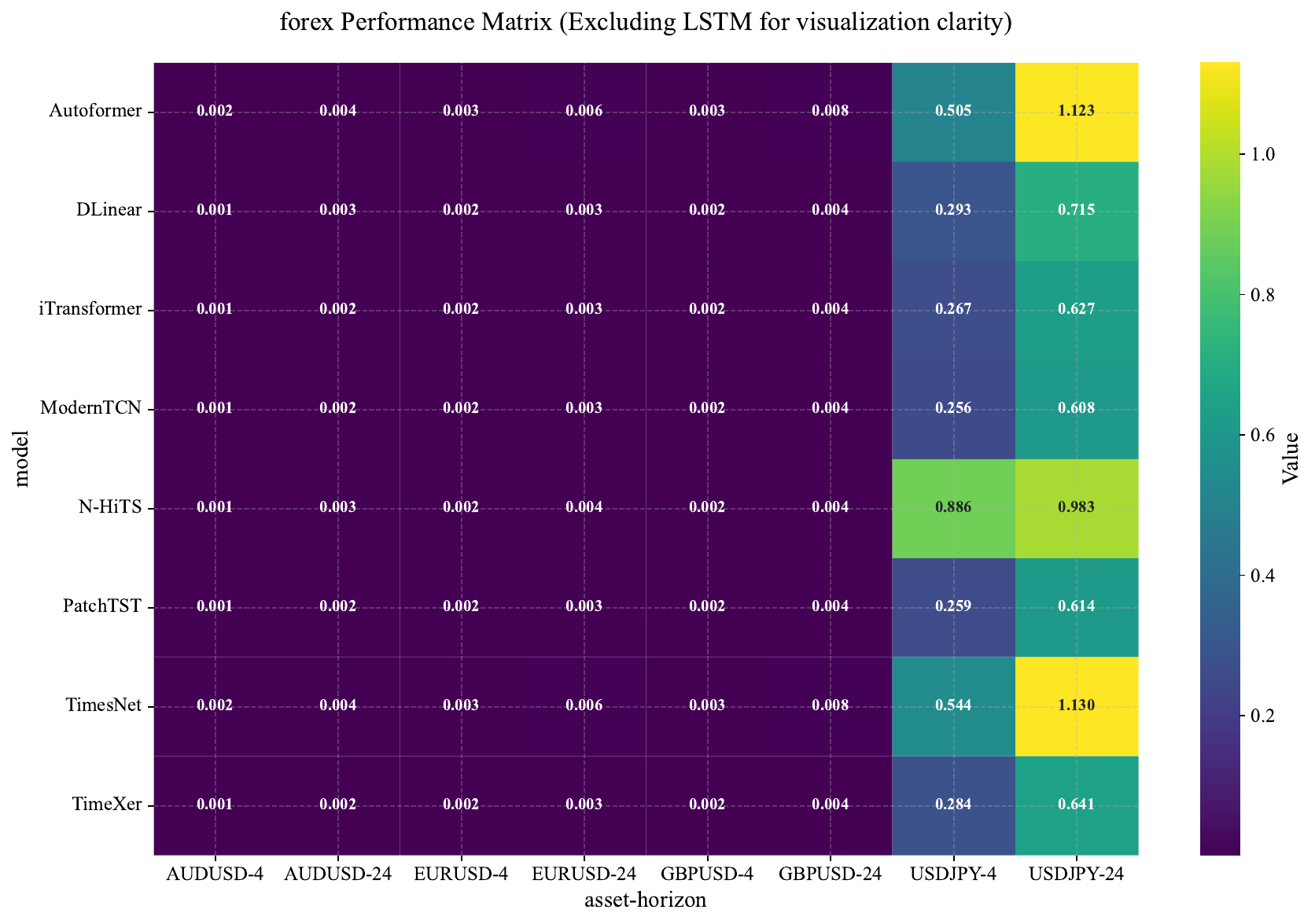}
    \caption{Forex.}
    \label{fig:app_forex_perf_matrix}
  \end{subfigure}

  \par\medskip
  \centering
  \begin{subfigure}[b]{0.65\textwidth}
    \includegraphics[width=\textwidth]{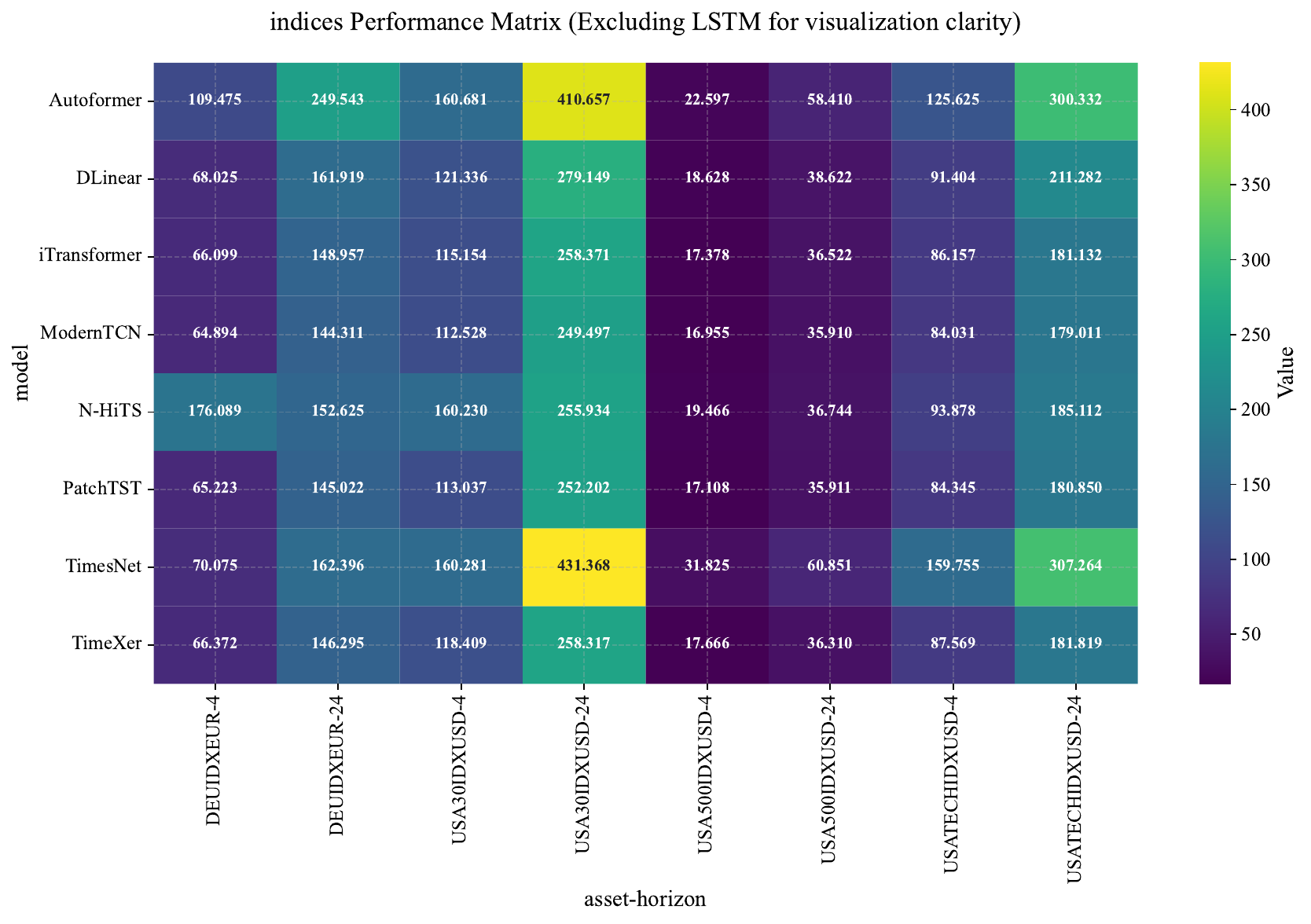}
    \caption{Equity Indices.}
    \label{fig:app_indices_perf_matrix}
  \end{subfigure}

  \caption{Category performance matrices for the eight modern architectures (\lstm excluded).
           Cell intensity encodes normalised \rmse relative to the best model per asset
           (darker = worse).  \moderntcn and \patchtst occupy the lightest cells
           consistently; \autoformer and \timesnet occupy the darkest, confirming the
           three-tier structure observed in the global leaderboard.}
  \label{fig:app_category_performance_matrices}
\end{figure}

Figures~\ref{fig:app_crypto_perf_matrix}--\ref{fig:app_indices_perf_matrix} show
category performance matrices displaying normalised \rmse values for each model--asset
combination within each class, providing a complementary view to the overall leaderboard.
Rows represent models and columns represent assets; cell intensity encodes the normalised
error relative to the best model on that asset.

\section{Methodological Details}
\label{sec:app_methods}

\subsection{Hyperparameter Search Spaces}
\label{sec:app_hpo_spaces}

Table~\ref{tab:hpo_search_spaces} in the main text summarises the HPO search dimensions for all
models.  The key varied parameters by model family are as follows.

\begin{itemize}[nosep]
  \item \textbf{Transformer models} (\autoformer, \patchtst, \itransformer, \timexer): model dimension,
        attention heads, encoder depth, feedforward width, dropout, learning rate, batch size, and
        architecture-specific parameters (patch configuration, moving-average windows).
  \item \textbf{MLP models} (\dlinear, \nhits): kernel size and decomposition mode (\dlinear); block
        count, hidden-layer width, depth, and pooling scales (\nhits).
  \item \textbf{CNN models} (\timesnet, \moderntcn): channel and depth parameters, dominant-frequency
        or patch-size settings, and model-specific design choices.
  \item \textbf{RNN model} (\lstm): hidden-state size, layer count, projection width, and
        directionality (unidirectional or bidirectional).
\end{itemize}

All shared training hyperparameters (learning-rate range, batch-size options) were held constant
across all models to ensure no implicit tuning advantage.  Complete YAML search-space definitions
for each model are available in the Online Supplementary Materials.

\subsection{Training Protocol}
\label{sec:app_training}

Final models were trained across three random seeds (123, 456, 789) using the frozen
best-hyperparameter configurations identified during the HPO stage; no retuning was performed.
Checkpoints were saved at each epoch to support full resumability from the last completed epoch.
All randomness sources (\texttt{random}, NumPy, PyTorch CPU and CUDA) were seeded identically per
run, with \texttt{cudnn.deterministic=True} and \texttt{cudnn.benchmark=False} enforced throughout.
Note: \texttt{torch.use\_deterministic\_algorithms(True)} was \emph{not} enabled; minor
hardware-dependent floating-point variation is possible across GPU models (see
Section~\ref{sec:determinism}).  Full training logs are archived in the Online
Supplementary Materials.

\subsection{Robustness Checks}
\label{sec:app_robustness}

\begin{figure}[htbp]
  \centering
  \begin{subfigure}[b]{0.48\textwidth}
    \includegraphics[width=\textwidth]{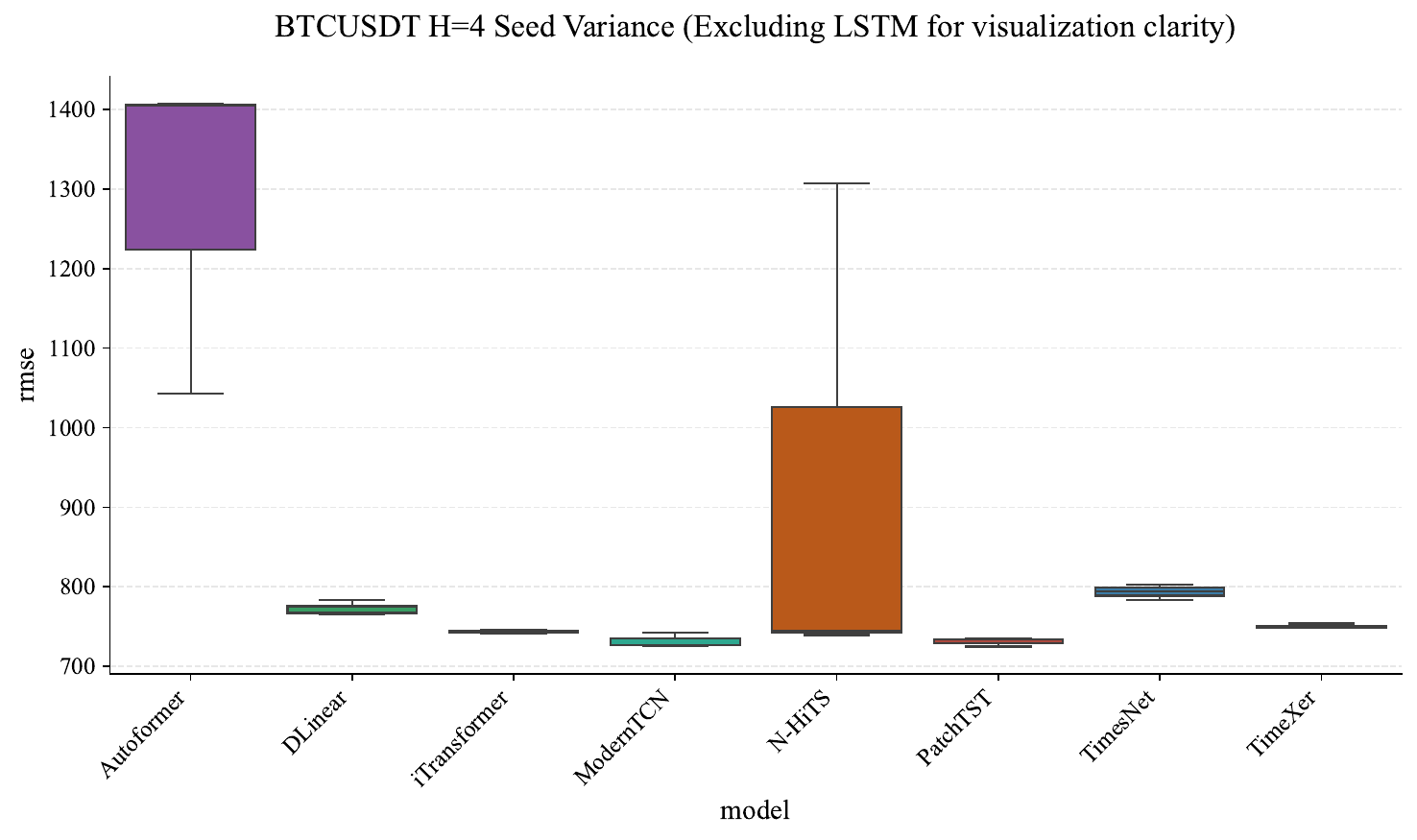}
    \caption{BTC/USDT.}
    \label{fig:app_seed_boxplot_btc_h4}
  \end{subfigure}
  \hfill
  \begin{subfigure}[b]{0.48\textwidth}
    \includegraphics[width=\textwidth]{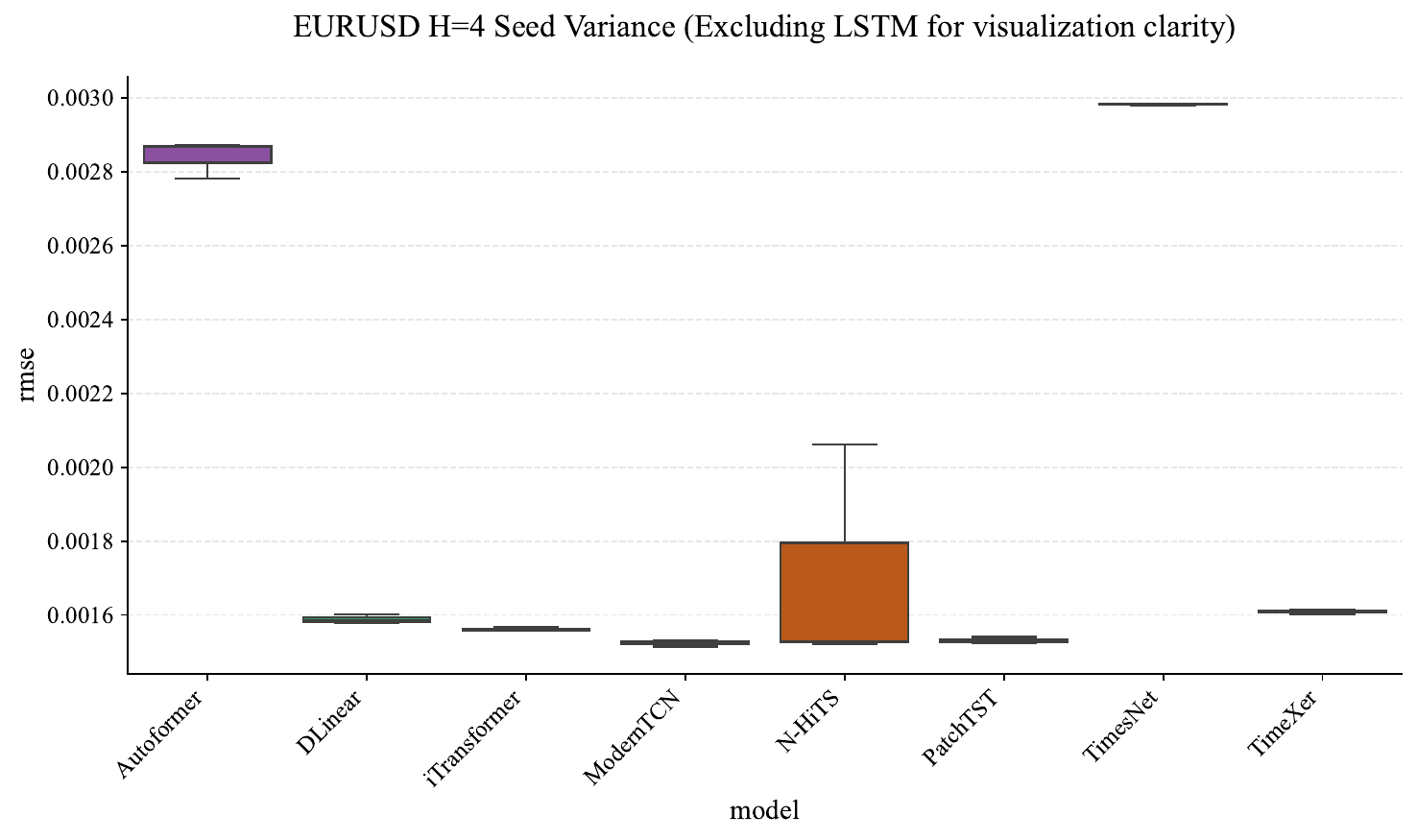}
    \caption{EUR/USD.}
    \label{fig:app_seed_boxplot_eur_h4}
  \end{subfigure}
  \par\medskip
  \centering
  \begin{subfigure}[b]{0.65\textwidth}
    \includegraphics[width=\textwidth]{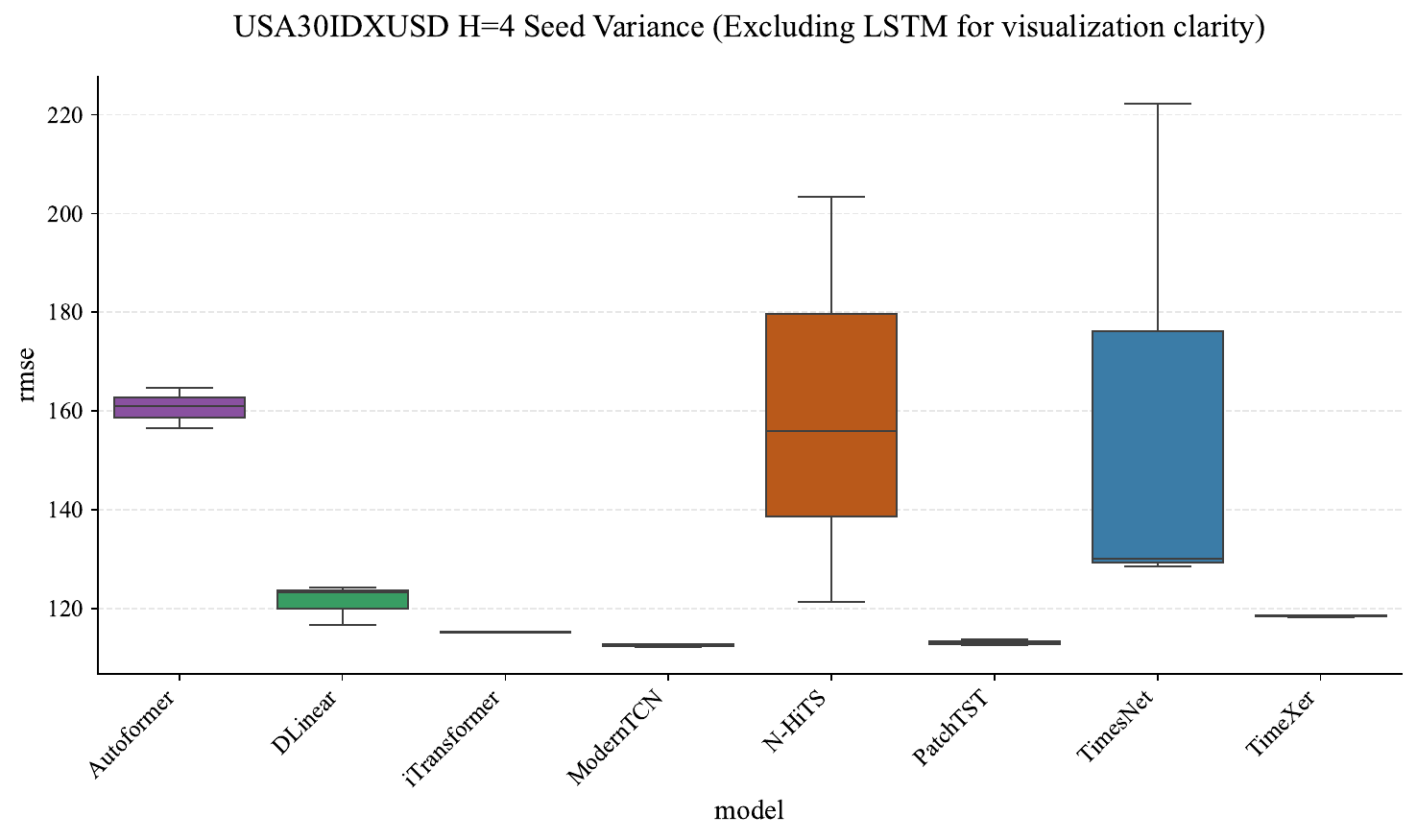}
    \caption{Dow Jones.}
    \label{fig:app_seed_boxplot_usa30_h4}
  \end{subfigure}
  \caption{Seed-variance box plots at $\hfour$ for the three representative assets
           (\lstm excluded).  Each box spans the interquartile range of \rmse values
           across seeds~123, 456, and 789.  Negligible box widths relative to
           inter-model differences confirm that seed accounts for $< 0.1\%$ of total forecast variance.}
  \label{fig:app_seed_boxplot_h4}
\end{figure}

\begin{figure}[htbp]
  \centering
  \begin{subfigure}[b]{0.48\textwidth}
    \includegraphics[width=\textwidth]{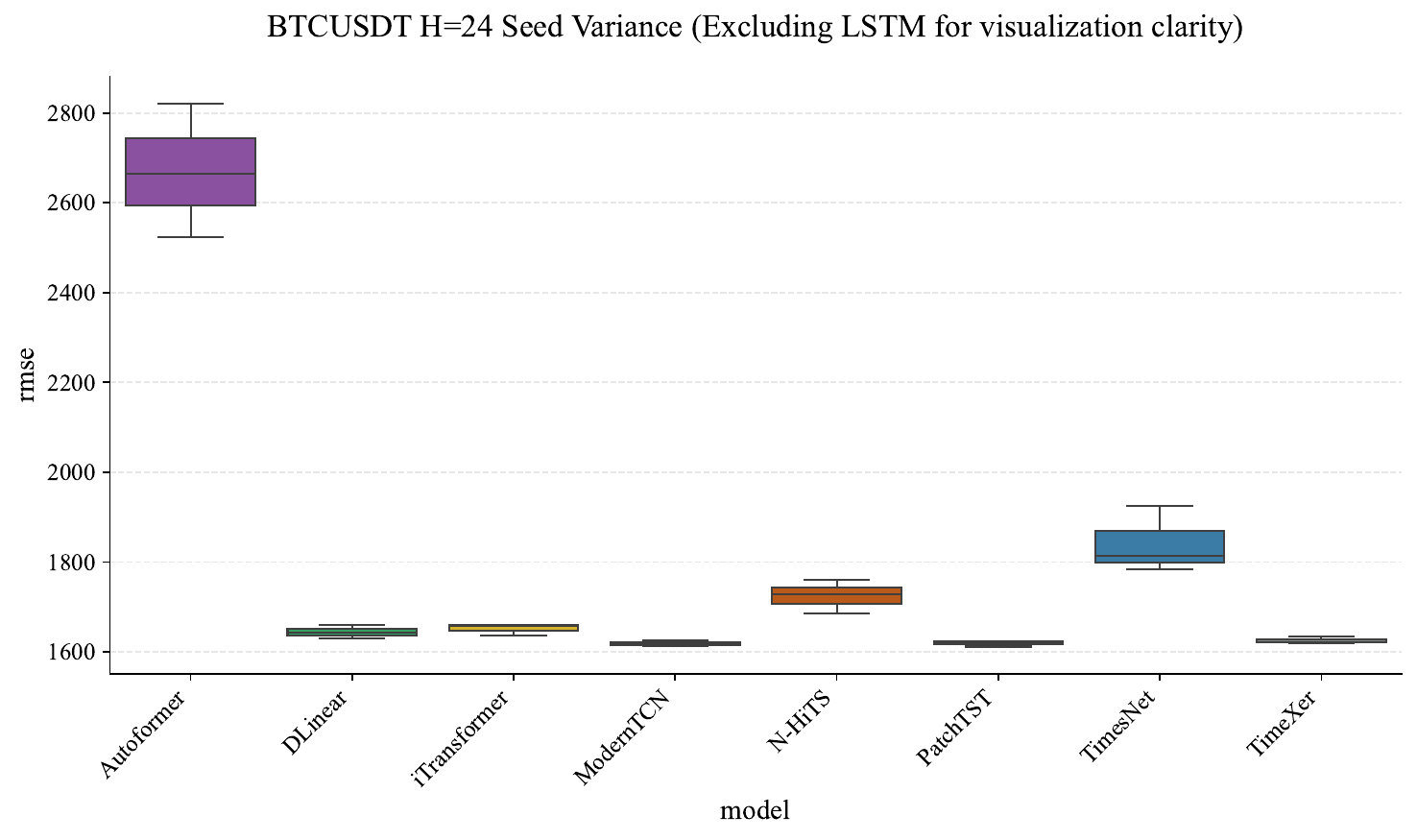}
    \caption{BTC/USDT.}
    \label{fig:app_seed_boxplot_btc_h24}
  \end{subfigure}
  \hfill
  \begin{subfigure}[b]{0.48\textwidth}
    \includegraphics[width=\textwidth]{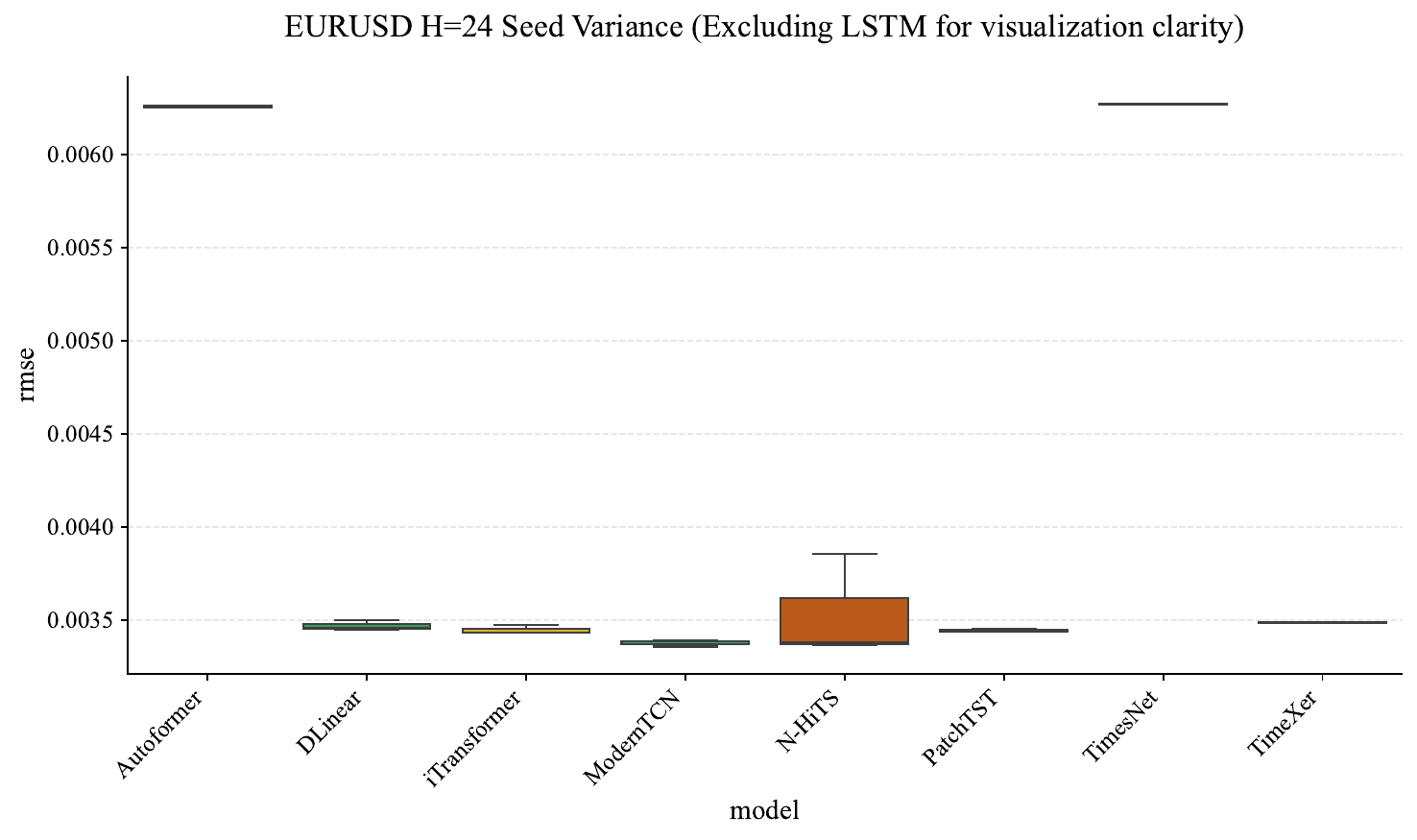}
    \caption{EUR/USD.}
    \label{fig:app_seed_boxplot_eur_h24}
  \end{subfigure}
  \par\medskip
  \centering
  \begin{subfigure}[b]{0.65\textwidth}
    \includegraphics[width=\textwidth]{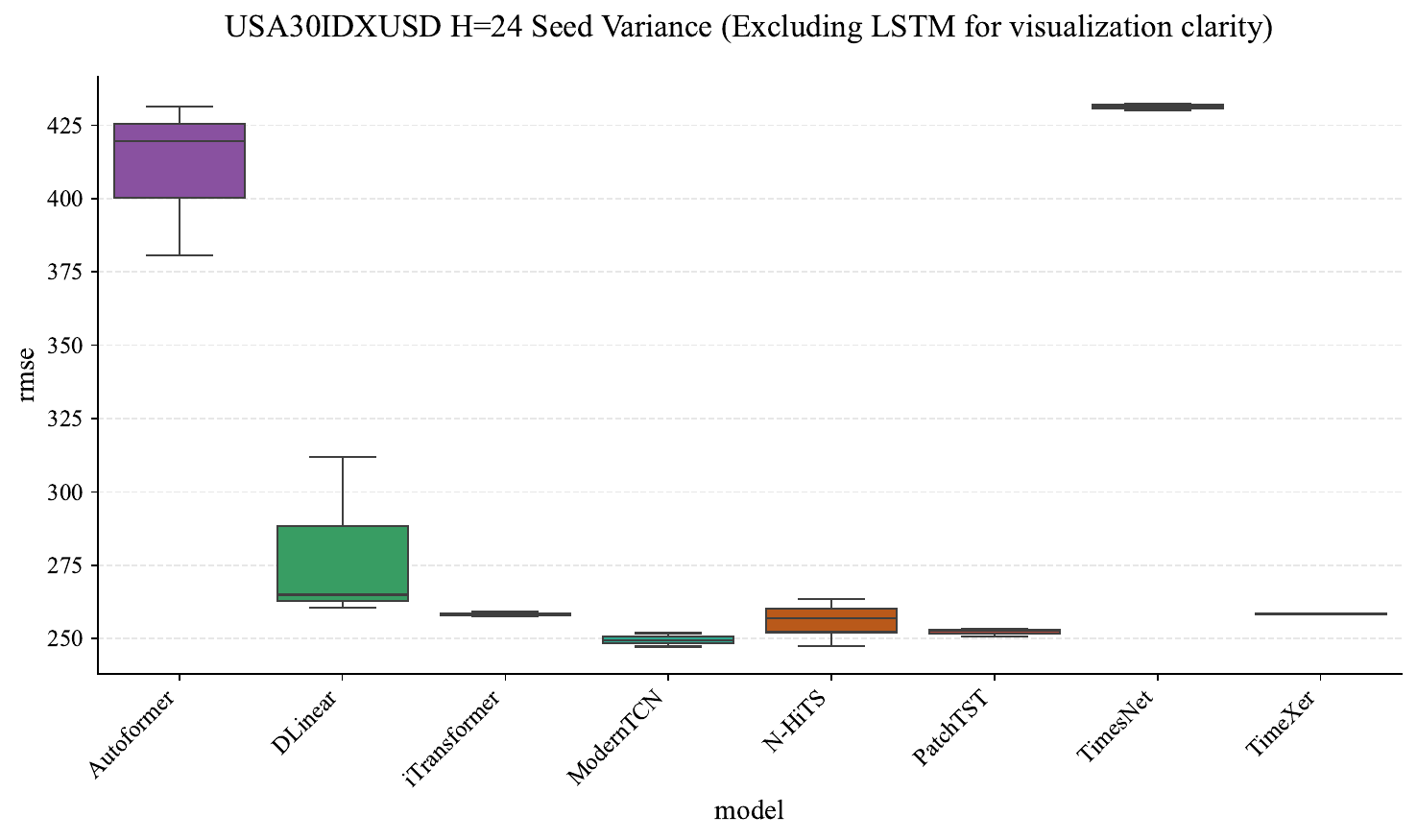}
    \caption{Dow Jones.}
    \label{fig:app_seed_boxplot_usa30_h24}
  \end{subfigure}
  \caption{Seed-variance box plots at $\htwentyfour$ for the three representative assets
           (\lstm excluded).  The pattern mirrors $\hfour$: box widths remain negligible
           at the longer horizon, confirming H3 holds at both forecast depths.}
  \label{fig:app_seed_boxplot_h24}
\end{figure}

\begin{figure}[htbp]
  \centering
  \begin{subfigure}[b]{0.48\textwidth}
    \includegraphics[width=\textwidth]{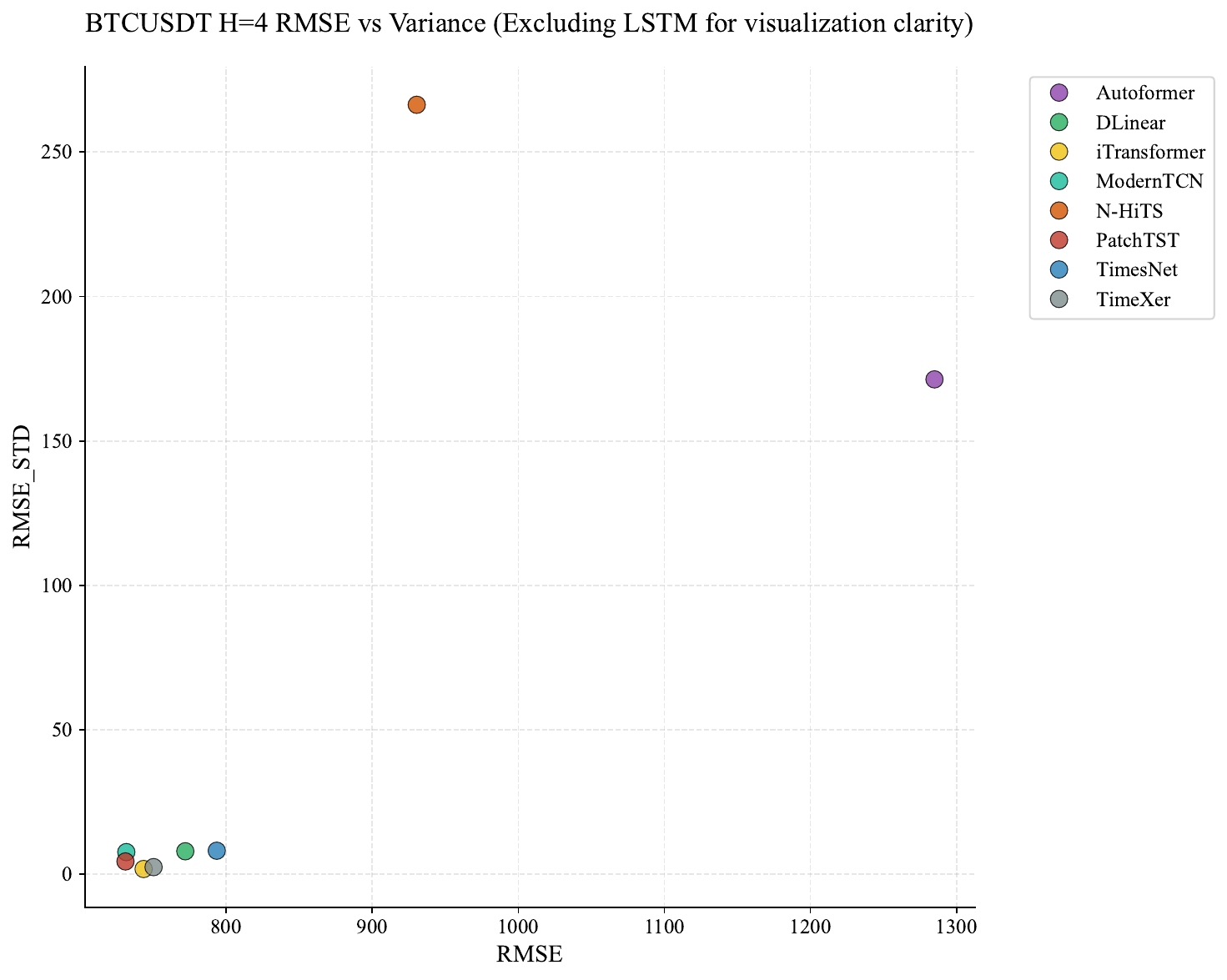}
    \caption{BTC/USDT.}
    \label{fig:app_scatter_btc_h4}
  \end{subfigure}
  \hfill
  \begin{subfigure}[b]{0.48\textwidth}
    \includegraphics[width=\textwidth]{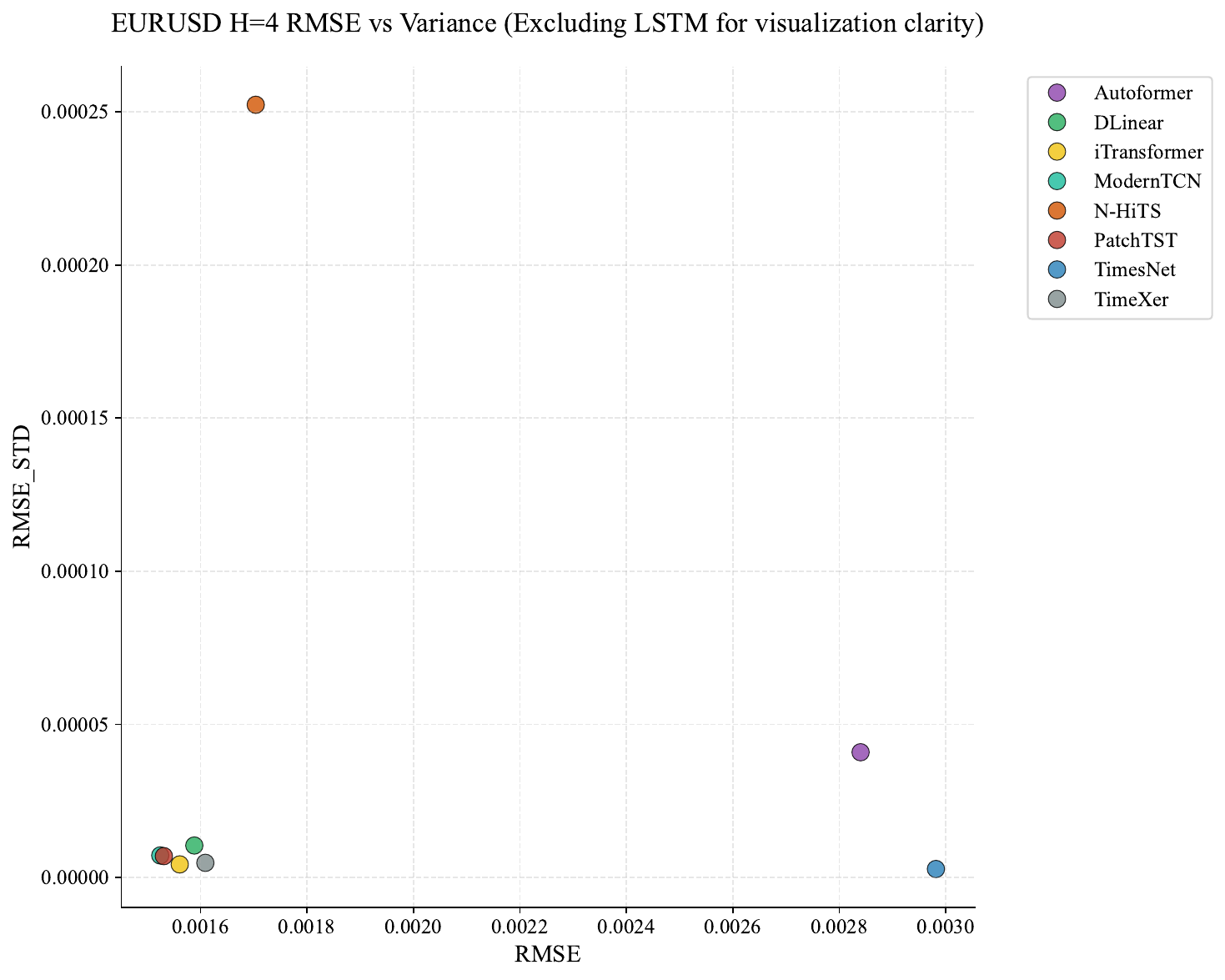}
    \caption{EUR/USD.}
    \label{fig:app_scatter_eur_h4}
  \end{subfigure}
  \par\medskip
  \centering
  \begin{subfigure}[b]{0.65\textwidth}
    \includegraphics[width=\textwidth]{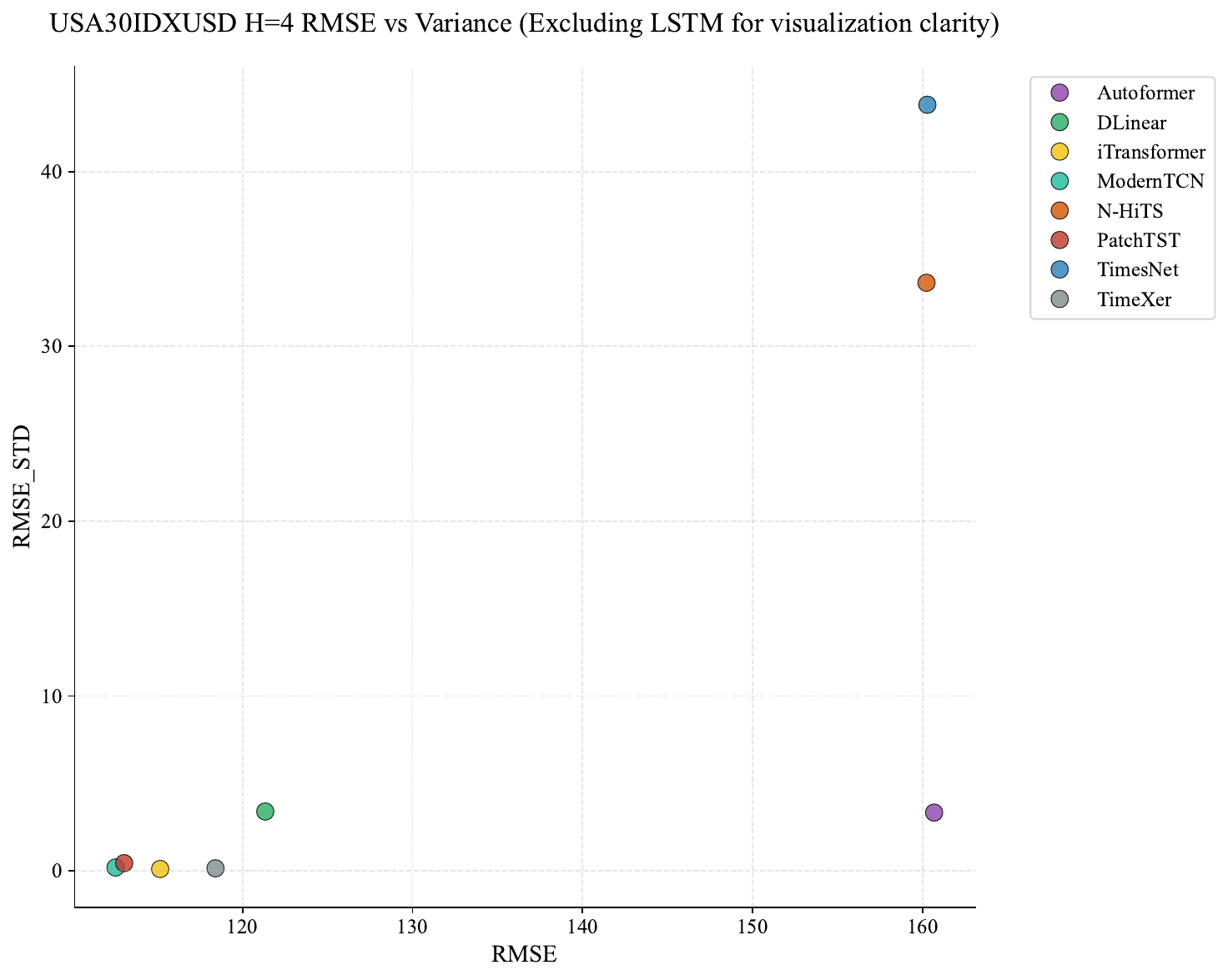}
    \caption{Dow Jones.}
    \label{fig:app_scatter_usa30_h4}
  \end{subfigure}
  \caption{Mean \rmse vs.\ seed variance scatter at $\hfour$ for eight modern architectures.
           Each point represents one model; the horizontal axis shows mean \rmse across
           seeds, and the vertical axis shows the across-seed variance.  High-performing
           models (low mean \rmse) also exhibit low seed variance, confirming that
           architectural quality and robustness co-vary positively.}
  \label{fig:app_scatter_seed_h4}
\end{figure}

\begin{figure}[htbp]
  \centering
  \begin{subfigure}[b]{0.48\textwidth}
    \includegraphics[width=\textwidth]{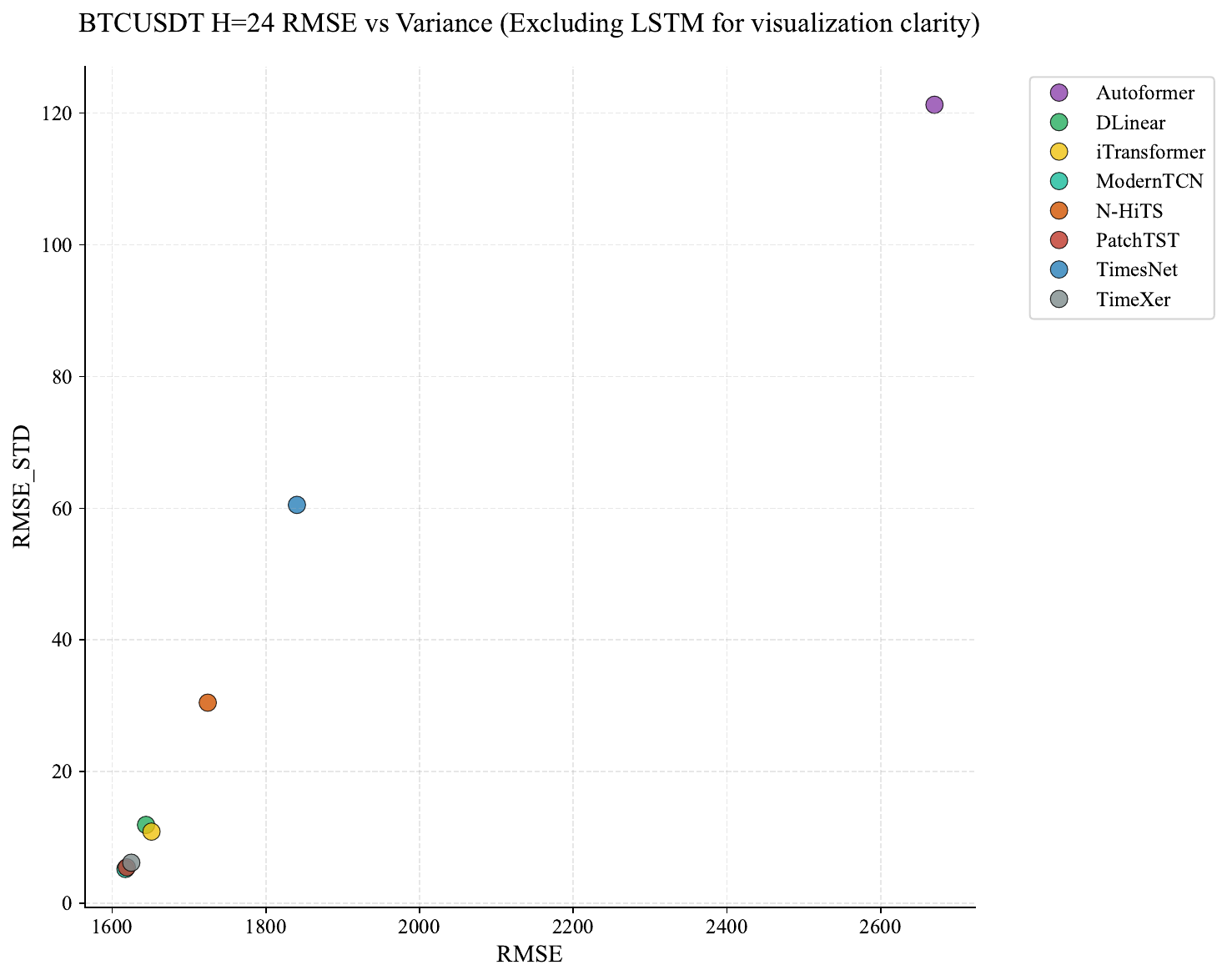}
    \caption{BTC/USDT.}
    \label{fig:app_scatter_btc_h24}
  \end{subfigure}
  \hfill
  \begin{subfigure}[b]{0.48\textwidth}
    \includegraphics[width=\textwidth]{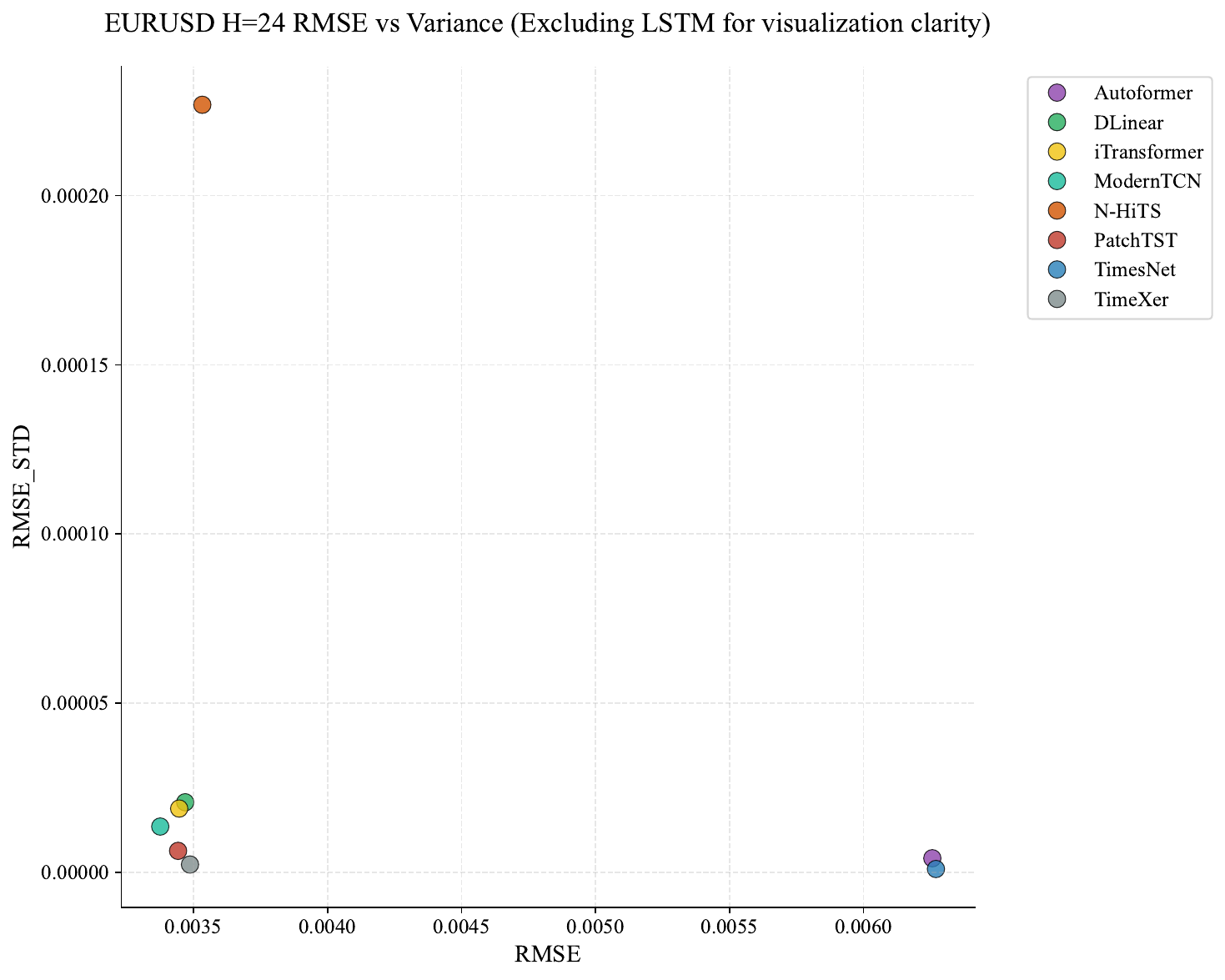}
    \caption{EUR/USD.}
    \label{fig:app_scatter_eur_h24}
  \end{subfigure}
  \par\medskip
  \centering
  \begin{subfigure}[b]{0.65\textwidth}
    \includegraphics[width=\textwidth]{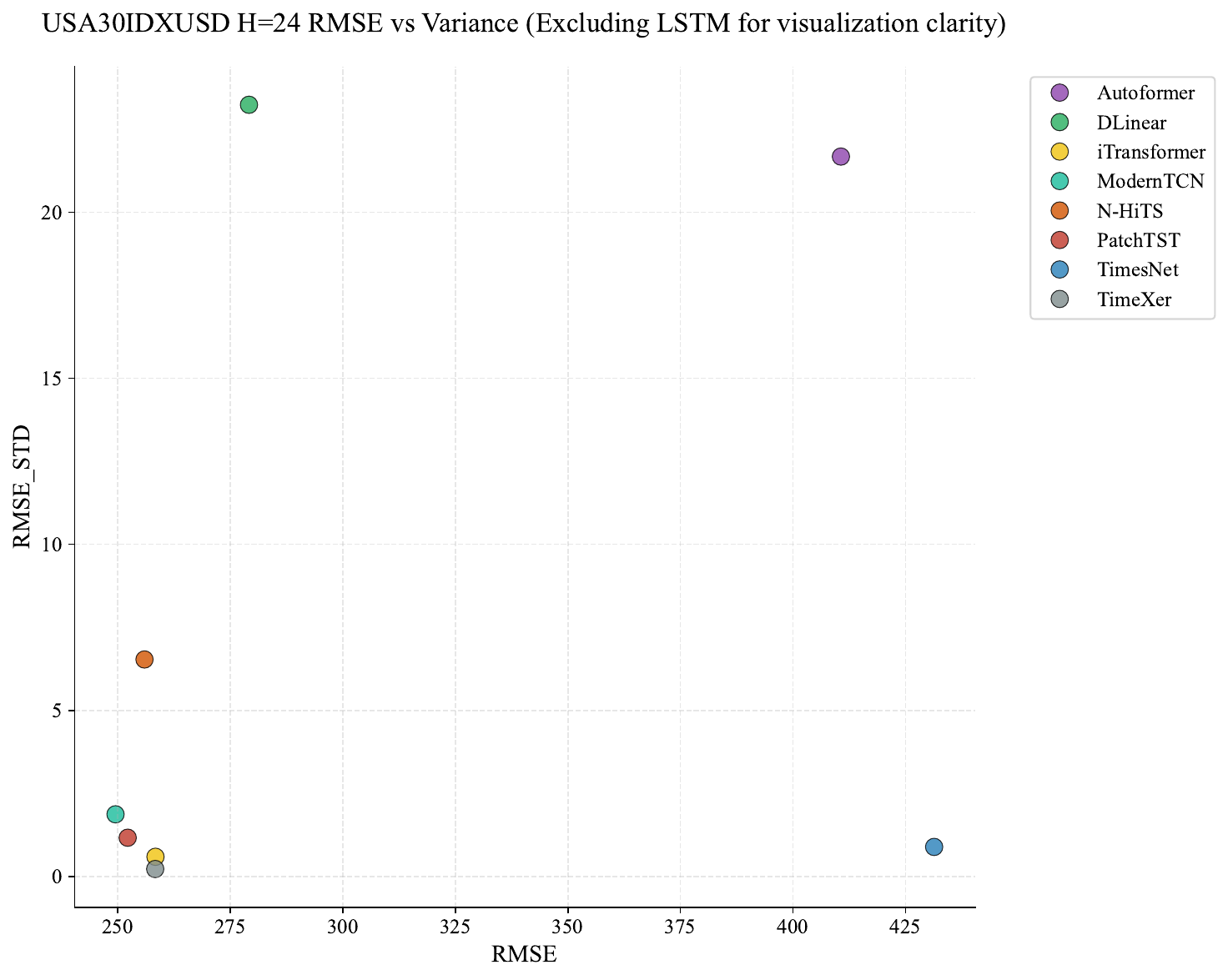}
    \caption{Dow Jones.}
    \label{fig:app_scatter_usa30_h24}
  \end{subfigure}
  \caption{Mean \rmse vs.\ seed variance scatter at $\htwentyfour$.  The positive
           co-variation between performance and seed stability is maintained
           at the longer horizon.}
  \label{fig:app_scatter_seed_h24}
\end{figure}

Seed-variance box plots (Figures~\ref{fig:app_seed_boxplot_h4}--\ref{fig:app_seed_boxplot_h24})
and \rmse-versus-seed-variance scatter plots
(Figures~\ref{fig:app_scatter_seed_h4}--\ref{fig:app_scatter_seed_h24})
confirm that inter-seed variance is negligible relative to inter-model variance across
all asset classes and horizons.  Results for BTC/USDT, EUR/USD, and Dow Jones are shown;
remaining assets exhibit qualitatively identical patterns and are archived in the Online
Supplementary Materials.

\subsection{Detailed Results and Figures}
\label{sec:app_repo}

While the main paper and this appendix present key empirical findings and summary visualisations, the full set of results from all 918 experimental runs is hosted in the project repository. This comprehensive archive ensures full transparency and supports independent verification of all reported metrics. The repository is available at:

\begin{center}
    \url{https://github.com/NabeelAhmad9/compare_forecasting_models}
\end{center}

The following content is available in the repository:
\begin{itemize}
    \item \textbf{Benchmark Results}: Complete \rmse, \mae, and \da scores for all 108 model--asset--horizon combinations across three seeds.
    \item \textbf{Figures}: High-resolution visualisations for all instruments, including per-seed actual vs.\ predicted plots.
    \item \textbf{Intermediate Outputs}: Full HPO trial logs, saved models and frozen best-hyperparameter configurations.
    \item \textbf{Training Artefacts}: All 648 trained model checkpoints and corresponding per-epoch CSV training logs.
    \item \textbf{Statistical Validation}: Raw outputs for all Friedman, \nem, and variance decomposition tests.
    \item \textbf{Full Project Code}: All scripts for models, training routines, benchmarking, evaluation, and utilities.
    \item \textbf{Notebooks}: Jupyter notebooks for running experiments, reproducing figures, and exploring intermediate results.
  \end{itemize}

\end{document}